\documentclass[aip,preprint]{revtex4-1}

\usepackage{graphicx}  
\usepackage{amsmath}
\usepackage{amssymb}
\usepackage{bm} 
\usepackage[english]{babel}
\usepackage{braket} 
\usepackage{pifont}
\usepackage{xcolor}

\usepackage{ulem}
\usepackage{hyperref}

\newcommand{\mvec}[1]{\bm{#1}}

\begin{document}

\title{Octopus, a computational framework for exploring light-driven phenomena and quantum dynamics in extended and finite systems}

\author{Nicolas Tancogne-Dejean}
\email{nicolas.tancogne-dejean@mpsd.mpg.de}
\affiliation{Max Planck Institute for the Structure and Dynamics of Matter, Luruper Chaussee 149, D-22761 Hamburg, Germany}

\author{Micael J.~T.~Oliveira}
\email{micael.oliveira@mpsd.mpg.de}
\affiliation{Max Planck Institute for the Structure and Dynamics of Matter, Luruper Chaussee 149, D-22761 Hamburg, Germany}

\author{Xavier Andrade}
\affiliation{Quantum Simulations Group, Lawrence Livermore National Laboratory, Livermore, California 94551, USA}

\author{Heiko Appel}
\affiliation{Max Planck Institute for the Structure and Dynamics of Matter, Luruper Chaussee 149, D-22761 Hamburg, Germany}

\author{Carlos H.~Borca}
\affiliation{Quantum Simulations Group, Lawrence Livermore National Laboratory, Livermore, California 94551, USA}

\author{Guillaume Le Breton}
\affiliation{D\'{e}partement de Physique, \'Ecole Normale Sup\'erieure de Lyon,  46 All\'ee d'Italie, Lyon Cedex 07, France}

\author{Florian Buchholz}
\affiliation{Max Planck Institute for the Structure and Dynamics of Matter, Luruper Chaussee 149, D-22761 Hamburg, Germany}

\author{Alberto Castro}
\affiliation{Institute for Biocomputation and Physics of Complex Systems, University of Zaragoza, Calle Mariano Esquillor, 50018 Zaragoza, Spain}
\affiliation{ARAID Foundation, Avda. de Ranillas 1-D, 50018 Zaragoza, Spain}

\author{Stefano Corni}
\affiliation{Dipartimento di Scienze Chimiche, Universit\`{a} degli studi di Padova, via F. Marzolo 1, 35131 Padova, Italy}
\affiliation{CNR -- Istituto Nanoscienze, via Campi 213a, 41125 Modena, Italy}

\author{Alfredo A. Correa}
\affiliation{Quantum Simulations Group, Lawrence Livermore National Laboratory, Livermore, California 94551, USA}

\author{Umberto De Giovannini}
\affiliation{Max Planck Institute for the Structure and Dynamics of Matter, Luruper Chaussee 149, D-22761 Hamburg, Germany}

\author{Alain Delgado}
\affiliation{Xanadu, 777 Bay Street, Toronto, Ontario, M5G 2C8, Canada}

\author{Florian G.~Eich}
\affiliation{Max Planck Institute for the Structure and Dynamics of Matter, Luruper Chaussee 149, D-22761 Hamburg, Germany}

\author{Johannes Flick}
\affiliation{John A. Paulson School of Engineering and Applied Sciences, Harvard University, Cambridge, MA 02138, USA}
\affiliation{Center for Computational Quantum Physics, Flatiron Institute, 162 5th Avenue, New York, NY 10010. }

\author{Gabriel Gil}
\affiliation{Dipartimento di Scienze Chimiche, Universit\`{a} degli studi di Padova, via F. Marzolo 1, 35131 Padova, Italy}
\affiliation{Instituto de Cibern\'etica, Matem\'atica y F\'isica, Calle E 309, 10400 La Habana, Cuba}

\author{Adri\'an Gomez}
\affiliation{Institute for Biocomputation and Physics of Complex Systems, University of Zaragoza, Calle Mariano Esquillor, 50018 Zaragoza, Spain}

\author{Nicole Helbig}
\affiliation{Nanomat/Qmat/CESAM and ETSF, Universit\'e de Li\`ege, B-4000 Sart-Tilman, Belgium}

\author{Hannes H\"ubener}
\affiliation{Max Planck Institute for the Structure and Dynamics of Matter, Luruper Chaussee 149, D-22761 Hamburg, Germany}

\author{Ren\'e Jest\"adt}
\affiliation{Max Planck Institute for the Structure and Dynamics of Matter, Luruper Chaussee 149, D-22761 Hamburg, Germany}

\author{Joaquim Jornet-Somoza}
\affiliation{Max Planck Institute for the Structure and Dynamics of Matter, Luruper Chaussee 149, D-22761 Hamburg, Germany}

\author{Ask H.~Larsen}
\affiliation{Nano-Bio Spectroscopy Group and ETSF, Universidad del Pa\'is Vasco, 20018 San Sebasti\'an, Spain}

\author{Irina V. Lebedeva}
\affiliation{Nano-Bio Spectroscopy Group and ETSF, Universidad del Pa\'is Vasco, 20018 San Sebasti\'an, Spain}

\author{Martin L\"uders}
\affiliation{Max Planck Institute for the Structure and Dynamics of Matter, Luruper Chaussee 149, D-22761 Hamburg, Germany}

\author{Miguel A.~L.~Marques}
\affiliation{Institut f\"ur Physik, Martin-Luther-Universit\"at Halle-Wittenberg, 06120 Halle (Saale), Germany}

\author{Sebastian T.~Ohlmann}
\affiliation{Max Planck Computing and Data Facility, Gie{\ss}enbachstra{\ss}e 2, 85741 Garching, Germany}

\author{Silvio Pipolo}
\affiliation{Université de Lille, CNRS, Centrale Lille, ENSCL, Université d’ Artois UMR 8181— UCCS Unité de Catalyse et Chimie du Solide, F-59000, Lille, France}

\author{Markus Rampp}
\affiliation{Max Planck Computing and Data Facility, Gie{\ss}enbachstra{\ss}e 2, 85741 Garching, Germany}

\author{Carlo A.~Rozzi}
\affiliation{CNR -- Istituto Nanoscienze, via Campi 213a, 41125 Modena, Italy}

\author{David A. Strubbe}
\affiliation{Department of Physics, School of Natural Sciences, University of California, Merced, CA, 95343, USA}

\author{Shunsuke A.~Sato}
\affiliation{Max Planck Institute for the Structure and Dynamics of Matter, Luruper Chaussee 149, D-22761 Hamburg, Germany}
\affiliation{Center for Computational Sciences, University of Tsukuba, Tsukuba 305-8577, Japan}

\author{Christian Sch\"afer}
\affiliation{Max Planck Institute for the Structure and Dynamics of Matter, Luruper Chaussee 149, D-22761 Hamburg, Germany}

\author{Iris Theophilou}
\affiliation{Max Planck Institute for the Structure and Dynamics of Matter, Luruper Chaussee 149, D-22761 Hamburg, Germany}

\author{Alicia Welden}
\affiliation{Quantum Simulations Group, Lawrence Livermore National Laboratory, Livermore, California 94551, USA}

\author{Angel Rubio}
\email{angel.rubio@mpsd.mpg.de}
\affiliation{Max Planck Institute for the Structure and Dynamics of Matter, Luruper Chaussee 149, D-22761 Hamburg, Germany}
\affiliation{Nano-Bio Spectroscopy Group and ETSF, Universidad del Pa\'is Vasco, 20018 San Sebasti\'an, Spain}
\affiliation{Center for Computational Quantum Physics, Flatiron Institute, 162 5th Avenue, New York, NY 10010. }


\begin{abstract}
Over the last years extraordinary advances in experimental and theoretical tools
have allowed us to monitor and control matter at short time and atomic scales
with a high-degree of precision. An appealing and challenging route towards
engineering materials with tailored properties is to find ways to design or
selectively manipulate materials, especially at the quantum level. To this end,
having a state-of-the-art \textit{ab initio} computer simulation tool that
enables a reliable and accurate simulation of light-induced changes in the
physical and chemical properties of complex systems is of utmost importance. The
first principles real-space-based Octopus project was born with that idea in
mind, providing an unique framework allowing to describe non-equilibrium
phenomena in molecular complexes, low dimensional materials, and extended
systems by accounting for electronic, ionic, and photon quantum mechanical
effects within a generalized time-dependent density functional theory
framework. The present article aims to present the new features that have been
implemented over the last few years, including technical developments related to
performance and massive parallelism. We also describe the major theoretical
developments to address ultrafast light-driven processes, like the new
theoretical framework of quantum electrodynamics density-functional formalism
(QEDFT) for the description of novel light-matter hybrid states. Those advances,
and other being released soon as part of the Octopus package, will enable the
scientific community to simulate and characterize spatial and time-resolved
spectroscopies, ultrafast phenomena in molecules and materials, and new emergent
states of matter (QED-materials).
\end{abstract}


\maketitle 

\section{Introduction}

It is a general challenge in the electronic structure community to develop
accurate and efficient methods of modeling materials of ever increasing complexity
in order to predict their properties. In this respect, time-dependent density
functional theory (TDDFT) and related methods have become a natural choice for
modeling materials and complex systems in and out of their equilibrium.

During the last years, novel directions of research emerged in the field of
chemistry, physics, and material science that required the development of novel
simulation tools needed to face the new challenges posed by those experimental
advances and emerging phenomena. Among these new fields of research, one could
cite some examples: the strong coupling between light and matter (including
materials embedded in cavities), new states of matter (hidden phases and
topological solids and molecules), and the strong field dynamics in periodic
systems.  Whereas the strong-field dynamics of atoms and molecules is now well
understood, the strong-field dynamics in solids is an active field of
research. Real-time TDDFT~\cite{marques2006time,marques2012fundamentals}
represents a natural tool to study highly non-linear phenomena in solids and
low-dimensional materials and the development of efficient numerical methods to
perform real-time TDDFT in such periodic or semi-periodic systems is crucial to
explore this new phenomena (including the quantum nature of light and
phonons). Indeed, it allows to describe highly nonlinear processes without
having to resort to the perturbation theory on top of equilibrium DFT
calculations.

Recent years have seen tremendous experimental progress in the field of strong
light-matter interactions,~\cite{ebbesen2016,ruggenthaler2014,flick2015} where
the strong coupling of light to chemical systems, quantum materials, or
nanoplasmonic systems, among others, has been demonstrated. In this regime,
light and matter meet on the same footing and the electron-photon interaction
has to be explicitly
considered.~\cite{ruggenthaler2014,flick2015,flick2018c,feist2017polaritonic,ribeiro2018polariton,kockum2019ultrastrong}
A theoretically novel approach accelerating this field is
quantum-electrodynamical density-functional theory
(QEDFT),~\cite{ruggenthaler2014,flick2015,tokatly2013,flick2016exact,ruggenthaler2017b,flick2017c,schafer2019modification}
which complements TDDFT with the photonic degrees of freedom and provides
reliable and predictive simulations in this emerging field of research.

In this paper we explore the recent advances in the Octopus code
project.~\cite{castro2004first,andrade2010linear,MARQUES200360,castro_octopus:_2006,andrade_real-space_2015} We
focus our attention on the recently added features, and particularly on the ones
that have not been described in the previous
papers.~\cite{MARQUES200360,castro_octopus:_2006,andrade_real-space_2015} These
new features include the implementation of new levels of self-consistent and
microscopic couplings of light and matter, the treatment of solvent effects
through the polarizable continuum model, the implementation of various methods
to treat van der Waals interactions, new methods to calculate magnons,
conductivities, and photoelectron spectroscopy from real-time TDDFT, and the
calculation of orbital magneto-optical responses. Advances in numerical
algorithms and methods, such as new propagators and the use of iterative
eigensolvers in the context of reduced density matrix functional theory, are
also discussed. Finally, recent improvements in the treatment of periodic
systems, as well as more technical code improvements, are also presented. For
a more detailed explanation about how to use these newly introduced features in
practice, we refer the readers to the Octopus webpage,~\footnote{The {Octopus}
  code, tutorials, and examples can be found on its website at
  \url{https://octopus-code.org/}.} as new tutorials and examples are regularly
being added to it.

This paper is organized as follow. First, we present in Sec.~\ref{sec:maxwell}
to Sec.~\ref{sec:local_domains} new implementations of physical theories and
algorithms that allow us to deal with new non-equilibrium phenomena in materials
and nanostructures. This is followed by a second set of sections, from
Sec.~\ref{sec:propagators} to Sec.~\ref{sec:technical}, dealing with technical
developments that are fundamental in improving the code performance and the
algorithm's stability. Finally, we draw our conclusions in
Sec.~\ref{sec:conclusion}. Unless otherwise stated, atomic units are used
throughout the paper.

\section{Coupled Maxwell-Kohn-Sham equations}
\label{sec:maxwell}

In most cases when light-matter interactions are considered, a decoupling of
light and matter is performed at the outset.  Either the electromagnetic fields
are prescribed and then properties of the matter subsystem are determined, as
frequently found in, e.g., quantum chemistry or solid-state physics, or the
properties of matter are prescribed and then the properties of the photon
subsystem are determined, as done in, e.g.,~quantum optics or
photonics.~\cite{ruggenthaler2017b,JRORA18}

The microscopic interaction of light and matter in Octopus has followed this
decoupling strategy and has been treated so far only in forward-coupling
direction. In this approximation an external classical laser pulse or kick is
prescribed and the response of the system is computed by evolving the Kohn-Sham
orbitals.~\cite{biry00:7998} The back-action of the matter subsystem on the
laser pulse, and the subsequent effect of this modified pulse on matter, and so
on, is ignored. This forward-coupling approximation is highly accurate when the
generated total current in the system is comparably small, such as in atoms or
smaller molecules. This has been exploited in Octopus and many different types
of spectroscopies have been computed successfully in the past using this
approach.

In contrast, in classical electromagnetic modeling the opposite view is taken.
Here the material properties are prescribed and the resulting electromagnetic
fields are computed. In practice, the material properties are routinely
approximated by a local continuum model or the dielectric function of the
system, such as a Debye, Lorentz, or Drude model and then Maxwell's equations
are solved for linear dielectric media arranged in appropriate
geometries.~\cite{taflove2005computational}

It is clear that both perspectives, a focus on matter dynamics alone or a focus
on electromagnetic field dynamics alone, break down when the total induced
currents become large and when electromagnetic near-field effects on the scale
of the material system are not negligible anymore. Prime examples for such cases
are nanoplasmonic systems, surface plasmon-polaritons, or tip-enhanced
spectroscopies.  In these cases the back-action of the material response on the
system itself has to be taken into account leading to screening and retardation
effects. The proper theoretical framework which encompasses all these effects is
quantum electrodynamics.  Starting with a generalized Pauli-Fierz field theory
for the combined system of electrons, nuclei, and photons, we have recently
derived different levels of self-consistent and microscopic couplings of light
and matter. In the classical limit this results in a coupled set of
Ehrenfest-Maxwell-Pauli-Kohn-Sham equations.~\cite{JRORA18} To implement these
equations, we have added a Maxwell solver to the Octopus code which we couple
self-consistently to the dynamics of the electrons and nuclei. In the following,
we briefly summarize the basic ingredients for this implementation and show an
example of self-consistent light-matter interactions for a nanoplasmonic
system. Further details of the implementation and nano-optical applications can
be found in Ref.~\onlinecite{JRORA18}.

Since over the years Octopus has been optimized heavily to solve time-dependent
Schr\"odinger and Kohn-Sham equations, we have exploited the fact that Maxwell's
equations can be formulated in Schr\"odinger form~\cite{bialynickibirula1994} to
benefit from the efficient time-evolution in the code. This reformulation is
based on the Riemann-Silberstein vector,~\cite{silberstein1907} which is a
combination of the electric $\mvec{E}(\mvec{r},t)$ and magnetic field
$\mvec{B}(\mvec{r},t)$
\begin{equation}
  \begin{alignedat}{2}
    \mvec{F}_{\pm}(\mvec{r},t)
    &= \sqrt{\frac{\epsilon_{0}}{2}} \mvec{E}(\mvec{r},t)
    \pm \mathrm{i} \sqrt{ \frac{1}{2 \mu_{0}} } \mvec{B}(\mvec{r},t)\,.
    \label{eq_RS_vector_pm_2}
  \end{alignedat}
\end{equation} 
The sign of the imaginary part of the Riemann-Silberstein vector corresponds to
different helicities.  The reformulation of Maxwell's equations in Schr\"odinger
form is purely algebraic and starts out with the microscopic Maxwell's equations
\begin{equation}\label{eq:max_div}
  \mathbf{\nabla}\cdot\mathbf{E}=
  \frac{\rho}{\epsilon_0},\qquad \mathbf{\nabla}\cdot\mathbf{B}=0\,,
\end{equation}
\begin{equation}\label{eq:max_curl}
  \mathbf{\nabla}\times\mathbf{B}=\frac{1}{c^2}
  \frac{\partial \mathbf{E}}{\partial t}+\mu_0\mathbf{J},\qquad
  \mathbf{\nabla}\times\mathbf{E}=-\frac{\partial \mathbf{B}}{\partial t}\,,
\end{equation}
where $\mathbf{E}$ and $\mathbf{B}$ are the classical electric and magnetic
fields, $\rho$ and $\mathbf{J}$ are the charge and current densities,
$\epsilon_0$ and $\mu_0$ are the vacuum permittivity and permeability, and
$c=\left(\epsilon_0\mu_0\right)^{-1/2}$ is the speed of light. Using the
Riemann-Silberstein vector, the electric and magnetic Gauss laws may now be
combined in real and imaginary part
\begin{equation}
  \mathbf{\nabla}\cdot\mathbf{F}=\frac{1}{\sqrt{2\epsilon_0}}\rho
\end{equation}
and, likewise, the Faraday and Ampere law can be combined into one evolution
equation for the Riemann-Silberstein vector
\begin{equation}
  i\hbar\frac{\partial \mathbf{F}}{\partial t}=
  c\left(\mathbf{S}\cdot\frac{\hbar}{i}\mathbf{\nabla}\right)\mathbf{F}-
  \frac{i\hbar}{\sqrt{2\epsilon_0}}\mathbf{J}\,.
  \label{eq:TDRS}
\end{equation}
Here $\mathbf{S}=(S_x,S_y,S_z)$ denotes a vector of spin one matrices
\begin{equation}
  S_x=\begin{pmatrix} 0 & 0 & 0 \\ 0 & 0 & -i \\ 0 & i & 0 \end{pmatrix}\,,\quad
  S_y=\begin{pmatrix} 0 & 0 & i \\ 0 & 0 & 0 \\ -i & 0 & 0 \end{pmatrix}\,,\quad
  S_z=\begin{pmatrix} 0 & -i & 0 \\ i & 0 & 0 \\ 0 & 0 & 0 \end{pmatrix}\,,
\end{equation}
which are analogous to the Pauli matrices and show the spin-one character of the
photon. Having cast Maxwell's equations as an inhomogeneous Schr\"odinger equation,
it is now straightforward to use the time-evolution algorithms in Octopus to
time-evolve the Riemann-Silberstein vector.  The only difference to the matter
propagation is that we are now dealing with the ``Maxwell Hamiltonian''
\begin{equation}
  H_\text{EM}=c\left(\mathbf{S}\cdot\frac{\hbar}{i}\mathbf{\nabla}\right)\,,
  \label{hem}
\end{equation}
which acts on six orbitals of the Riemann-Silberstein vector corresponding to
the three components of the electric and magnetic field vectors. As in the
matter case in Octopus, the discretization of the gradient in the Maxwell
Hamiltonian is performed with finite-difference stencils and the domain
parallelization of Octopus can be used seamlessly for the Maxwell case as well.
While also finite-difference discretizations are used for the Maxwell solver in
Octopus, the difference to finite-difference time-domain (FDTD) codes based on
the Yee algorithm is that we employ not two shifted grids for the electric and
magnetic fields,~\cite{taflove2005computational} but rather a single grid for
the Riemann-Silberstein vector.  This simplifies the coupling to matter and
allows us to use higher-order finite-difference discretizations for the
gradient. Since the spatial discretization is connected to the temporal
discretization through the Courant condition, this in turn allows to take larger
time-steps and, from our experience, a unified grid also improves the stability
compared to FDTD.

Instead of using the constitutive relations, we couple Maxwell's equations
directly to the microscopic current density of the matter subsystem, consisting
of the usual paramagnetic current term, the diamagnetic current term, and the
magnetization current term. For the coupling of the electromagnetic fields to
the matter subsystem, we are relying on the Power-Zienau-Woolley
transformation,~\cite{loudon1988,craig1998molecular} which leads to a multipole
expansion. We have implemented the first two orders of this expansion: the
dipole approximation in lowest order and electrical-quadrupole and
magnetic-dipole coupling in the next order. In addition, we are currently
working on implementing the full minimal coupling with a full position
dependence of the vector potential.

\begin{figure*}[t]
  \begin{minipage}{0.47 \textwidth}
    {\scriptsize \bf Only forward coupling}
    \includegraphics[scale=0.40]{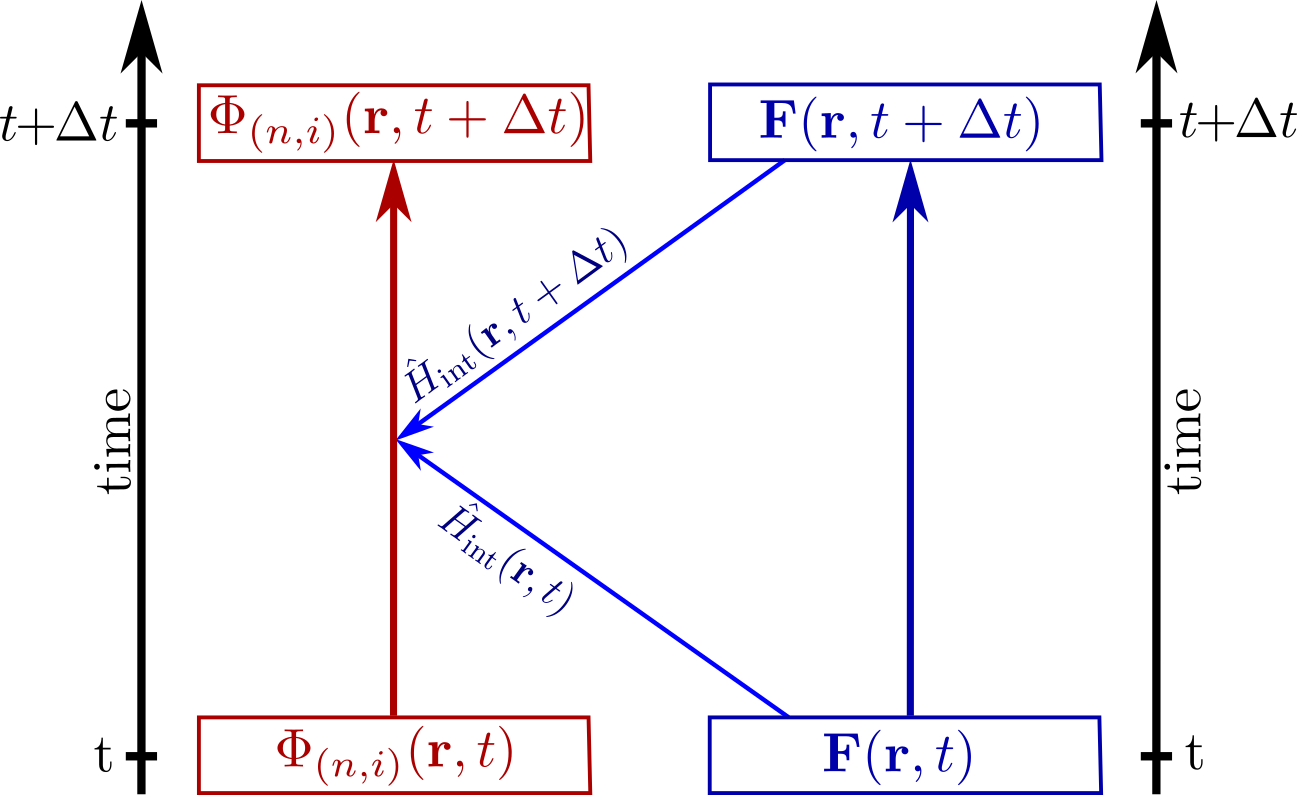}
  \end{minipage}
  \qquad
  \begin{minipage}{0.47 \textwidth}
    {\scriptsize \bf Self-consistent forward and backward coupling}
    \includegraphics[scale=0.40]{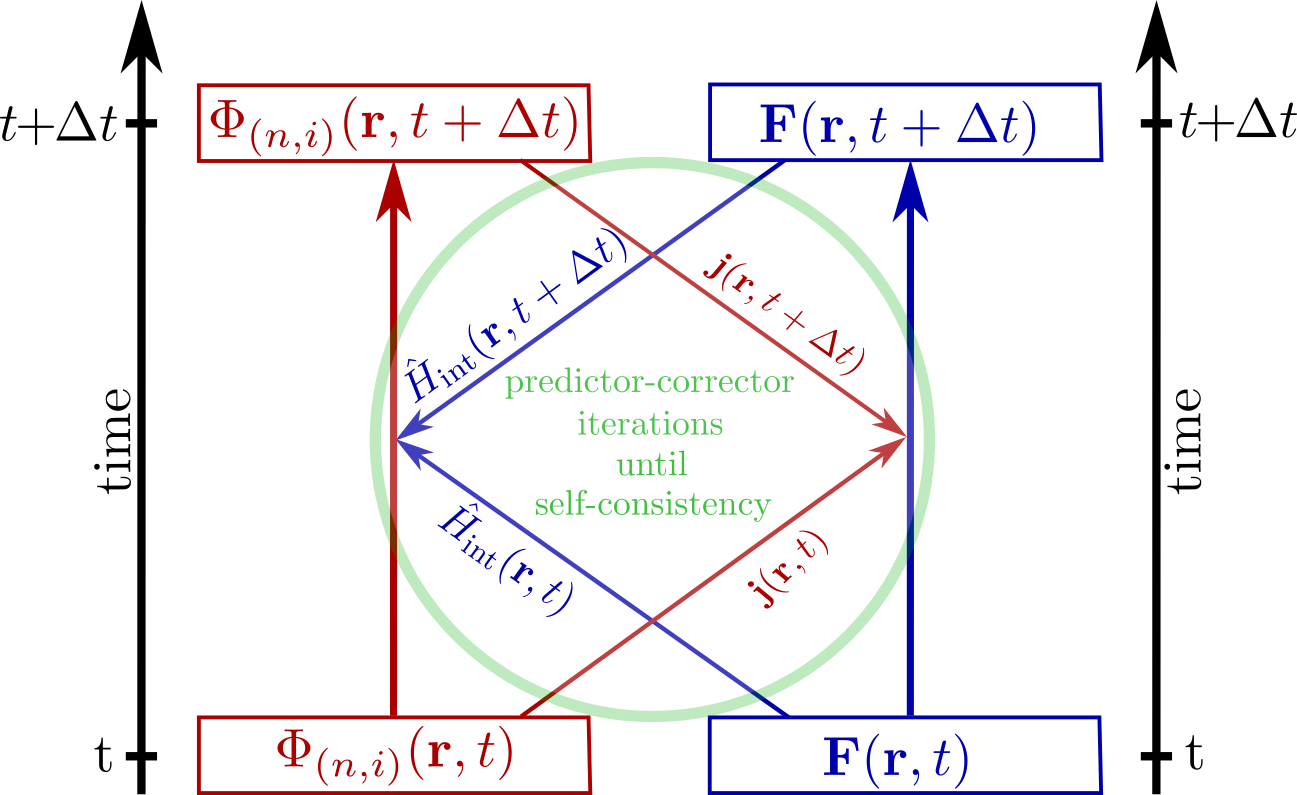}
  \end{minipage}
  \caption{ The figure on the left-hand side illustrates the standard
    forward-coupling approximation: the electromagnetic fields (in blue)
    propagate freely and only influence the propagation of the matter (in red).
    The back reaction of the matter currents on the electromagnetic fields is
    neglected. \\   
    In the figure on the right-hand side we illustrate a fully self-consistent
    predictor-corrector scheme for a coupled Maxwell-Pauli-Kohn-Sham
    time-stepping.  As before, the electromagnetic fields influence the
    propagation of the matter (forward coupling). However, in addition, here the
    currents from the matter propagation also influence the propagation of the
    electromagnetic fields (backward coupling).  A given time-step for the
    matter wavefunctions and the electromagnetic fields is repeated until
    self-consistency is found (self-consistent forward-backward coupling).}
  \label{fig_predictor_corrector_method}
\end{figure*}

The time evolution of the Kohn-Sham orbitals and the Maxwell fields is performed
side-by-side and the two subsystems are coupled self-consistently in each
time-step, as illustrated in Fig.~\ref{fig_predictor_corrector_method}.  To
propagate different subsystems with different Hamiltonians and different sets of
orbitals simultaneously, we have implemented a multi-system framework in Octopus
(more details about this can be found in Subsection~\ref{subsec:multisystem}).

Similar to the matter propagation, also in the Maxwell case outgoing waves that
reach the box boundary of the simulation box have to be absorbed to avoid
artificial reflections and backscattering at the boundaries. Our first attempt
was to use also the mask functions that are used for the matter propagation in
Octopus.  However, in the electromagnetic case the mask absorption of outgoing
waves turned out to be not efficient enough so that we implemented a perfectly
matched layer (PML)~\cite{taflove2005computational} for the Maxwell propagation.

When considering incoming electromagnetic fields with optical wavelengths, the
coupling to atomistic or nano-scale systems leads to a multi-scale problem. The
optical wavelength of the radiation is in this case much larger than the de
Broglie wavelength of the matter. Likewise, the electromagnetic waves are
traveling with the speed of light, which requires sub-attosecond time-steps. We
have therefore implemented different multi-scale couplings in space and time.
For example, the Maxwell simulation box can be on the same scale as the matter
box. In this case the electromagnetic waves are represented as incoming
analytical time-dependent boundary conditions and are propagated numerically
inside the simulation box. Alternatively, the Maxwell simulation box can be much
larger than the matter box to fully encompass laser pulses with optical
wavelengths. In this case prolongations and interpolations have to be used
similar to multi-grid methods.  Since the electronic and nuclear motion is much
slower than the time-evolution of the electromagnetic waves, we also have
implemented a multi-scale approach for the real-time propagation. The
Riemann-Silberstein vector is propagated with frozen electronic current from the
last point of interaction for many intermediate time-steps before a coupling to
the matter subsystem takes place. The number of intermediate steps is a
convergence parameter and depends on the physical situation at hand.

Since we now include the description of classical electromagnetic fields
explicitly in our real-time simulations, we have directly access to the outgoing
electromagnetic radiation.  This allows to define electromagnetic detectors at
the box boundaries which accumulate the outgoing electromagnetic waves. We have
implemented such electromagnetic detectors in Octopus and this allows to run
simulations in close analogy with experiments and to directly observe the
outgoing radiation.  For example, it is then no longer needed to Fourier
transform the time signal of the matter dipole to get optical spectra, but we
rather have access to the spectrum directly on the Maxwell grid.

\begin{figure*}[t!]
  \begin{minipage}{0.95 \textwidth}
    \includegraphics[scale=0.48]{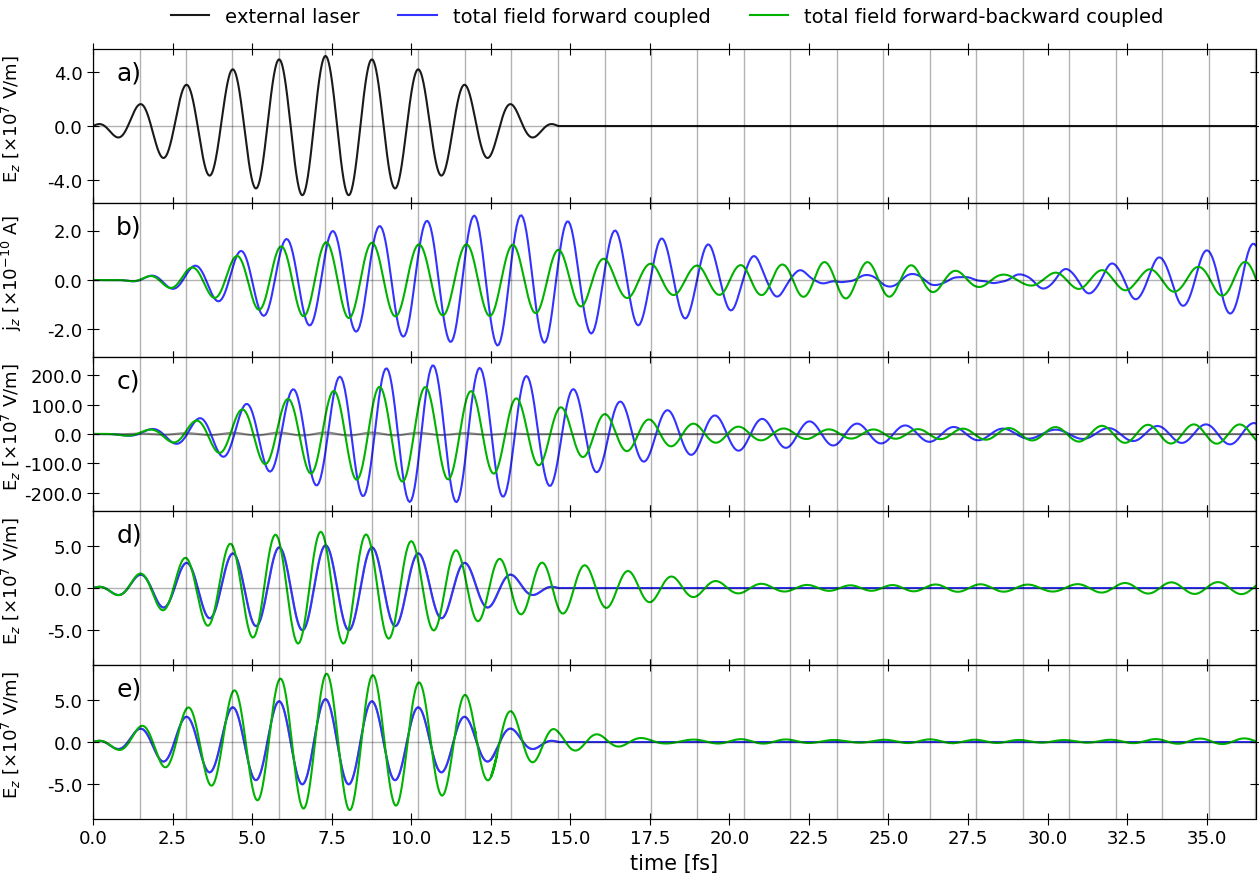}
    \caption{Electric field values and current density in the $z$-direction for
      a dimer of sodium nanoparticles. The centers of the two nanoparticles are
      located along the $z$-axis and the distance between the two effective
      spheres of the nanoparticles is 0.5~nm. The electric field values are
      calculated at two different points: the center point
      $\mvec{r}_{\mathrm{cp}}$ between the two nanoparticles, located at the
      origin, and an off-center point $\mvec{r}_{\mathrm{ocpx}}$ located along
      the $x$-axis at a distance of 1.957~nm from $\mvec{r}_{\mathrm{cp}}$.  The
      first panel $a)$ illustrates the incident cosinusoidal laser pulse with
      frequency $\omega_1 = 3.05$~eV ($0.112 $~a.u.), $\lambda_1 = 406.5$~nm
      ($7681.84$~a.u.), and amplitude of $E_z^{0} = 5.142\times10^{7} $~V/m
      ($10^{-4}$~a.u.), which drives the system. The second panel $b)$ displays
      the current density at $\mvec{r}_{\mathrm{cp}}$. In panels $c)-e)$, we
      show the electric field enhancement at $\mvec{r}_{\mathrm{cp}}$, the
      electric field enhancement at $\mvec{r}_{\mathrm{ocpx}}$, and the average
      of the electric field over the detector surface close to the box boundary,
      respectively. The curve in bright gray in panel $c)$ has been added to
      simplify the comparison and is identical to the laser pulse in panel
      $a)$. The period $T_1 = 1.36$~fs corresponding to the laser frequency $
      \omega_1 $ is indicated with grey vertical lines.}
    \label{fig_Na_297_dimer_0_5_nm_electric_field_fixed_ions}
  \end{minipage}
\end{figure*}

As an example of a coupled Ehrenfest-Maxwell-Pauli-Kohn-Sham propagation with
Octopus we have selected a nano-optical application. We consider in this example
two almost spherical sodium nanoparticles with 297 sodium atoms each, which are
arranged in a dimer configuration. This system is excited with an incoming laser
pulse which excites either the internal dipole or quadrupole plasmon motion of
the dimer. In Fig.~\ref{fig_Na_297_dimer_0_5_nm_electric_field_fixed_ions} we
show the resulting electromagnetic field enhancements for different levels of
light-matter coupling. In panels a) and b) we show the temporal profile of the
incoming laser pulse and the resulting current density at the center point
between the two nanoparticles. The field enhancement can be seen in panels
c)-e). Including a self-consistent back reaction in the light-matter coupling,
the field enhancement is reduced at the center point between the two
nanoparticles, as can be seen in panel c), while far away from the dimer, as
shown in panels d) and e), the field enhancement is larger than in the forward
coupled case. Furthermore, frequency shifts can also be observed, which are more
pronounced in the near-field. We have found that the field enhancement is also
sensitive to the coupling terms of the multipole expansion which are included
and that the quantitative difference of switching from LDA to PBE for the
exchange-correlation functional is in this case smaller than including the back
reaction in the light-matter coupling, cf. Ref.~\onlinecite{JRORA18}.

To conclude, with our new efficient implementation for coupled
Ehrenfest-Maxwell-Pauli-Kohn-Sham equations in Octopus we have now a very
versatile tool at hand which allows to compute fully self-consistent
forward-backward light-matter coupling in real-time and real-space for a vast
set of applications and in close analogy to experiments.  As the coupling to
Maxwell's equations of motion represents the classical limit of the light-matter
interaction, this development leads to the classical limit of QEDFT.

\section{Strong electron-photon interactions in real space: Quantum-electrodynamical density-functional theory (QEDFT)}
\label{sec:qedft}


The nascent field of strong light-matter interaction expanded over the past
decades from small atomic structures to chemistry~\cite{ebbesen2016} and
material science.~\cite{baumberg2019extreme} This development necessitates
predictive first-principles methods capable to describe light and matter on the
same footing.
We have introduced for the first time a time-dependent density functional theory
for quantum electrodynamics (QEDFT)~\cite{tokatly2013,ruggenthaler2014} to treat
\textit{ab-initio} weak and strong light-matter interactions, and its applications to
chemistry and materials, that provides a unique framework to explore, predict,
and control new states of matter out of equilibrium. This generalization of the time-dependent density-functional method allows us, for the first time, to explore the
effects of dressing electronic states with photons, while retaining the
electronic properties of real materials.

A general non-relativistic Hamiltonian for light-matter systems treating $N$
interacting electrons coupled to ${N}_p$ photonic modes in the case of the
so-called length-gauge, and after employing the long-wavelength (dipole)
approximation,~\cite{flick2017c} reads as follows:
\begin{align}
  \label{eq:Hamiltonian_dipole}
  \hat{H} = \sum_{k=1}^N \left[ -\tfrac{1}{2} \partial_{\mvec{r}_k}^2 + v(\mvec{r}_k,t) \right] + \tfrac{1}{2}\sum_{k \neq l} w(\mvec{r}_k,\mvec{r}_l) + \frac{1}{2} \sum^{N_p}_{\alpha=1}\left[ -\tfrac{\partial^2}{\partial q_\alpha^2} + \left(\omega_\alpha q_\alpha - \bm{\lambda}_\alpha \cdot \sum_{k=1}^N \mvec{r}_k \right)^2\, + 2\frac{j^{(\alpha)}_\text{ext}(t)}{\omega_\alpha} q_\alpha \right]\,.
\end{align}
Here the first two terms on the right-hand side correspond to the usual
electronic many-body Hamiltonian, while the last term describes the photon mode,
which is characterized for each photon mode $\alpha$ by its elongation $q_\alpha$, frequency $\omega_\alpha$,
and electron-photon coupling strength vector $\bm{\lambda}_\alpha$ that includes
the polarization of the photon mode and introduces the coupling to the total
dipole $\sum_{k=1}^N \mvec{r}_k$ of the electronic system. The external variable
for the photon system is the time-dependent current
$j^{(\alpha)}_\text{ext}(t)$.

QEDFT~\cite{tokatly2013,ruggenthaler2014} is structurally similar to
time-dependent density-functional theory in that it is based on a one-to-one
correspondence between internal and external variables. If now photons are also
considered, the set of internal variables has to be expanded. In the frame of
Eq.~\ref{eq:Hamiltonian_dipole}, the internal variables become the density
$n(\textbf{r},t)$ and the mode-resolved contributions to the electric
displacement field $q_\alpha(t)$. By introducing and exploiting the bijective
mapping of these internal and the external variables
($v_\text{ext}(\textbf{r},t)$ and $j_\text{ext}(t)$), the auxiliary Kohn-Sham
system can be set up which is characterized by the electronic Kohn-Sham
equations as well as Maxwell's equations,~\cite{flick2015} leading to no
exchange-correlation contribution in the photon subsystem
($j^{(\alpha)}_\text{xc}(t)=0$).  This reformulation subsumes the 'quantumness'
of the light-matter interaction solely into the local exchange-correlation
potential that now features a component due to the electron-photon interactions,
in addition to the part due to electron-electron interaction part, i.e.,
$v_{xc\sigma}(\textbf{r},t) = v^{ee}_{xc\sigma}(\textbf{r},t) +
v^{ep}_{xc\sigma}(\textbf{r},t)$.~\cite{tokatly2013,ruggenthaler2014,ruggenthaler2017b}
Extending the coupled Maxwell-Kohn-Sham equations to consider quantum photons is
thus condensed into the calculation of the local exchange-correlation potential.

In practice, QEDFT requires the construction of an additional
exchange-correlation potential that describes the electron-photon
interaction. First attempts to construct this potential are based on many-body
perturbation theory within the exact exchange
approximation~\cite{pellegrini2015,flick2017c} by utilizing the
optimized-effective potential (OEP) method.~\cite{kuemmel2008}

\subsection{The optimized effective potential (OEP)}

The OEP photon energy has been introduced in Ref.~\onlinecite{pellegrini2015}
and depends on occupied and unoccupied orbitals of the system. Alternatively,
the OEP photon energy can also be formulated using occupied orbitals and
orbitals shifts only.~\cite{flick2017c} The full energy expression is given by
\begin{align}
  E_{xc} = E^{(ee)}_{xc}+\sum^{N_p}_{\alpha=1} E^{(\alpha)}_{x}\,,
  \label{eq:total-ex}
\end{align}
where $E^{(ee)}_{xc}$ describes the electronic exchange-correlation energy and $E^{(\alpha)}_{x}$ the exchange energy due to the interaction of the electrons with the photon mode $\alpha$.
Avoiding unoccupied orbitals is computationally much more favorable for larger
systems and the photonic induced exchange energy can be correspondingly
expressed as sum over the $N_\sigma$ occupied orbitals of spin channel $\sigma$
\begin{align}
  E^{(\alpha)}_{x} &= 
  \sum_{\sigma=\uparrow,\downarrow}\sum_{i=1}^{N_\sigma}\sqrt{\frac{\omega_\alpha}{8}} \bra{\Phi_{{i\sigma},\alpha}^{(1)}} \hat{d}_\alpha \ket{\varphi_{i\sigma}} 
  +\frac{1}{4}  \bra{\Phi^{(2)}_{{i\sigma},\alpha}}\hat{d}_\alpha \ket{\varphi_{i\sigma}} + c.c.\,,
  \label{eq:exchange_energy}
\end{align}
where $\omega_\alpha$ describes the $\alpha$th mode of the electromagnetic field
and $\hat{d}_\alpha = \boldsymbol\lambda_\alpha\cdot \textbf{r}$ describes the
dipole operator and the electron-photon coupling strength. We can now reformulate the problem
in terms of two electron-photon orbital shifts. The Kohn-Sham
orbitals $\varphi_{i\sigma}$ contribute to both electron-photon orbital shifts
$\Phi_{{i\sigma},\alpha}^{(1)}$ and $\Phi_{{i\sigma},\alpha}^{(2)}$ that can be
calculated using the Sternheimer equations.~\cite{flick2017c} The first electron-photon orbital
shift can be obtained explicitly by the solution of a linear equation, i.e., a
Sternheimer equation
\begin{align}
  \left[\hat{h}_{s\sigma} - \left(\epsilon_{i\sigma} - \omega_\alpha\right)\right] {\Phi^{(1)}_{i\sigma,\alpha}}(\textbf{r}) = -\sqrt{\frac{\omega_\alpha}{2}}\hat{d}_\alpha {\varphi_{i\sigma}}(\textbf{r}) 
  + \sqrt{\frac{\omega_\alpha}{2}}\sum_{k=1}^{N_\sigma}{d}^{(\alpha)}_{ki\sigma}{\varphi_{k\sigma}}\,(\textbf{r})\,,
  \label{eq:1st-occ}
\end{align}
with the matrix element
$d^{(\alpha)}_{ij\sigma}=\bra{\varphi_{i\sigma}}\hat{d}_\alpha\ket{\varphi_{j\sigma}}$. In
contrast, the second electron-photon orbital shift ${\Phi^{(2)}_{i\sigma,\alpha}}(\textbf{r})$ can be defined explicitly
as follows
\begin{align}
 {\Phi^{(2)}_{i\sigma,\alpha}}(\textbf{r}) = \hat{d}_\alpha {\varphi_{i\sigma}}(\textbf{r}) - \sum_{k=1}^{N_\sigma}{d}^{(\alpha)}_{ki\sigma}{\varphi_{k\sigma}}\,(\textbf{r}).
  \label{eq:2nd-occ}
\end{align}

From Eq.~\eqref{eq:exchange_energy}, we can now deduce the potential using
\begin{align}
  v_{xc\sigma}(\textbf{r}) = \frac{\delta E_{xc}}{\delta n_\sigma(\textbf{r})}.
  \label{eq:vx-ex-dn}
\end{align}
Doing these reformulations, we find for the final OEP equation including
electron-electron effects as well as electron-photon effects
\begin{align}
  \sum_{i=1}^{N_\sigma}\psi^*_{i\sigma}(\textbf{r})\varphi_{i\sigma}(\textbf{r}) - \Lambda_{i\sigma}(\textbf{r}) + c.c. = 0\,.
  \label{eq:soep}
\end{align}
where the homogeneity $\Lambda_{i\sigma}(\textbf{r})$ is given by
\begin{align}
  \Lambda_{i\sigma}(\textbf{r}) = \frac{1}{2}\sum^{N_p}_{\alpha=1}&\Biggl[{\Phi}^{(1)*}_{i\sigma,\alpha}(\textbf{r}){\Phi}^{(1)}_{i\sigma,\alpha}(\textbf{r})- \braket{{\Phi}^{(1)}_{i\sigma,\alpha}|{\Phi}^{(1)}_{i\sigma,\alpha}}\varphi^{*}_{i\sigma}(\textbf{r})\varphi_{i\sigma}(\textbf{r}) \Biggr]\nonumber
\end{align}
In Eq.~\ref{eq:soep} we defined a third orbital shift, the exchange-correlation
orbital shift, that will be used to obtain the corresponding exchange
correlation potential and is defined along the lines of the orbital shift
usually used in OEP calculations.~\cite{kuemmel2008} We can obtain
$\psi^*_{i\sigma}(\textbf{r})$ using a Sternheimer equation
\begin{align}
  \left(\hat{h}_{s\sigma} - \epsilon_{i\sigma}\right)\psi^*_{i\sigma}(\textbf{r}) = M_{i\sigma}^*(\textbf{r}) - \langle M_{i\sigma}|\varphi_{i\sigma}\rangle\varphi^*_{i\sigma}(\textbf{r})\,,
  \label{eq:soep2}
\end{align}
where $M_{i\sigma}^*(\textbf{r})$ now consist of
the electron-photon orbital shifts and the Kohn-Sham orbitals, as described in
Ref.~\onlinecite{flick2017c}. Accordingly we define this quantity as
\begin{align}
  \label{eq:soep3}
  M_{i\sigma}^*(\textbf{r}) =& -\left(v_{x\sigma}(\textbf{r}) - u_{xi\sigma}(\textbf{r})\right)\varphi^*_{i\sigma}(\textbf{r})\\
  &+\sum_{\alpha=1}^{N_p}\Biggl[\hat{d}_\alpha\Bigg(\sqrt{\frac{\omega_\alpha}{2}}\Phi^{(1)*}_{i\sigma,\alpha}(\textbf{r})+\frac{1}{2}\hat{d}_\alpha \varphi^*_{i\sigma}(\textbf{r})\Bigg)
    \nonumber\\
    &-\sum_{k=1}^{N_\sigma}{d}^{(\alpha)}_{ik\sigma}\Bigg(\sqrt{\frac{\omega_\alpha}{2}}{\Phi}^{(1)*}_{k\sigma,\alpha}(\textbf{r}) + \;\hat{d}_{\alpha} \varphi^*_{k\sigma}(\textbf{r})\Bigg)\Biggr].\nonumber
\end{align}
and include here effects of the electron-electron interaction in the quantity $u_{xi\sigma}(\textbf{r})$. For instance in exchange-only calculations this quantity is defined as 
$u_{xi\sigma}(\textbf{r})=\frac{1}{\varphi^*_{i\sigma}(\textbf{r})}\frac{\delta E^{(ee)}_{x}[\{\varphi_{j\tau}\}]}{\delta \varphi_{i\sigma}(\textbf{r})}$, where $E^{(ee)}_{x}$ is the usual Fock exchange energy. 

Eq.~\ref{eq:soep2} has to be solved self-consistently with
Eq.~\eqref{eq:1st-occ}. By this reformulation, we have replaced the problem of
calculating the OEP equation using all unoccupied states by a problem of solving
$N_p$+1 Sternheimer equations that each only invoke occupied orbitals. In this
way, the formulation of the problem becomes similar to the one of
Ref.~\onlinecite{kuemmel2003} for electrons only, and which can be easily
extended.

\begin{figure}[t]
  \centerline{\includegraphics[width=0.9\textwidth]{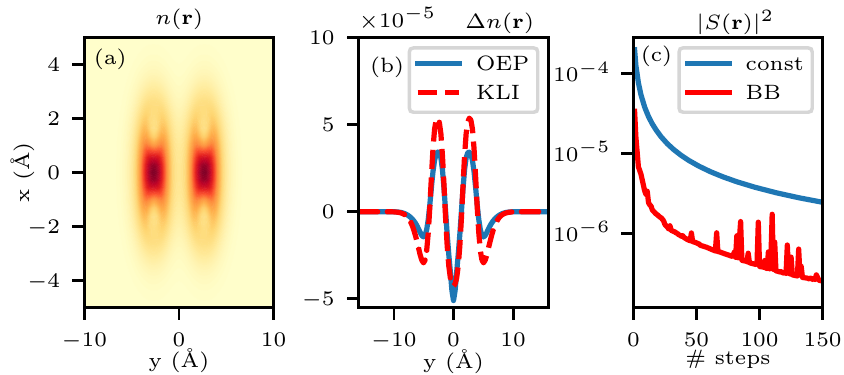}}
  \caption{
    Ground-state density of two sodium dimers affected by the
    vacuum-field of a cavity: In (a) we show the electron density, in (b) the
    difference of the electron density inside and outside the cavity for the OEP
    and the KLI approximations (see main text for details). We find for the exchange energy
    $E^{(\alpha)}_x=6.52$~meV and the number of photons in the correlated light-matter ground-state $n_\text{pt}=\langle a_\alpha^\dagger a_\alpha \rangle = 2.03\times10^{-3}$ for the OEP
    case, and $E^{(\alpha)}_x=6.67$~meV and $n_\text{pt}=2.22\times10^{-3}$ for
    KLI. Panel (c) shows the convergence of $|S(\textbf{r})|^2$ as
    defined by Eq.~\ref{eq:oep-S} using a constant with $c=20$, and using the
    Barzilai-Borwein (BB) scheme. The simulation is set up as described in
    Refs.~\onlinecite{flick2017c,ruggenthaler2017b}, with $\hbar\omega_\alpha =
    2.19$~eV, $\lambda_\alpha=2.95$~eV$^{1/2}$nm$^{-1}$, $\boldsymbol
    \lambda_\alpha = \lambda_\alpha \textbf{e}_y$, and the real-space grid is
    sampled as $31.75 \times 31.75 \times 31.75$~\AA$^3$, with a grid spacing of
    $0.265$~\AA.}
  \label{fig:oep-results}
\end{figure}

For the practical implementation, we reformulate the OEP equation in the
following form, as it is commonly done to construct the electronic
OEP~\cite{kuemmel2008}:
\begin{align}
  S_\sigma(\textbf{r}) = \sum_{i=1}^{N_\sigma}\psi^*_{i\sigma}(\textbf{r})\varphi_{i\sigma}(\textbf{r}) - \Lambda_{i\sigma}(\textbf{r}) + c.c.
  \label{eq:oep-S}
\end{align}
and update the potential with
\begin{align}
  v_{x\sigma}^{(new)}(\textbf{r}) = v_{x\sigma}^{(old)}(\textbf{r}) + c(\textbf{r})S_\sigma(\textbf{r})\,.
  \label{eq:oep-vxc-new}
\end{align}
The quantity $S_\sigma(\textbf{r})$ becomes a measure for convergence, since it
is vanishing in the case of convergence (compare Eq.~\eqref{eq:oep-S} and
Eq.~\eqref{eq:soep}). For the function $c(\textbf{r})$, we have different
possibilities, like using a constant or using the inverse of the electron
density, as used in Ref.~\onlinecite{kuemmel2003}. Other methods are also
possible, such as the Barzilai-Borwein method.~\cite{hollins2012} We found
stable algorithms when using a constant and when using the Barzilai-Borwein
method.

While we will show the computational feasibility of this approach in the
following, often a simplified solution is beneficial as starting point for the
self-consistency procedure. Such a simplified approximation can be deduced by
reformulating Eq.~\eqref{eq:soep2} into the following equivalent form:
\begin{align}
  \begin{split}
    v_{x\sigma}(\textbf{r}) =& \frac{1}{2n_\sigma(\textbf{r})} \sum_{i=1}^{N_\sigma} \bigg(
    \bra{\varphi_{i\sigma}} v_{x\sigma} \ket{\varphi_{i\sigma}} |\varphi_{i\sigma}(\textbf{r})|^2
    + \bigg[\{M^*_{i\sigma}(\textbf{r}) + v_{x\sigma}(\textbf{r})\varphi^*_{i\sigma}(\textbf{r})\}\\
      &- \bra{\{M_{i\sigma}+v_{x\sigma}\varphi_{i\sigma}\} }\varphi_{i\sigma}\rangle\varphi^*_{i\sigma}(\textbf{r})
      - (\hat{h}_{s\sigma}(\textbf{r}) -\epsilon_{i\sigma})\psi^*_{i\sigma}(\textbf{r})
      \bigg] \varphi_{i\sigma}(\textbf{r})
    \bigg) + c.c.
  \end{split}
  \label{eq:kli}
\end{align}
and subsequently assuming $(\hat{h}_{s\sigma}(\textbf{r})
-\epsilon_{i\sigma})\psi^*_{i\sigma}(\textbf{r}) = 0$ to start the iterative
process. In situations where $\Lambda_{i\sigma}(\textbf{r}) = 0$, such as pure
electronic exact exchange, this approximation is exact for a single electron and
is referred to as the Krieger-Li-Iafrate (KLI)
approximation.~\cite{krieger1990,krieger1992construction,krieger1992systematic,kuemmel2003,flick2017c}
By multiplying Eq.~\eqref{eq:kli} by $|\varphi_{j\sigma}(\textbf{r})|^2$ and
integrating over the spatial coordinates, we arrive at a linear equation which
in turn can be solved for the approximate $v_{x\sigma}$. This approximation
scheme has proven to often deliver sufficiently accurate results for electronic
structure calculation with significantly lower computational effort and reliable
stability. As this approximation can be seen as a diagonal approximation to the
response function, it unavoidably fails in accurately describing polarization
features. In the context of light-matter correlated ground-states, this leads to
a slight unbalancing when approximating components including photonic
excitations (introduced by $\Phi^{(1)}$) and self-polarization interaction
(introduced by
$\Phi^{(2)}$).~\cite{pellegrini2015,flick2017c,schafer2018insights,rokaj2017}
This results in a violation of translational invariance and introduces an
artificial dependence on the permanent dipole. When performing KLI calculations
including light-matter interaction, we thus suggest moving the set of
coordinates into the electronic center-of-charge instead of the center-of-mass
frame. To reduce the effect of this dependence during a self-consistent
calculation, the optional input parameter KLIpt\_coc has been introduced in the
code. When activated, this option defines the dipole operator $\hat{d}_{\alpha}$
with respect to the electronic center-of-charge and can improve the stability of
the algorithm.

Finally, in Fig.~\ref{fig:oep-results} we show the capabilities of the new
implementation, where we calculate two sodium dimers in the weak coupling regime
under light-matter coupling. Figure~\ref{fig:oep-results}~(a) shows the electron
density and Fig.~\ref{fig:oep-results}~(b) shows the comparison of OEP and KLI
results. As shown in Ref.~\onlinecite{flick2017c}, in the weak-coupling regime
the KLI is close to the OEP result. In Fig.~\ref{fig:oep-results}~(c) we show
the convergence behavior when using a constant in Eq.~\eqref{eq:oep-vxc-new} and
when using the Barzilai-Borwein method.~\cite{hollins2012}

Extensions of QEDFT to the regime of vibrational strong
coupling,~\cite{flick2018b} the linear-response regime,~\cite{flick2018} as well
as multitrajectory methods that capture quantum fluctuations~\cite{hoffmann2019}
are currently work in progress and will further strengthen the capabilities of
the Octopus code for the real-space description of strong light-matter coupled
systems. To describe effects of the ultra-strong coupling regime, one can use an
alternative method that is presented in the next section.

\section{Dressed Reduced Density Matrix Functional Theory for Ultra-strongly Coupled Light-Matter Systems}
\newcommand{\td}{\mathrm{d}}

The accurate description of the (ultra-)strong coupling regime of light-matter
systems is a formidable task. In many cases the known functionals for QEDFT (see
Sec. \ref{sec:qedft}) are inaccurate and for complex electronic systems typical
few-level approximations become unreliable.~\cite{Galego2019,schafer2018ab} In this section,
we present the Octopus implementation of an alternative real-space ab-initio
method for coupled light-matter systems. Dressed Reduced Density Matrix
Functional Theory (dressed RDMFT)~\footnote{Note that Hartree-Fock theory is
  also included within RDMFT by fixing the orbital occupations to 0 and 1.}
extends standard electronic RDMFT to coupled light-matter systems similarly to how QEDFT extends DFT. First tests on simple model systems suggest that
dressed RDMFT remains accurate from the weak to the ultra-strong coupling
regime. A proper introduction of the theory, examples, and convergence studies
can be found in Ref.~\onlinecite{Buchholz2018}.

This approach allows for the description of an interacting $N$-electron system
coupled to one photonic mode within the dipole approximation. The respective
Hamiltonian is given in Eq.~\eqref{eq:Hamiltonian_dipole} of
Sec.~\ref{sec:qedft}. Note that we set $j_{ext}=0$ throughout this section and
that the ground state $\Psi$ of Eq.~\eqref{eq:Hamiltonian_dipole} depends on
$4N+1$ coordinates, \textit{i.e.},
$\Psi=\Psi(\mvec{r}_1,\sigma_1,...\mvec{r}_N,\sigma_N,p)$, where $\{\sigma_i\}$
denote the spin degrees of freedom and p is the elongation of the photon mode.

Within dressed RDMFT, the original Hamiltonian \eqref{eq:Hamiltonian_dipole} of
$N$ electrons in $d$ dimensions and one mode is replaced with an extended
auxiliary Hamiltonian of $N$ dressed fermions in $d+1$ dimensions with
coordinates $\mvec{z}=(\mvec{r},q)\in \mathbb{R}^{d+1}$. This auxiliary
Hamiltonian reads
\begin{align}
  \hat{H}' &= \sum_{k=1}^{N}\left[- \tfrac{1}{2} \Delta_k' + v'(\mvec{z}_k) \right] + \tfrac{1}{2} \sum_{k \neq l} w'(\mvec{z}_k,\mvec{z}_l)
  \label{AuxiliaryHamiltonian}
\end{align}
and gives access to the same physics (see Ref.~\onlinecite{Buchholz2018},
Sec. 4). For a $d=1$ matter subsystem, the operators introduced in
Eq.~\eqref{AuxiliaryHamiltonian} read: the dressed Laplacian
$\Delta'=\tfrac{\partial^2}{\partial x^2}+\tfrac{\partial^2}{\partial q^2}$, the
dressed local potential
\begin{equation}
  v'(\mvec{z}) = v(x)+ \left[ \tfrac{1}{2}\omega^2q^2 - \tfrac{\omega}{\sqrt{N}} \lambda q x + \tfrac{1}{2}(\lambda x)^2 \right]\,,
\end{equation}
and the dressed interaction kernel
\begin{equation}
  w'(\mvec{z},\mvec{z}')= w(x, x') + \left[ - \tfrac{\omega}{\sqrt{N}} \lambda q x' - \tfrac{\omega}{\sqrt{N}} \lambda q' x + \lambda^2 x x' \right]\,.
\end{equation}
The ground state
$\Psi'(\mvec{z}_1,\sigma_1,...,\mvec{z}_N,\sigma_N)=\Psi(x_1,\sigma_1,...,x_N,\sigma_n)\otimes\chi(p_2,...,p_N)$
of $\hat{H}'$ is a product of the original physical ground state $\Psi$ and the
ground state of $\chi$, which in turn is the product of $N-1$ harmonic
oscillator ground states. The auxiliary Hamiltonian~\eqref{AuxiliaryHamiltonian}
contains only one-body and two-body terms in terms of the dressed coordinates,
which makes in principle every standard electronic structure method applicable
(see also Ref.~\onlinecite{Buchholz2018}, Secs. 4 and 5). We use this
construction to develop dressed RDMFT and dressed Hartree-Fock (HF). For that,
we define the dressed (spin-summed) first order reduced density matrix (1RDM)
\begin{align}
  \gamma'(\mvec{z},\mvec{z}') = N \sum_{\sigma_1,...,\sigma_N}\int \td^{2(N-1)}  z \, {\Psi'}^*(\mvec{z}' \sigma_1,\mvec{z}_2 \sigma_2,...,\mvec{z}_{N} \sigma_{N}){\Psi'}(\mvec{z} \sigma_1,\mvec{z}_2 \sigma_2,...,\mvec{z}_{N} \sigma_{N})\,.
  \label{eq:1RDM_polaritonic}
\end{align}
To apply RDMFT theory on the auxiliary system, we have to replace the total
energy functional of electronic RDMFT, given in Ref.~\onlinecite{Andrade2015},
with the newly introduced quantities of the dressed system, i.e., the auxiliary
Hamiltonian of Eq.~\eqref{AuxiliaryHamiltonian} (approximately) evaluated by the
dressed 1RDM $\gamma'$ of Eq.~\eqref{eq:1RDM_polaritonic}. By that, common
approximations for the two-body energy expression in terms of the 1RDM can be
directly transferred from electronic theory to the dressed system.~\footnote{At
  the current state of the code, we have only implemented the so-called
  M{\"u}ller functional,~\cite{Mueller1984} but one could potentially use any
  RDMFT functional to approximate two-body expression.}  The minimization is
performed like in the electronic case and is based on the RDMFT implementation
of Octopus, though the convergence of the dressed system requires a more
complicated protocol that can be found in the Supplement of
Ref.~\onlinecite{Buchholz2018}. The current implementation in Octopus
approximates the conditions under which the dressed 1RDM corresponds to a wave
function by ensuring the fermionic ensemble N-representability
conditions.~\cite{C1963} However, the auxiliary wave function exhibits also
another exchange symmetry with respect to the auxiliary coordinates, which is
currently neglected. For the practical validity of this approach, the reader is
referred to Ref.~\onlinecite{Buchholz2018} (Sec.~5 and the Supplement).

\begin{figure}
  \includegraphics[width=0.49\columnwidth]{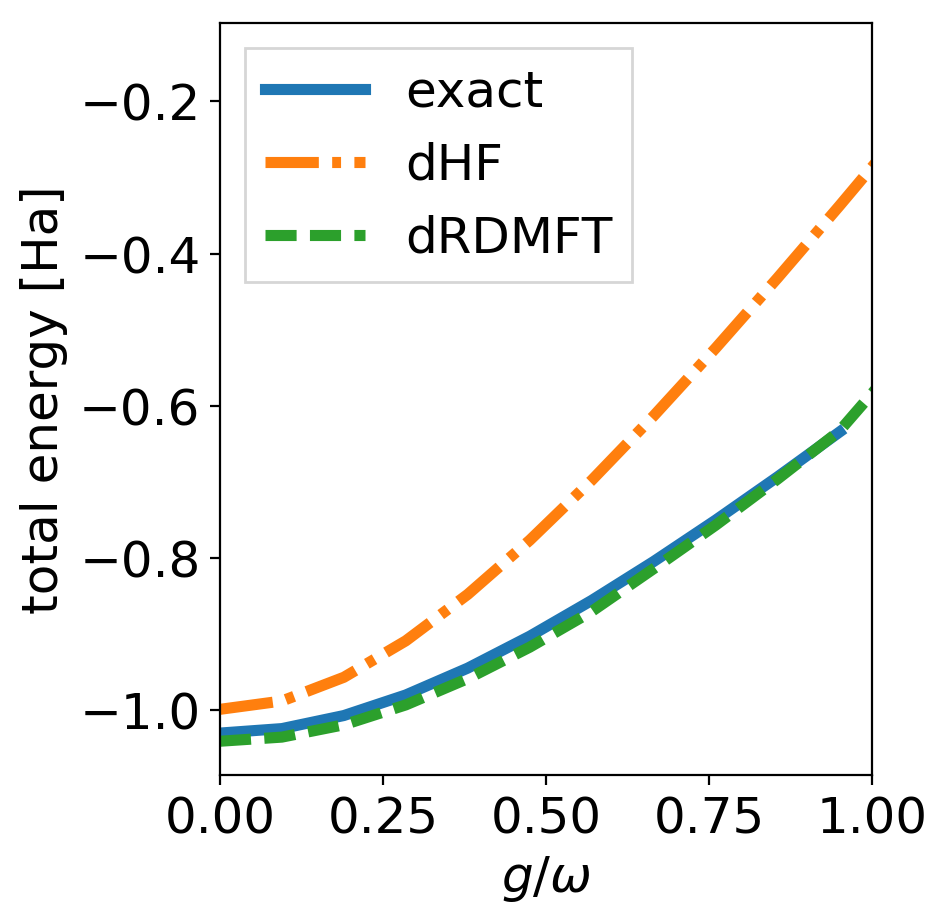} \hfill
  \includegraphics[width=0.49\columnwidth]{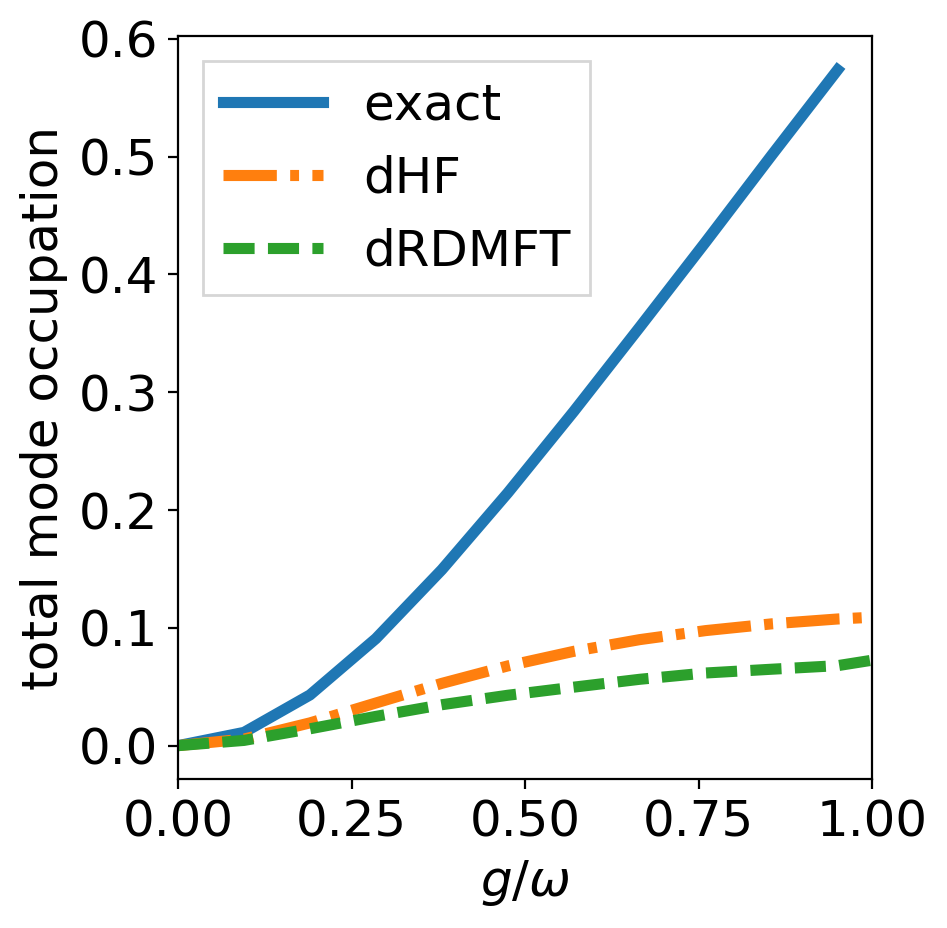}
  \caption{Differences of dressed HF (dHF) and dressed RDMFT (dRDMFT) from the
    exact ground state energies (in hartree, left) and from the exact photon
    number (right) as a function of the coupling $g/\omega$ for the stretched
    H$_2$ molecule in the dressed orbital description. Dressed RDMFT improves
    considerably upon dressed HF for the energy, which is especially due to the
    better description of the electronic correlation. The photon number, an
    example of a photonic observable, is captured similarly with both methods.}
  \label{fig:He_E_nphot_lambda}
\end{figure}

As an example, we consider the one-dimensional H$_2$ molecule in a soft-Coulomb potential with a slightly
stretched bond-length of $b=2.0$ bohr~\footnote{The respective results for the
  equilibrium distance $b=1.61$ bohr are shown in
  Ref.~\onlinecite{Buchholz2018}, Sec. 7.} that is modeled by the local
potential
\begin{equation}
  v_{\mathrm{H}_2}(x) = -\frac{1}{\sqrt{(x-b/2)^2+1}}-\frac{1}{\sqrt{(x+b/2)^2+1}} + \frac{1}{\sqrt{b^2+1}}\,,
\end{equation}
and the soft Coulomb interaction
\begin{equation}
w(x,x') = \frac{1}{\sqrt{|x-x'|^2+1}}\,.
\end{equation}
In Fig.~\ref{fig:He_E_nphot_lambda}, we show the total energy and the total
photon number of the dressed RDMFT, the dressed HF, and the exact many-body
calculations~\footnote{Performed with the many-body routine of Octopus, see
  Ref.~\onlinecite{andrade_real-space_2015}, Sec. 13 for details.} for different
coupling strengths. We see that for small couplings, both observables are
captured well by dressed RDMFT and dressed HF. With increasing coupling
strength, both approximations fail to capture the strongly increasing photon
number. For the total energy instead, dressed RDMFT remains very close to the
exact result, whereas the deviations to dressed HF increase with increasing
coupling strength. This shows the potential of dressed RDMFT to describe
correlated electron systems that are strongly coupled to a cavity mode. In the
future, we plan to investigate better approximations to the polaritonic
N-representability conditions that also account for the symmetry of the
many-body wave function with respect to the exchange of photon coordinates.

\section{Towards dynamics of strongly correlated systems: The TDDFT+{\it U} functional}
It was soon recognized that the standard local and semilocal functionals of DFT
tend to over-delocalize the electrons, as usual approximations are based on the
homogeneous electron gas. This leads to several failures of DFT for materials in
which the localization of electrons plays a critical role in dictating the system's
properties. This is, for instance, the case for transition metals oxides. The
DFT+$U$ method was originally proposed to compensate for some of the failures of
the LDA for such
materials.~\cite{anisimov_band_1991,anisimov_density-functional_1993,liechtenstein_density-functional_1995,anisimov1997first}
In essence, the DFT+$U$ method aims at a better description of the local
electron-electron interaction, which is achieved by adding the mean-field
Hubbard model on a chosen localized subspace to the DFT total energy. The double
counting of electron interaction in this localized subspace is then removed. The
DFT+$U$ total energy functional reads
\begin{equation}
  E_{\mathrm{DFT+U}}[n,\{n_{mm'}^{I,\sigma}\}] = E_{\mathrm{DFT}}[n] + E_{ee}[\{n_{mm'}^{I,\sigma}\}] - E_{dc}[\{n_{mm'}^{I,\sigma}\}]\,,
  \label{eq:E_DFT_U}
\end{equation}
where $E_{ee}$ is the usual electron-electron interaction energy, and $E_{dc}$
accounts for the double counting of the electron-electron interaction already
present in $E_\mathrm{DFT}$.  This double-counting term is not known in the
general case and this is a general problem to all +U methods. Several
approximated forms have been proposed along the
years.~\cite{PhysRevLett.115.196403,PhysRevB.67.153106} In the Octopus code, we
implemented the most commonly used double-counting terms: the fully localized
limit (FLL)~\cite{PhysRevB.57.1505} and the around-mean field (AMF)
double-counting terms.~\cite{PhysRevB.49.14211} They respectively read
as~\cite{PhysRevB.67.153106}
\begin{equation}
  E_{dc}^{\mathrm{FLL}}[\{n_{mm'}^{I,\sigma}\}] = \frac{U}{2}N(N-1) - \frac{J}{2}\sum_{\sigma}N^\sigma(N^\sigma-1)
\end{equation}
and
\begin{equation}
  E_{dc}^{\mathrm{AMF}}[\{n_{mm'}^{I,\sigma}\}] = UN_{\uparrow}N_{\downarrow} - (U-J)\frac{2l}{2l+1}\sum_{\sigma}N_\sigma^2\,,
\end{equation}
where $N_\sigma = \sum_{m}n_{mm}^\sigma$ and $N=N_{\uparrow}+N_{\downarrow}$.
The $E_{ee}$ and $E_{dc}$ energies depend on the density matrix of a localized
orbital basis set $\{\phi_{I,m}\}$, which are localized orbitals attached to
the atom $I$. In the following we refer to the elements of the density matrix of
the localized basis as occupation matrices and we denote them
$\{n_{mm'}^{I,\sigma}\}$. In the rotational-invariant form of DFT+U proposed by
Dudarev \textit{et al.},~\cite{PhysRevB.57.1505} and for the FLL double-counting
term, we obtain the $E_U$ energy to be added to the DFT total energy, which only
depends on an effective Hubbard U parameter $U^{\mathrm{eff}}=U-J$:
\begin{equation}
  E_U[\{n_{mm'}^{I,\sigma}\}] = E_{ee}[\{n_{mm'}^{I,\sigma}\}] - E_{dc}[\{n_{mm'}^{I,\sigma}\}] 
  = \sum_{I,n,l} \frac{U^{\mathrm{eff}}_{I,n,l}}{2} \sum_{m,\sigma}\Big(n_{mm}^{I,n,l,\sigma} - \sum_{m'}n_{mm'}^{I,n,l,\sigma}n_{m'm}^{I,n,l,\sigma} \Big)\,,
  \label{eq:E_Dudarev}
\end{equation}
where $I$ is an atom index, $\sigma$ is the spin index, and $n$, $l$, and $m$
refer to the principal, azimuthal, and angular quantum numbers, respectively.
In the case of a periodic system, the occupation matrices
$n^{I,n,l,\sigma}_{mm'}$ are given by
\begin{equation}
  n^{I,n,l,\sigma}_{mm'} = \sum_{n}\sum_{\mathbf{k}}^{\mathrm{BZ}} w_\mathbf{k}f_{n\mathbf{k}}^\sigma \braket{\psi_{n,\mathbf{k}}^{\sigma} | \phi_{I,n,l,m}}  \braket{\phi_{I,n,l,m'}|\psi_{n,\mathbf{k}}^{\sigma}}\,,
  \label{eq:occ_matrices}
\end{equation}
where $w_{{\mathbf k}}$ is the $\mathbf{k}$-point weight and
$f_{n\mathbf{k}}^{\sigma}$ is the occupation of the Bloch state
$\ket{\psi_{n,\mathbf{k}}^{\sigma}}$.  Here, $\ket{\phi_{I,n,l,m}}$ are the
localized orbitals that form the basis used to describe the electron
localization. Details of the implementation can be found in
Ref.~\onlinecite{PhysRevB.96.245133}. We recently extended our original
implementation to be able to construct a localized subspace from localized
states, such as Wannier states,~\cite{xian2019multiflat} and to treat the
intersite interaction.~\cite{intersite}

In its usual formulation, the DFT+$U$ method is an empirical method, in which
the effective $U$ is a parameter of the calculation. However, it recently became
possible to evaluate $U$ and $J$ fully \textit{ab initio} and self-consistently,
using the ACBN0 functional.~\cite{Agapito_PRX} We also implemented this method
in Octopus and extended it to the time-dependent case~\cite{PhysRevB.96.245133}
in order to be able to investigate strongly correlated materials out-of
equilibrium. We showed that the absorption spectra of transition metal oxides,
such as NiO or MnO, are well reproduced by our TDDFT+$U$
simulations.~\cite{PhysRevB.96.245133}

\begin{figure}[t]
  \begin{center}
    \includegraphics[width=0.6\columnwidth]{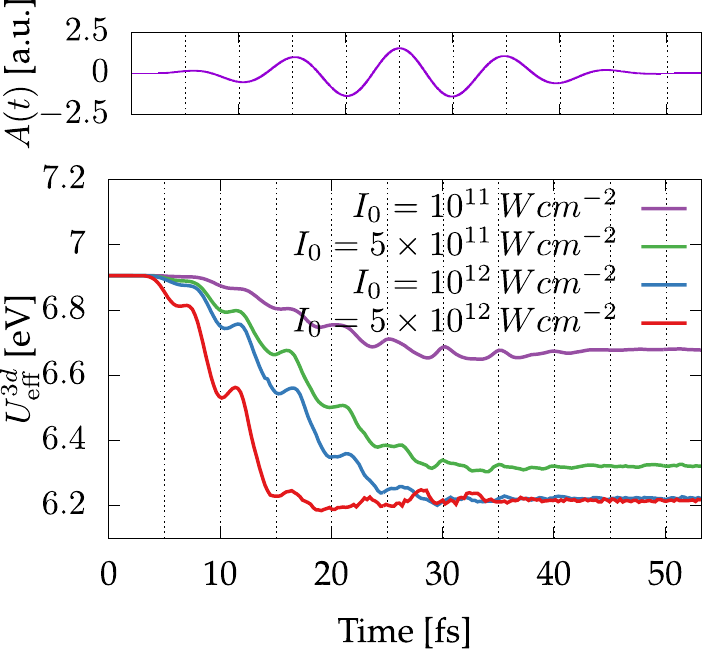}
    \caption{\label{fig:TD_U} Self-consistent dynamics of Hubbard $U$ for the Ni
      $3d$ orbitals (bottom panel) for pump intensities as indicated. The top
      panel represents the time-dependent vector potential and the vertical
      dashed lines indicates the extrema of the vector potential, \textit{i.e.}, the
      minima of the driving electric field. Results from
      Ref.~\onlinecite{PhysRevLett.121.097402}.}
  \end{center}
\end{figure}

Figure \ref{fig:TD_U} shows the calculated time profile of the effective Hubbard
$U_{\mathrm{eff}}=U-J$ for the $3d$ orbitals of Ni, for light-driven NiO. The
top panel shows the time profile of the driving vector potential. This shows
that strongly-driven correlated materials cannot be described by simply assuming
that the effective electronic parameters (here, the effective Hubbard $U$) remain
constant out of equilibrium. Moreover, the possibility to tune these effective
electronic parameters offers opportunities for light-driven phase transitions,
such as light-induced magnetic Weyl semimetals.~\cite{topp2018all}

\section{van der Waals interactions}

The van der Waals (vdW) interactions arise from correlations between electrons
and are in principle described by the exact energy functional through the
exchange and correlation energy $E_{xc}[n(\mathbf{r})]$. However, the vdW
interactions are inherently non-local and long-range and, by construction,
cannot be described by usual local and semi-local
functionals.~\cite{berland2015van} Therefore, much work has been devoted to
finding consistent ways to enhance available functionals to correctly describe
the necessary correlations yielding vdW forces.

Within a pure DFT approach, the inclusion of vdW interactions should be done
through the exchange and correlation functional. To this end, a family of
non-local density functionals has been proposed.~\cite{BerCooLee15} The
so-called vdW density functionals (vdW-DF) are derived \textit{ab-initio} from the
screened response of the homogeneous electron gas.~\cite{berland2015van} Another
route to include vdW interactions in DFT calculations is by adding explicit
corrections to the energy and forces based on atomic parametrizations. The most
well-known method of this type is probably the vdW-D3 scheme of dispersion
corrections from Grimme.~\cite{Grime_2010} Although computationally more
favorable than the vdW-DF method, the major disadvantage of this type of
approach, particularly from the perspective of time-dependent simulations, is
that it fails to correctly describe effects that cannot simply be understood
from atomic configurations. A trivial example would be a lone electron traveling
through space. Another scheme based on explicit corrections to the energy and
forces was proposed by Tkatchenko and Scheffler
(vdW-TS).~\cite{PhysRevLett.102.073005} This scheme has the advantage of
retaining much of the low computational cost of the vdW-D3 scheme, while making
the atomic parametrization dependent on the electronic density. Recently, these
three schemes (vdW-DF, vdW-TS, and vdW-D3) were implemented in the Octopus code
to deal with vdW interactions in isolated and periodic systems.

\subsection{vdW-DF}
\newcommand{\dee}{\mathrm d}

Octopus supports vdW-DF functionals~\cite{BerCooLee15} through the libvdwxc
library.~\cite{libvdwxc} The vdW-DF functionals are expressed as a sum,
\begin{equation}
  E_{\mathrm{xc}}^{\textrm{vdW-DF}}[n] = E_{\mathrm c}^{\mathrm{LDA}}[n] + E_{\mathrm x}^{\mathrm{GGA}}[n] + E_{\mathrm c}^{\mathrm{nl}}[n]\,,
  \label{vdw-functional-full}
\end{equation}
of the LDA correlation energy,~\cite{PerWan92} a GGA exchange energy, and a
fully non-local correlation term. The latter is the integral over a kernel
function $\phi(q_0, q_0', r)$,
\begin{equation}
  E_{\mathrm c}^{\mathrm{nl}} = \frac 12 \iint n(\mathbf r) \phi(q_0(\mathbf r), q_0(\mathbf r'), |\mathbf r - \mathbf r'|) n(\mathbf r')
  \, \dee \mathbf r \, \dee \mathbf r'\,,
\end{equation}
where $q_0(\mathbf r)$ depends on the local density and its gradient. Explicit
evaluation of this 6-dimensional integral is very expensive and scales as volume
squared, $\mathcal O(N^2)$.  Roman-P\'erez and Soler proposed an efficient
method which approximates it as a sum of 3-dimensional
integrals.~\cite{RomSol09} The method works by expressing the integrand as a
convolution using a limited set of helper functions, then applying the Fourier
convolution theorem for a scaling of $\mathcal O(N \log N)$ and much lower
prefactor. This has since become the standard method for evaluating vdW-DF
functionals.

As implemented in Octopus, the density is redistributed from its normal uniform
grid of arbitrary shape onto a uniform 3D grid forming a cube or parallelepiped,
which is suitable for 3D Fourier transforms. libvdwxc then evaluates the
non-local energy and contributions to the potential, relying on the FFTW
library~\cite{FFTW05} for efficient parallel Fourier transforms.  After
calculating the energy and potential, the potential is redistributed back to the
original form.

Octopus supports the standard functionals vdW-DF1,~\cite{DioRydSch04,Dion05}
vdW-DF2,~\cite{LeeMurKon10} and vdW-DF-cx,~\cite{BerHyl14} as well as other
common forms that differ by substituting a different GGA exchange functional in
Eq.~\eqref{vdw-functional-full}. Some common supported variations are
vdW-DF-optPBE, vdW-DF-optB88,~\cite{KliBowMic10} and
vdW-DF-C09.~\cite{cooper10p161104}

\subsection{vdW-TS}

\begin{figure}
  \centering
  \includegraphics[width=0.75\columnwidth]{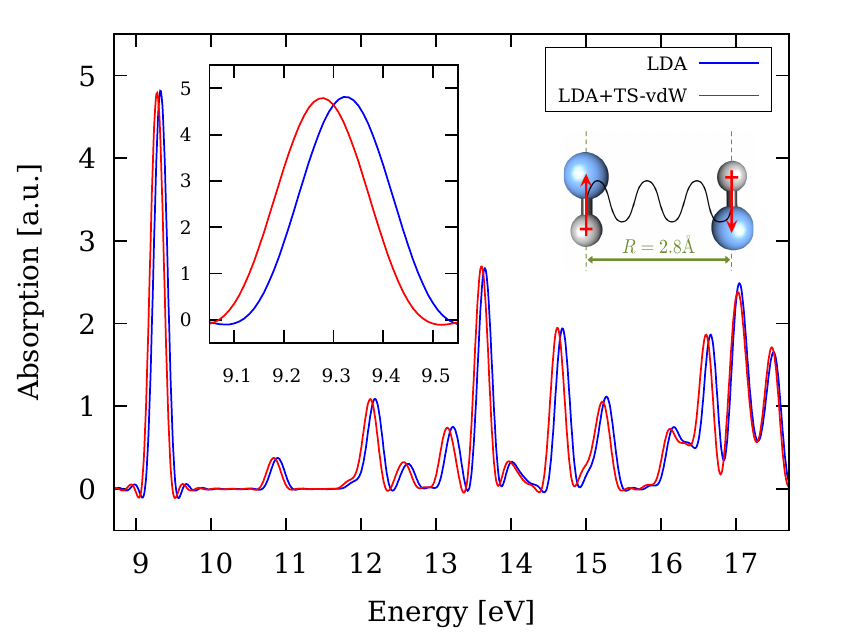} 
  \caption{Absorption cross section spectrum of the hydrogen fluoride dimer with
    and without van der Waals effects. A zoom-in of the absorption feature at
    around 133 nm illustrates a small red shift when van der Waals effects are
    considered. Both hydrogen fluoride molecules are placed on the same plane at
    a separation of 2.8~\AA.}
  \label{dft-d_fig_ACSvsE}
\end{figure}

Since the vdW-TS approach depends on the density, the effect of the van der Waals interaction can be observed in properties other than the forces.
In particular, we expect to observe an effect in the excited states of systems that interact through vdW forces.
We use the hydrogen fluoride dimer as a simple model system for a
proof-of-concept application of the modular implementation of the TS-vdW
functional correction on TDDFT calculations. The dimer geometry is setup as
shown in Fig.~\ref{dft-d_fig_ACSvsE}. The hydrogen fluoride monomers are placed
in anti-parallel fashion, each one with its main symmetry axis oriented along
the $y$-axis. The hydrogen fluoride bond in each monomer is 0.92~\AA\ long and
the molecules are separated by 2.8~\AA\ along the $z$-axis. At this distance,
the van der Waals interaction between the monomers is the strongest, according
to the TS-vdW model. We choose this dimer model because it is a conveniently
small system in which the effects of including the dispersion correction can be
shown at an affordable computationally cost.

To calculate absorption, we excite the system with an infinitesimal
electric-field pulse, and then the time-dependent Kohn-Sham equations are
propagated for 30.38535~${\hbar}$/eV. The singlet dipole spectrum is evaluated
from the time-dependent dipole moment. The strength of the perturbation is set
to 0.01~\AA$^{-1}$ and it is polarized in the $z$-axis. The time evolution is
carried out using the Enforced Time-Reversal Symmetry (ETRS) propagator, with
(default) time steps of 0.03352~${\hbar}$/eV.

The results on Fig.~\ref{dft-d_fig_ACSvsE} show a small van der Waals-induced
bathochromic-like (red) shift in the optical spectrum of the hydrogen fluoride
dimer calculated with the LDA. This example opens the door for a new series of
applications in supra-molecular chemistry, structural biology, polymer science,
etc., that incorporate van der Waals effects on real-time electron dynamics.

\subsection{vdW-D3}

Octopus also supports the DFT-D3 van der Waals correction.~\cite{Grime_2010}
This correction depends only on the atomic positions and does not depend on the
atomic density. It is implemented in Octopus by linking to the DFT-D3 library
provided by the authors. As we have validated Octopus results with the reference
data provided with the library, the results are guaranteed to be consistent with
other codes that implement this correction.

The only modification we did to the library was to move the very large set of
coefficients from a source file to a stand alone text file that is parsed only
when necessary. This speeds up compilation and reduces the size of the
binaries. The modified library is distributed with Octopus, so users need not
compile it separately.

\section{Polarizable Continuum Model}

The Polarizable Continuum Model (PCM)~\cite{tomasi2005quantum} comprises a
family of implicit-solvent approaches to tackle quantum-mechanical calculations
of molecules in solution. PCM assumes that (i) the solvent is a continuum and
infinite dielectric medium characterized by a frequency-dependent dielectric
function; and that (ii) a void cavity of appropriate shape and size encapsulates
the solute molecule, separating it from the solvent by a sharp interface. The
numerical implementation of PCM relies on the Apparent Surface Charge (ASC)
approach and the Boundary Element Method (BEM).~\cite{tomasi1994molecular}
In this framework, the solvent polarization response induced by the molecule's
charge density is modeled by a reaction potential defined
by a set of point charges ${\bf q}=\{q_1, q_2, ..., q_T\}$ that spread over a
tessellated cavity surface consisting of $T$ finite surface elements or {\it
  tesserae}.~\cite{Tomasi2007} The reaction potential is the object through
which the complex dielectric environment is accounted for.~\cite{alain_PCM} The
implementation of the PCM in Octopus rests on the Integral Equation Formalism
(IEF).~\cite{Cances1997} The key IEF-PCM equation for computing the polarization
charges is
\begin{equation}
  \label{pcm_eq}
  \mathbf{q}=\mathbf{Q}\mathbf{V}\,,
\end{equation}
where $\mathbf{q}$ and $\mathbf{V}$ are $T\times1$ column vectors storing the
induced polarization charges and the molecule's electrostatic potential at the
tesserae representative points, respectively. $\mathbf{Q}$ is the $T \times T$
PCM response matrix, which depends on the geometry of the cavity and the
dielectric function of the solvent.~\cite{tomasi2005quantum}

A realistic description of the molecular cavity is key to capture accurate
electronic and optical properties of molecules in solution. In principle, the
cavity should (i) exclude the solvent, (ii) comprise most of the solute
electronic density, and (iii) conform with the molecular shape. Octopus uses the
GEPOL algorithm,~\cite{pascual1994gepol} which builds up the van der Waals
cavity from the union of interlocking spheres with element-specific radii
centered at each atom position (by default no spheres are built around hydrogen atoms). Within GEPOL, the BEM tessellation of the solute-solvent interface is done starting from a 60- or 240-face circumscribed polyhedron per sphere, selecting only exposed tesserae and properly reshaping those that are partially exposed. The BEM tessellation is a surface grid constructed independently from the real-space three-dimensional grid used in Octopus to represent both the Kohn-Sham electronic Hamiltonian and molecular orbitals.  The mismatch between the two discrete representations might cause numerical problems arising from the Coulomb singularities whenever tesserae and grid points are close to each other. The implementation of the PCM in Octopus regularizes such singularities by using normalized spherical Gaussian functions to smooth the discretized polarization charges ${\bf q}$.~\cite{alain_PCM}

Having briefly described the implementation, we now look at specific and
relevant cases for chemical reactions, namely solvation energies and
ground-state stabilization. The most evident effect of the presence of a solvent
is to change the total energy of a system compared to its value in vacuum.
In the framework of DFT, the Kohn-Sham Hamiltonian of the solvated molecule
contains the solvent reaction potential, which is a functional of the electronic
and nuclear densities of the solute molecule.~\cite{alain_PCM} The latter
implies that the Kohn-Sham and IEF-PCM equations become coupled and the
polarization charges ${\bf q}_e = {\bf Q} {\bf V}_\mathrm{Hartree}[\rho^e]$,
induced by the molecule's electronic density, have to be computed at each
Kohn-Sham iteration until convergence is reached. The resulting electronic
density can be used to compute the electrostatic contribution to the solvation
free energy $\Delta
G_\mathrm{el}=G[\rho_\mathrm{solv}^e]-E[\rho_\mathrm{vac}^e]$, where $G$ and $E$
denote the free and total energy functionals of the molecule in solvent and in
vacuum, respectively. The current PCM implementation in Octopus~\cite{alain_PCM}
has been thoroughly tested on a benchmark set of organic molecules and compared
with analogous calculations performed with the quantum chemistry package {\sc
  Gamess}.~\cite{gamess} As an example of the solvent stabilization effect in
the ground state, we studied the nitrobenzene molecule in water (static
dielectric constant set to $\epsilon_s=80$).



\begin{figure}[t]
  \centering
  \includegraphics[width=1\textwidth]{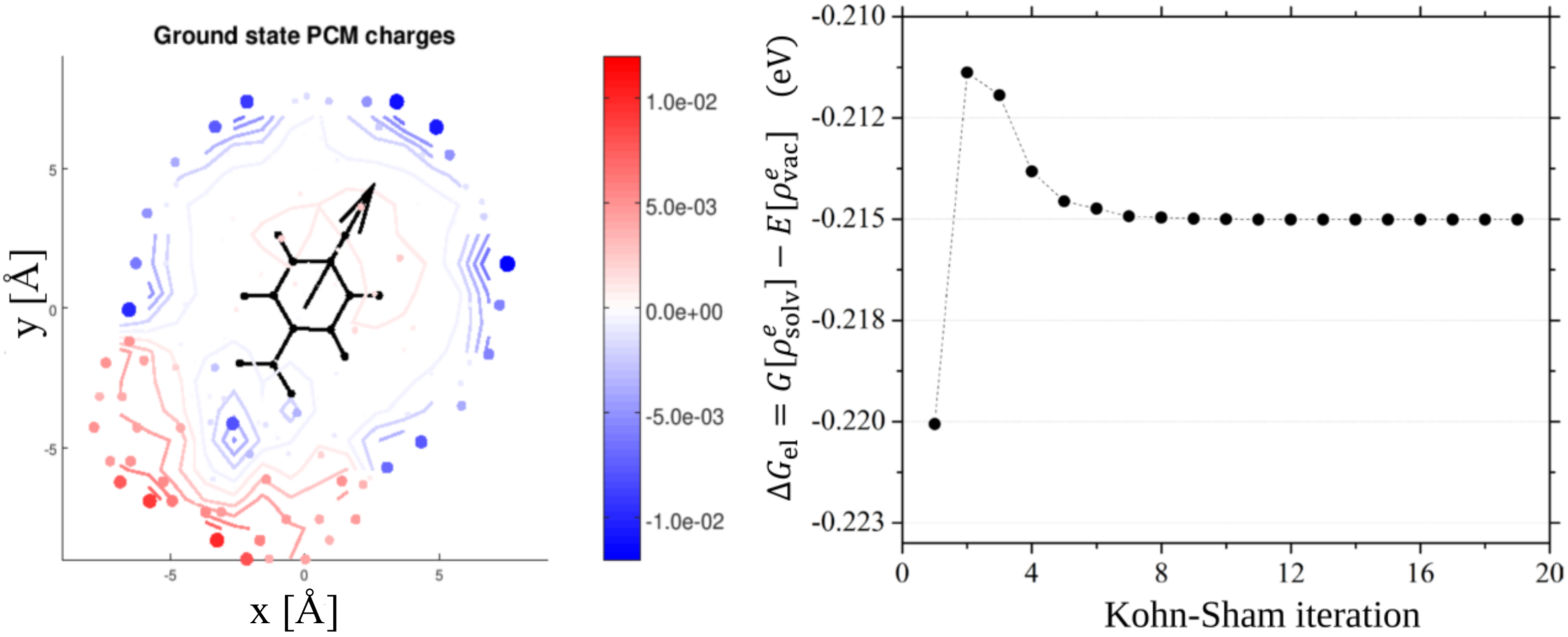}
  \caption{(Left panel) Nitrobenzene molecule in water ($\epsilon_s=80$)
    surrounded by the ground state PCM charges distributed on the solvation
    cavity. The actual distribution of the charges in 3D can be seen in the
    supplementary animation SM1. (Right panel) Electrostatic contribution to the
    solvation free energy of the computed at each Kohn-Sham iteration. Kohn-Sham
    equations were solved in real-space as implemented in Octopus using the
    GGA-PBE approximation~\cite{perdew1998perdew} to the exchange-correlation
    energy. The simulation box was built using spheres of radius 5~\AA\ and an
    uniform spacing of 0.19~\AA\ between grid points. The radii of the spheres
    used to build the cavity surface of the solute molecule are 2.4~\AA\ for
    carbon, 1.8~\AA\ for oxygen and 1.9~\AA\ for nitrogen.}
  \label{fig:pcm_solvation_energy}
\end{figure}

In Fig.~\ref{fig:pcm_solvation_energy}, we show the convergence of the solvation
free energy as a function of the Kohn-Sham
iteration. The solvation free energy is stabilized already after 10 Kohn-Sham
iterations out of the 19 required to optimize the molecular orbitals of the
solvated molecule. We have also verified that the numerical error inherent to
the discretization of the solute cavity surface is very small. For example, the
total polarization charge $Q_\mathrm{PCM}=\sum_{i=1}^T q_i^e$ at each Kohn-Sham
iteration only deviates by only $0.03 \%$ from the actual number of valence
electrons in the nitrobenzene molecule (the relation between $\mathbf{q}$ and
solute charge is determined by Gauss's
theorem~\cite{tomasi2005quantum}). However, the magnitude of this error will
depend in general on the size and the geometry of the molecule.

Now we move forward and describe the extension of the previous PCM method to the
time domain, using real-time PCM and non-equilibrium solvation. Ground state PCM
is unable to capture the complex dynamical interactions between the solvent and
the solute when the latter is in an excited state. Fortunately, the PCM admits a
generalization to account for solvation dynamics.~\cite{Caricato2005} The
time-dependent PCM (TD-PCM) implementation in Octopus comes in three flavors of
increasing complexity, all coupled with real-time TDDFT calculations of the
solute molecules. The first one, called equilibrium TD-PCM,~\cite{alain_PCM}
assumes that the solvent is fast enough to instantaneously equilibrate the
solute charge density fluctuations and to polarize accordingly. This approach is
physically sound for weakly polar solvents, having similar values for the static
and dynamic dielectric constants. The second is a nonequilibrium approach called
inertial TD-PCM and amounts to partitioning the solvent response in a fast
(dynamic) and a slow (inertial) part.~\cite{Cammi1995} Faster degrees of freedom
respond instantaneously to the changes of the applied potential (either of
molecular or external in origin), whereas slower degrees of freedom remain
``frozen'' and in equilibrium with the initial value of the field. This
approximation works well when the solvent relaxation times are large enough with
respect to the electronic excitations in the solute molecule (e.g., within the
picosecond scale). The third TD-PCM approach, called equation of motion (EOM)
TD-PCM, considers the full history-dependent evolution of the solvent
polarization through a set of equations of motion for the polarization
charges.~\cite{Corni2014} Nonequilibrium polarization effects of this sort
originate from the frequency-dependent dielectric response of the solvent,
encoding the fact that it takes a different non-negligible time to adjust
to fast or slow electrostatic perturbations. Solvation dynamics affect strongly
and non-trivially the absorption spectrum of molecules, especially for fast and
polar solvents, by inducing solvatochromic shifts of the peaks in the UV-Vis
absorption spectrum and modifying their relative amplitudes. The details about
the implementation of all of these schemes and a detailed discussion of their
effects can be found in Refs.~\onlinecite{alain_PCM} and
\onlinecite{gabriel_PCM}.

\begin{figure}[b]
  \centering
  \includegraphics[width=0.7\textwidth]{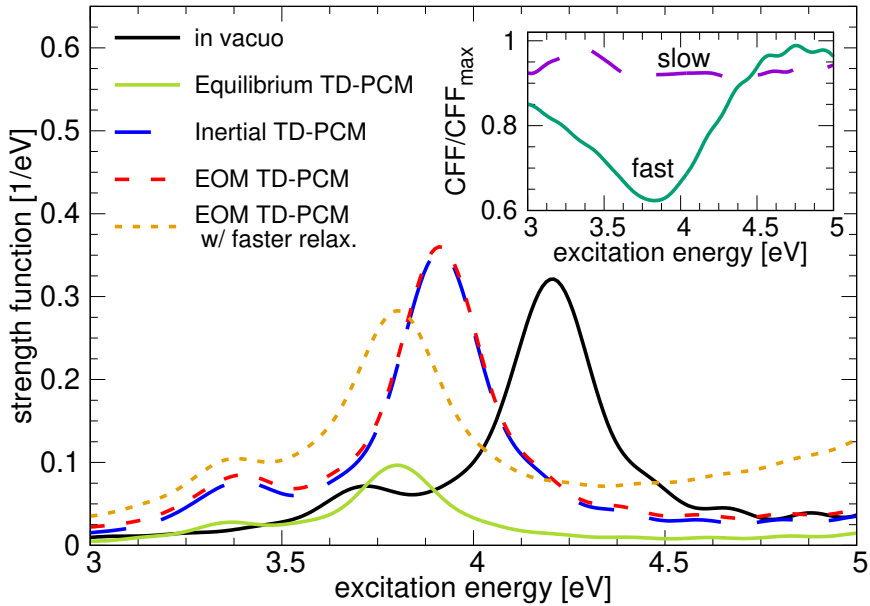}
  \caption{Absorption spectrum of nitrobenzene in vacuo and in water
    ($\epsilon_s=80$, $\epsilon_d = 1.786$) for all the TD-PCM schemes. The
    inset shows the normalized cavity field factor (CFF) vs. the excitation
    energy for the two EOM TD-PCMs, with solvent relaxation times of 3.37~ps
    (slow) and 1 fs (fast) corresponding to $\mathrm{CFF}_{\mathrm{max}}=1.27$
    and 2.55, respectively. Real-time TDDFT simulations of 20~fs with a time
    step of 1.7$\times10^{-3}$~fs were performed to obtain the absorption
    spectra. An electric dipole perturbation within the linear regime and
    oriented along $x$ was applied as an initial perturbation. The rest of the
    computational details are shared with the ground-state DFT calculations
    leading to Fig.~\ref{fig:pcm_solvation_energy}.}
  \label{fig:pcm_tdpcm}
\end{figure}

Here we show the differences among the TD-PCM flavors described above for the
case of the nitrobenzene molecule in aqueous solution, excited with an
electrical dipole perturbation in the $x$ direction. We take the dynamic
dielectric constant of water, entering in the inertial and EOM TD-PCMs, as
$\epsilon_d = 1.786$. Figure \ref{fig:pcm_tdpcm} shows the photo-absorption
spectrum of the gas-phase and solvated molecule. We can see that, with respect
to the absorption {\it in vacuo}, all of the TD-PCM methods produce shifts of
the features (in this case, toward lower excitation energies). The shifts are
not rigid overall, but depend on the excited state and on the specific solvation
scheme (equilibrium, inertial, EOM). Within Debye's model, the equilibrium and
inertial TD-PCMs are limiting cases for the dynamics when the relaxation time is
zero or infinite, respectively. The EOM TD-PCM, with a finite relaxation time,
interpolates between these limits. Water has a large relaxation time of 3.37~ps
and, therefore, the absorption spectrum is almost coincident for the EOM and the
inertial TD-PCMs, as can be seen in Fig.~\ref{fig:pcm_tdpcm}. Whenever the
solvent is as slow as water, there is not much gain in selecting EOM over
inertial TD-PCM and the latter is the method of choice in terms of computational
performance. Instead, for faster solvents, the EOM results depart from those of the inertial algorithm, approaching to those of the equilibrium TD-PCM. In Fig.~\ref{fig:pcm_tdpcm}, we have considered an effective solvent having the same static and dynamic dielectric constants as water, but with a 1~fs relaxation. The EOM TD-PCM for such a model solvent produces almost the same excitation energy shifts of equilibrium TD-PCM and peak intensities in between the inertial and the
equilibrium TD-PCMs, as expected.


A real-time representation of the molecular dipole coupled with the PCM surface
charges can be appreciated in the supplementary movie SM2, which also highlights
how the different TD-PCM approaches affect the time-evolution of the dipole.

When studying the evolution of a solvated molecule under the effect of an
external time-dependent electromagnetic field, a separate treatment is required
to take into account that there are two solvent polarization contributions
interacting with the solute molecule, namely the reaction and the cavity
fields.~\cite{onsager1936electric} The reaction field comes from the
polarization induced by solute molecule itself, while the cavity field arises as
a polarization induced by the external electric field. Both reaction and cavity
field effects can be accounted for in static and real-time quantum-mechanical
calculations of molecules within PCM and TD-PCM simulations in
Octopus.~\cite{gabriel_PCM}

All of the TD-PCM calculations shown here were performed using both reaction-
and cavity-field effects, although the redshift of the peaks in each TD-PCM
scheme is mainly a feature of the reaction-field, as ascertained in test
calculations (not shown) and as expected.~\cite{corni2005electronic} Cavity
field effects impact directly on the peak intensities, by making
the absorption more favorable depending on how large the effective local field acting on the
molecule is allowed to be by solvent dielectric properties and the geometry of
the cavity (normally, reassembling the molecular shape). Still, cavity field
effects can have a non-trivial influence in the absorption spectrum shape when
considering non-equilibrium solvation dynamics (EOM TD-PCM), by changing the
relative peak intensities. This effect can be seen in the inset of the
Fig.~\ref{fig:pcm_tdpcm}, where the normalized cavity field factor (CFF) -- the
ratio between absorption cross-section with and without cavity field effects --
is plotted against the excitation energy for the EOM TD-PCM simulations with
water and the aforementioned faster water-like solvent. The large difference in
absorption peak intensities between the equilibrium and the rest of the TD-PCMs
in Fig.~\ref{fig:pcm_tdpcm} is also related to the modification of the probing
electromagnetic field inside the dielectric. The photo-absorption cross-section
is, by definition, the ratio between the absorbed and the incoming power of
light. In our case, both increase with the dielectric constant: (i) the larger
the dielectric constant of a medium the larger the CFF; but also (ii) a larger
dielectric constant implies a smaller phase velocity of light, therefore
increasing the power of the traveling electromagnetic wave. The impact of the
latter effect ($\propto \mathrm{~refractive ~index}=
\sqrt{(|\epsilon(\omega)|+\Re{\epsilon(\omega)})/2}$) is stronger than the former
($\propto \mathrm{~CFF}$, e.g., for a spherical
cavity~\cite{onsager1936electric} $\propto
\left|\frac{3\epsilon(\omega)}{2\epsilon(\omega)+1}\right|^2$) for a large
enough dielectric constant such as the one corresponding to the equilibrium
TD-PCM for water ($\epsilon_s=80$).


In conclusion, Octopus is now capable of including an implicit dielectric
continuum model as an environment for quantum mechanical calculations of excited
states in the time-domain. The different versions of the TD-PCM scheme
implemented allow us to select the most suitable to capture the relevant physics
accounting for a full range of different relaxation and response times.

\section{Magnons from real-time TDDFT}

In the last couple of years the first studies investigating magnetization
dynamics from first principles in real time have
emerged.~\cite{BuczekSandratskii:11b,NiesertPhD:11,RousseauBergara:12,GorniBaroni:18,CaoGiustino:18,SinghSharma:19}
Here we are interested in transverse magnetic excitations, specifically magnons,
which are long wavelength collective excitations with a typical energy of tens
to hundreds of milli electronvolts. We recently developed an alternative to the
linear-response TDDFT
formulation~\cite{Savrasov:98,CaoGiustino:18,GorniBaroni:18} to compute the spin
susceptibilities of magnetic systems, based on real-time TDDFT.~\cite{magnons}
Our approach follows the work of Bertsch \textit{et al.} for optical
excitations. In the original work of Bertsch \textit{et al.} the system is
perturbed by a sudden change in the vector potential, which induces a charge-
and current-density response. To investigate magnons, we employ a ``transverse
magnetic kick'', which induces magnetic fluctuations in the system.  To be more
precise, our perturbation corresponds to an infinitely short application of a
Zeeman term,
\begin{equation}
  \delta \hat{H}_{\mathbf{q}}(t) = \frac{\mathcal{B}}{2} \delta(t)\int d^3\mathbf{r} \big[ e^{-i \mathbf{q}\cdot\mathbf{r}} \hat{\sigma}^+(\mathbf{r}) + e^{i \mathbf{q}\cdot\mathbf{r}} \hat{\sigma}^-(\mathbf{r}) \big]\,,
\end{equation}
where $\mathbf{q}$ is the momentum of the spin wave we are exciting,
$\mathcal{B}$ is the strength of the perturbation, and $\sigma^{\pm} =
\sigma_{x}\pm i\sigma_{y}$ are Pauli matrices. In this expression, the $z$ axis
is taken to be along the direction of the magnetization of the system before
excitation. In case the system has a preferred magnetization direction due to
the presence of spin-orbit coupling, usually referred as an ``easy axis'', we
perform a kick in the transverse direction with respect to this ``easy axis'' of
the system.

The subsequent time evolution of the spin magnetization
$\mathbf{m}(\mathbf{r},t)$, governed by the time-dependent Schr\"odinger
equation, is then computed and analyzed in Fourier space to spin
susceptibilities.  If we perturb our system from its ground state and we assume
linear response, we directly have that
\begin{equation}
  m_+(\mathbf{q};\omega) = \chi_{+-}(\mathbf{q}; \omega)  \frac{\mathcal{B}}{2}\,,
\end{equation}
where $\chi_{+-}(\mathbf{q}; \omega)$ is the spin susceptibility we want to
extract. A similar expression is obtained for $\chi_{-+}(\mathbf{q};
\omega)$.

In order to access finite momenta which are a fraction of the Brillouin zone,
one typically has to employ large supercells to perform the dynamics, which can
be computationally very expensive when a few meV energy resolution is needed.
Fortunately, there is a way to circumvent the construction of supercells, the
so-called generalized Bloch theorem (GBT). The GBT has been introduced by
Sandradskii~\cite{Sandratskii:86} for the calculation of ground-state spin waves and requires the
implementation of specific boundary conditions. Therefore, we also implemented
the GBT for the real-time calculation of magnons, taking advantage that Octopus
is a real-space finite differences code, for which we can easily specify any
type of boundary conditions. However, it is important to note that this applies
only in the absence of spin-orbit coupling.  The boundary condition, described
below, depends on the momentum ${\mathbf q}$ of the perturbation and acts
differently depending on whether a state was originally ``up'' or ``down'' with respect
to the unperturbed magnetization. This is determined by the sign of
$\braket{\Phi_{n{\mathbf k}} | S_z| \Phi_{n{\mathbf k}}} $ just before the
perturbation, for each spinor state $|\Phi_{n\mathbf{k}}\rangle$. If we label
these states $\alpha$ and $\beta$, the boundary condition reads
\begin{equation} 
  \Phi_{\alpha, \mathbf{k} n}(\mathbf{r}, t) = e^{i \mathbf{k} \cdot \mathbf{r}} 
  \begin{pmatrix}
    u^{\uparrow}_{\alpha, \mathbf{k} n}(\mathbf{r}, t) \\ 
    e^{i \mathbf{q} \cdot \mathbf{r}} u^\downarrow_{\alpha, \mathbf{k} n}(\mathbf{r}, t)
  \end{pmatrix}\,,
\end{equation}
\begin{equation} 
  \Phi_{\beta, \mathbf{k} n}(\mathbf{r}, t) = e^{i \mathbf{k} \cdot \mathbf{r}} 
  \begin{pmatrix}
    e^{-i \mathbf{q} \cdot \mathbf{r}} u^\uparrow_{\beta, \mathbf{k} n}(\mathbf{r}, t) \\ 
    u^{\downarrow}_{\beta, \mathbf{k} n}(\mathbf{r}, t) \nonumber
  \end{pmatrix} \,.
\end{equation}

\begin{figure}[t]
  \begin{center}
    \includegraphics[width=0.6\columnwidth]{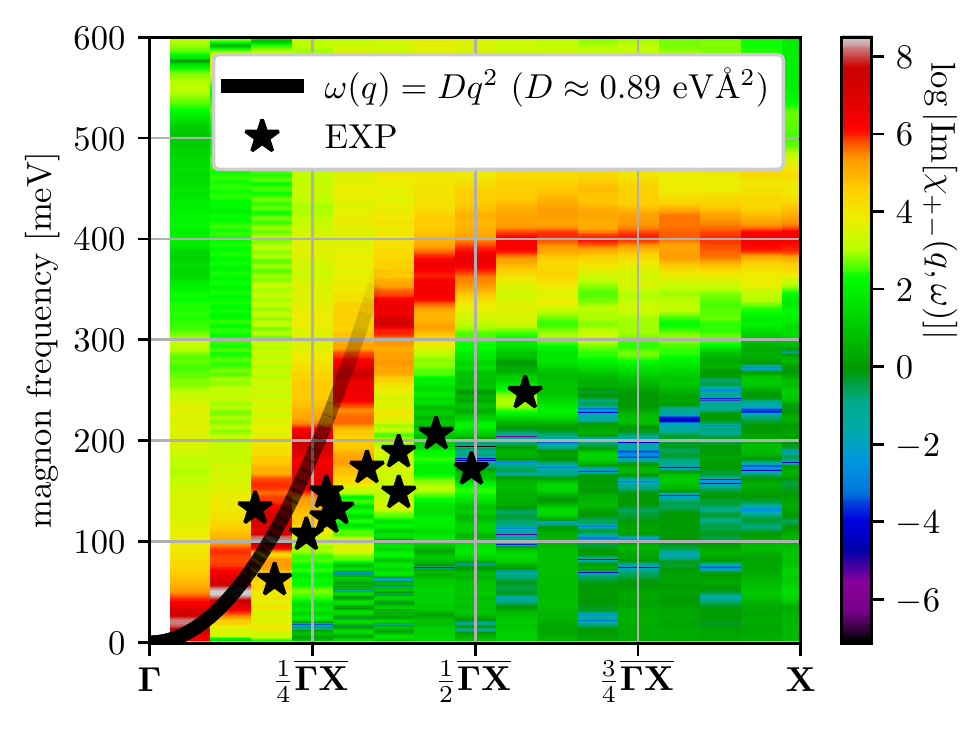}
    \caption{\label{fig:magnonDOS_NI_log} Calculated magnon dispersion: Spin
      susceptibility of bulk Ni along the $\overline{\Gamma X}$ direction,
      displayed in logarithmic scale. Numerical details are given in the main
      text. The experimental data are taken from
      Ref.~\onlinecite{PhysRevLett.54.227}.}
  \end{center}
\end{figure}

We checked that performing the simulation using the GBT or the supercell
approach was leading to the same results, up to numerical precision.  We tested
our approach against cubic Ni, Fe, and Co, which have been widely studied using
linear response TDDFT, and we found that our results are in very good agreement
with previous works,~\cite{NiesertPhD:11} thus validating our
implementation. Figure \ref{fig:magnonDOS_NI_log} shows the results obtained for
bulk nickel using the adiabatic local-density approximation. We used here a
real-space grid spacing of 0.27~bohr, norm-conserving pseudopotentials, and we
employed a $16\times16\times16$ $\mathbf{k}$-point grid shifted 4 times in order
to resolve momenta $\mvec{q}$ which are multiples of $2\pi/(16a)$, where $a$ is
the lattice parameter of Ni, which we took to be $3.436$~\AA. In order to obtain
the response within linear response, we took $\mathcal{B}=0.02$ and we
propagated during 435.4~fs, using a time-step of 1.81~as. To reduce the
numerical burden, symmetries were employed.

Let us now comment on the interest of the proposed approach. So far, we only
used this new method for investigating weak magnetic kicks, where we recover the
results from linear response theory. However, our approach does not rely on the
assumptions of small perturbations and can be used to investigate nonlinear
phenomena induced by strong magnetic field, as well as out-of-equilibrium
situations where the system is kicked from an excited state, which is a strength
of a real-time
method.~\cite{de2013simulating,walkenhorst2016tailored,de2018pump,sato2018ab}
Moreover, we can directly investigate the coupling with other degrees of
freedom, such as phonons or photons, without any new theory or code development.
Using the real-space method, we benefit from the favorable scaling of the time
propagation, which is linear in the number of states, whereas sum-over-states
approaches usually scale quadratically with the number of states.  Finally, our
approach offers the great advantage of not requiring the use of an
exchange-correlation kernel, as only the exchange-correlation potential is
needed to perform a time propagation. This is very interesting in order to test
new functionals and theory levels, for which deriving the expression of the
exchange-correlation kernel can become complicated.  These different aspects
will be investigated in future works.

\section{Orbital magneto-optical response of solids and molecules from a Sternheimer approach}

In the present section we review the implemented routines for magneto-optical
phenomena, which arise from the loss of symmetry between left and right
circularly polarized light in the presence of a magnetic
field.~\cite{barron2004,agranovich1991} While magneto-optical spectra can be
straightforwardly computed for
molecules,~\cite{solheim2008a,solheim2008b,lee2011,seth2008,seth2008a} the
theory for solids has been developed only recently.~\cite{lebedeva2019} The
reason is that external electromagnetic fields break the translational symmetry
of periodic systems, which is formally expressed through the unboundness of the
position operator.~\cite{kingsmith1993,vanderbilt1993,resta1994} The orbital
response to magnetic fields is especially complicated to describe, as such
fields lead to non-perturbative changes in wavefunctions and introduce vector
coupling to electron dynamics.~\cite{essin2010,chen2011,gonze2011} Here we will
focus on calculations of changes in the linear optical response in the presence
of the magnetic field within the Sternheimer
approach.~\cite{andrade_real-space_2015,andrade2007,strubbe_tesis,strubbe2012}

To treat uniform magnetic fields in periodic systems, we use the approach based
on perturbation theory for the one-particle density
matrix.~\cite{essin2010,gonze2011,lebedeva2019} Such an approach allows us to
work under purely periodic boundary conditions and to automatically take into
account gauge invariance. A periodic and gauge-invariant counterpart
$\tilde{\mathcal{O}}$ is distinguished for any operator
$\mathcal{O}=\mathcal{O}_{\mathbf{r}_1\mathbf{r}_2}$ defined for two points
$\mathbf {r}_1$ and $\mathbf {r}_2$ in real space
\begin{equation}
  \mathcal{O}_{\mathbf{r}_1\mathbf{r}_2} =  \tilde{\mathcal{O}}_{\mathbf{r}_1\mathbf{r}_2}\mathrm{exp}\left(-\frac{i}{c}\int_{\mathbf {r}_2}^{\mathbf {r}_1}\mathbf{A}(\mathbf {r}) \mathrm{d} \mathbf {r} \right)\,,
  \label{eq_mo4}
\end{equation}
where $\mathbf{A}$ is the vector potential associated with external electric
($\mathbf{E} = -c^{-1} \partial_t \mathbf{A}$, $c$ is the speed of light) and
magnetic ($\mathbf{B} = \nabla \times \mathbf{A}$) fields and the integral is
taken along the straight line between points $\mathbf {r}_2$ and $\mathbf{r}_1$.

The time-dependent Liouville equation for the gauge-invariant counterpart
$\tilde{\rho}$ of the one-particle density matrix to the first order in
$\mathbf{E}$ and $\mathbf{B}$ takes the form~\cite{lebedeva2019}
\begin{equation}
  -i \partial_t  \tilde{\rho} + [H_0,\tilde{\rho}]= -\frac{1}{2} \left\{
  \mathbf{E}+\frac{1}{c} \mathbf{V} \times \mathbf{B}, [\mathbf{r},\tilde{\rho}] \right\} -  [\delta \tilde{H},\tilde{\rho}]\,.
  \label{eq_mo5}
\end{equation}
Here the commutator and anticommutator of operators $\mathcal{O}^{(1)}$ and
$\mathcal{O}^{(2)}$ are denoted as $[\mathcal{O}^{(1)}, \mathcal{O}^{(2)}]$ and
$\{\mathcal{O}^{(1)}, \mathcal{O}^{(2)}\}$, respectively, the velocity operator
$\mathbf{V}=-i[\mathbf{r},\tilde{H}]$ is computed with account of all non-local
contributions to the Hamiltonian, such as from non-local pseudopotentials, and
the Hamiltonian is represented as $\tilde{H}=H_0+\delta \tilde{H}$, where the
difference $\delta \tilde{H}$ between the gauge-invariant counterpart
$\tilde{H}$ of the Hamiltonian and unperturbed Hamiltonian $H_0$ is related to
the local-field effects. Unlike the singular position operator $\mathbf{r}$, the
commutator $ [\mathbf{r},\tilde{\rho}]$ of the position operator with the
periodic function $\tilde{\rho}$ is well defined in Eq.~\eqref{eq_mo5} and
corresponds to the derivative with respect to the wave vector, $
i\partial_{\mathbf{k}} \tilde{\rho}_{\mathbf{k}}$, in reciprocal
space. Differentiating Eq.~\eqref{eq_mo5}, one finds the derivatives of the
density matrix $\tilde{\rho}^{(P)} = \partial \tilde{\rho}/\partial P$ with
respect to external perturbations $P$ ($\mathbf{E}$, $\mathbf{B}$, etc.).

For TDDFT calculations in Octopus, $\rho$ is the Kohn-Sham density matrix.  The
$n$-th order derivative $\tilde{\rho}^{(P)}$ describing the joint response to
the perturbations $P=P_1P_2...P_n$ is divided into four blocks within and
between the occupied (V) and unoccupied subspaces (C):
$\tilde{\rho}^{(P)}_{\mathrm{VV}}=P_v\tilde{\rho}^{(P)}P_v$,
$\tilde{\rho}^{(P)}_{\mathrm{CC}}=P_c\tilde{\rho}^{(P)}P_c$,
$\tilde{\rho}^{(P)}_{\mathrm{VC}}=P_v\tilde{\rho}^{(P)}P_c$, and
$\tilde{\rho}^{(P)}_{\mathrm{CV}}=P_c\tilde{\rho}^{(P)}P_v$, where $P_{v}
=\rho^{(0)}$ and $P_{c} = 1 - P_{v} $ are the projectors onto the occupied and
unoccupied bands. In accordance with the density matrix perturbation
theory,~\cite{lazzeri2003} to get the elements
$\tilde{\rho}^{(P)}_{\mathrm{CV}}$, Eq.~\eqref{eq_mo5} is projected onto
unperturbed Kohn-Sham wavefunctions $ | \psi_{v\mathbf{k}}^{(0)} \rangle$ of
occupied bands $v$ to give an equation for the response function $|
\eta_{v\mathbf{k}}^{(P)} \rangle=\tilde{\rho}^{(P)}_{\mathrm{CV}} ( \Omega) |
\psi_{v\mathbf{k}}^{(0)} \rangle$
\begin{equation}
  L_{v\mathbf{k}} (\Omega)  | \eta_{v\mathbf{k}}^{(P)} \rangle = P_c R^{(P)} [\tilde \rho^{(n-1)},...,\rho^{(0)},n^{(P)}]  | \psi_{v\mathbf{k}}^{(0)} \rangle\,.
  \label{eq_mo8}
\end{equation}
Here the operator on the left-hand side is given by $L_{v\mathbf{k}} (\Omega) =
\Omega + H_0 - \epsilon_{v \mathbf{k}}$, where $\Omega$ is the frequency
considered and $\epsilon_{v \mathbf{k}}$ is the energy of the unperturbed state
$ | \psi_{v\mathbf{k}}^{(0)} \rangle$. The operator $R$ comes from the
right-hand side of Eq.~\eqref{eq_mo5} and is determined by the derivatives of
the density matrix of the previous orders. If the local-field effects are taken
into account, the right-hand side $R$ also depends on the derivative of the
electron density
$n^{(P)}(\mathbf{r}_1)=\rho^{(P)}(\mathbf{r}_1,\mathbf{r}_2)\delta(\mathbf{r}_1-\mathbf{r}_2)$. In
this case, Eq.~\eqref{eq_mo8} needs to be solved self-consistently. The
calculations in practice work with the periodic parts of the wavefunctions, $|
u_{v\mathbf{k}}^{(0)} \rangle$. The commutator $[\mathbf{r},\tilde{\rho}]$
corresponding to $ i\partial_{\mathbf{k}} \tilde{\rho}_{\mathbf{k}}$ in
reciprocal space is computed within the $\mathbf{k} \cdot \mathbf{p}$
theory.~\cite{andrade_real-space_2015,strubbe_tesis,strubbe2012}
Equation \eqref{eq_mo8} is solved using the efficient Sternheimer
approach,~\cite{andrade_real-space_2015,andrade2007,strubbe_tesis,strubbe2012}
where the function $| \eta_{v\mathbf{k}}^{(P)}(\Omega) \rangle$ that fits into
Eq.~\eqref{eq_mo8} is found iteratively at each frequency $\Omega$.  To avoid
divergences at resonances, a small but finite imaginary frequency $i\delta$ is
added to the frequency $\Omega_0$ of the external perturbation so that $\Omega =
\Omega_0 + i\delta$.

Once the solution of Eq.~\ref{eq_mo8} is known, the elements
$\tilde{\rho}^{(P)}_{\mathrm{CV}}$ are obtained as
\begin{equation}
  \tilde{\rho}^{(P)}_{\mathrm{CV}} ( \Omega) = \int_\mathrm{BZ}  \frac{\mathrm{d}\mathbf{k}}{(2\pi)^{3}} \sum_{v} | \eta^{(P)}_{v\mathbf{k}} ( \Omega)  \rangle \langle \psi_{v\mathbf{k}}^{(0)} |\,.
  \label{eq_mo14}
\end{equation}
The elements of $\tilde{\rho}^{(P)}$ between the occupied and unoccupied
subspaces are found using the relation $\tilde{\rho}^{(P)}_{\mathrm{VC}} (
\Omega) = (\tilde{\rho}^{(P)}_{\mathrm{CV}} (-\Omega^*))^*$ and, for that,
Eq.~\ref{eq_mo8} is also solved for the frequency $-\Omega^*$.

To find the elements within the occupied $\tilde{\rho}^{(P)}_ {\mathrm{VV}}$ and
unoccupied $ \tilde{\rho}^{(P)}_{\mathrm{CC}}$ subspaces, the idempotency
condition $\rho = \rho \rho $ is used. In terms of the periodic counterpart
$\tilde{\rho}$ of the density matrix and to the first order in the magnetic
field, it is written as~\cite{essin2010,gonze2011}
\begin{equation}
  \tilde{\rho}=\tilde{\rho}\tilde{\rho}+
  \frac{i}{2c} \mathbf{B} \cdot [\mathbf{r},\tilde{\rho}] \times [\mathbf{r},\tilde{\rho}].
  \label{eq_mo15}
\end{equation}
The contribution $\alpha_{\nu \mu, \gamma}$ to the polarizability in the
presence of the magnetic field ($\alpha_{\nu \mu}=\alpha_{0 \nu \mu}+\alpha_{\nu
  \mu, \gamma} B_\gamma$) is finally obtained from the current response as
\begin{equation}
  \alpha_{\nu \mu, \gamma} (\Omega)  = \frac{i}{\Omega} \mathrm{Tr} \left[V_{\nu} \tilde\rho^{(E_\mu B_\gamma)} (\Omega) \right]\,,
  \label{eq_mo16}
\end{equation}
where indices $\nu$, $\mu$, and $\gamma$ are used to denote components of the
vectors $\mathbf{V}$, $\mathbf{E}$, and $\mathbf{B}$. The dielectric tensor in
the presence of the magnetic field is computed as $\epsilon_{\nu
  \mu}=\delta_{\nu \mu}+4\pi\alpha_{\nu \mu}/w$, where $w$ is the unit cell
volume.  Note that according to the ``$2n+1$"
theorem,~\cite{gonze1989,corso1996} there is no need to calculate explicitly the
second-order derivative $\rho^{(E_\mu B_\gamma)}$. Instead, $\alpha_{\nu \mu,
  \gamma}$ is expressed through the first-order derivatives to the perturbations
$P = E_\mu$, $B_\gamma$ and a supplementary perturbation corresponding to a
vector potential $P=A_\nu$ at frequency $-\Omega$.~\cite{lebedeva2019}

\begin{figure}
  \centering
  \includegraphics[width=0.7\textwidth]{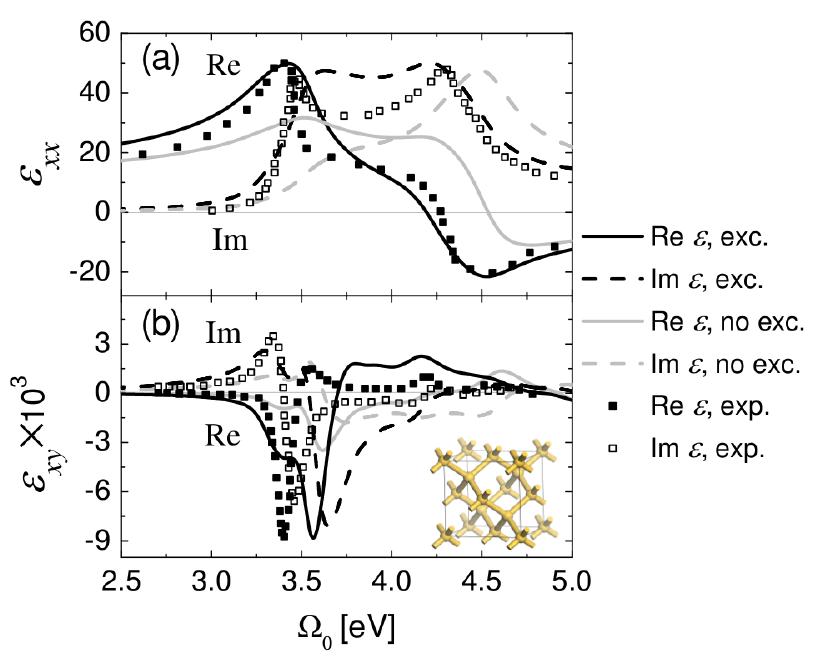}
  \caption{Calculated components (a) $\epsilon_{xx}$ and (b) $\epsilon_{xy}$ of
    the dielectric tensor of silicon for different frequencies of light
    $\Omega_0$ (in eV) and the magnetic field of 1~T along the $z$ axis (with
    and without account of excitonic effects). The linewidth $\delta=0.1$~eV is
    used. The blue shift by 0.7~eV is applied to take into account the GW
    correction to the band gap.~\cite{albrecht1998, botti2004} The experimental
    data for $\epsilon_{xx}$~\cite{lautenschlager1987} and
    $\epsilon_{xy}$~\cite{agranovich1991} (scaled by 1/2) are indicated by
    symbols.}
    \label{fig:si}
\end{figure}

To test the developed formalism for solids, it has been applied to bulk silicon
and the corresponding results are shown in Fig.~\ref{fig:si}. The full details
of the calculations can be found in Ref.~\onlinecite{lebedeva2019}. It is seen
in Fig.~\ref{fig:si}~(b) that even without account of excitonic effects, the
calculated spectra $\mathrm{Re}/\mathrm{Im}\ \epsilon_{xy}$ for the transverse
component of the dielectric tensor are already qualitatively similar to the
experimental curves~\cite{agranovich1991} at the direct absorption edge. To
model excitonic effects, we have used the model from
Ref.~\onlinecite{berger2015}. The spectra computed with account of excitonic
effects show a better agreement with the experimental results
(Figs.~\ref{fig:si}(a) and \ref{fig:si}(b)). Although the magnitudes of the
peaks in the magneto-optical spectra are about a factor of two smaller than in
the magneto-optical measurements,~\cite{agranovich1991} they can be corrected by
reducing the linewidth $\delta$ assumed in the calculations.

In the limit of a large supercell, this formalism becomes equivalent to the
simpler standard formulation for finite systems, which we have also implemented
for reference. In this case, the Liouville equation for the density matrix is
written in the Coulomb gauge as
\begin{equation}
  \Omega  \rho + [H_0,\rho]= \left[
    \mathbf{d} \cdot \mathbf{E} +\mathbf{m} \cdot \mathbf{B} - \delta H,\rho \right]\,,
  \label{eq_mo19}
\end{equation}
where $\mathbf {d}= -\mathbf{r}$ is the electric dipole moment, $\mathbf {m}=
-\mathbf{r} \times \mathbf{V}/2c$ is the orbital magnetic dipole moment, and
$\delta H$ describes local-field effects. The change in the polarizability
$\alpha_{\nu \mu, \gamma}B_\gamma$ in the presence of the magnetic field is
calculated from the dipole response
\begin{equation}
  \alpha_{\nu \mu, \gamma} (\Omega)  = \mathrm{Tr} \left[d_{\nu} \rho^{(E_\mu B_\gamma)} (\Omega) \right]\,.
  \label{eq_mo20}
\end{equation}

As in the periodic case, based on the ``$2n+1$"
theorem,\cite{gonze1989,corso1996} explicit calculation of the second-order
derivative $\rho^{(E_\mu B_\gamma)}$ is avoided. Instead, a supplementary
electric field at frequency $-\Omega$, $E'_\nu$, is
introduced.~\cite{lebedeva2019} In the case of real wavefunctions, as can be
chosen for finite systems in real space in the presence of time-reversal
symmetry, $| \eta_{v}^{(E'_\nu)} (-\Omega) \rangle $ = $(| \eta_{v}^{(E_\nu)}
(-\Omega^*) \rangle)^*$ and there is no need to solve additionally the Liouville
equation for this supplementary electric field.

The efficiency of the present scheme to compute magneto-optics is comparable to
standard linear-response calculations of simple optical polarizability in the
absence of the magnetic field.~\cite{lebedeva2019} When local-field effects are
included self-consistently, the calculations of magneto-optical spectra for
solids take only twice as long as those of polarizability. For finite systems,
the computational effort is the same as for the simple optical polarizability.

\section{Time-dependent angular resolved photoelectron spectroscopy}
The real-space representation of the dynamics of the electronic structure allows
for a seamless and straightforward description of dynamics processes outside the
material, i.e., processes where electrons are excited into the vacuum. Beyond
simply describing the ionization process, Octopus has routines implemented that
compute photoelectron spectroscopy in different flavors. Photoelectron
spectroscopy is particularly ubiquitous for the characterization of the
electronic structure in solids, because it provides a direct observable of the
energy and momentum distribution of electronic states, known as the
bandstructure. In contrast to other electronic structure methods that compute
quantities linked to the photoelectron spectrum, most commonly the single-quasiparticle spectral function obtained through the $GW$ approximation to the
many-body self-energy, the approach described here does not consider unit cells
of bulk materials, but instead directly computes the energy and momentum
resolved ionization probability.~\cite{de2018pump,DeGiovannini:2018bl}
Besides the study of the electronic structure of solids, photoelectron spectra
of atoms and molecules are of equal interest and the implementation in Octopus
is capable to describe accurately all such systems on equal footing.

The most detailed quantity available in the experiments is the momentum-resolved
photoelectron probability $\mathcal{P}({\bf p})$, i.e., the probability to
detect an electron with a given momentum ${\bf p}$. In some cases the
experimental setup offers the possibility to measure the spin polarization along
a given axis. The formalism we are going to outline can easily accommodate a
non-collinear spin structure and, therefore, calculate spin-resolved quantities,
but for the sake of simplicity, in the following, we are going to restrict
ourselves to closed shell systems with collinear spins, i.e., where all the
orbitals are doubly occupied with electrons having opposite spins. The reader
interested in the most general case can find a detailed description in
Ref.~\onlinecite{DeGiovannini:2016bb}. From $\mathcal{P}({\bf p})$ one can
obtain other derived quantities by simple manipulation. For instance, the energy-resolved spectrum $\mathcal{P}(E)$, used to identify the occupied energy levels,
can be obtained with a change of variable using the free electron energy
dispersion relation $E=p^2/2$. The angle-resolved photoelectron spectrum
(ARPES), normally employed to measure the quasiparticle bandstructure, is simply
obtained by taking the energy resolved spectrum as a function of the electron
momentum parallel to the surface $\mathcal{P}({\bf p}_\parallel,E)$.

The t-SURFF method was first proposed by Scrinzi~\cite{Tao:2012ev} for
one-electron systems and later extended to many electrons with TDDFT for
periodic~\cite{DeGiovannini:2016bb} and non-periodic
systems.~\cite{Wopperer:2017bm}
It is based on the assumption that the Kohn-Sham
Hamiltonian describing the full experimental process, i.e., including the
ionization and detection, can be decomposed into the sum of two Hamiltonians
acting into complementary spatial regions, \textit{inner} and \textit{outer}, and that 
can be approximated in different ways. In the \textit{inner} region surrounding the system
the electron dynamics is governed by the interacting Kohn-Sham Hamiltonian $\hat{H}_{\rm
  KS}({\bf r},t)$ while in the \textit{outer} region, 
electrons are free from the Coulomb tails of the parent system and behave as 
independent particles driven by an external field and therefore are described by the Volkov Hamiltonian
$\hat{H}_{\rm V}({\bf r},t)=1/2\left( -i \nabla - {\bf A(t)}/c \right)^2$.  
We express the field with a time dependent vector potential 
potential ${\bf A(t)}$ in the dipole approximation, i.e. by discarding the spatial dependence of the field ${\bf
  A({\bf r},t)}\approx {\bf A(t)} $. 
The advantage of this approach is provided
by the fact that the time-dependent Schr\"odinger equation associated with the
Volkov Hamiltonian, can be solved analytically and that the solutions, the Volkov
waves
\begin{equation}
  \chi_{\bf p}({\bf r},t) = \frac{1}{(2\pi)^{\frac{3}{2}}} e^{i {\bf p}\cdot{\bf r}} e^{i \Phi({\bf p},t)}\,,
  \label{eq:volkowchi}
\end{equation}
with
\begin{equation}
  \Phi({\bf p},t)=\int_0^t {\rm d}\tau \left( {\bf p} - \frac{{\bf A}(t)}{c} \right)^2\,,
\end{equation}
are eigenstates of the momentum operator. We can therefore expand the Kohn-Sham
orbitals $\varphi_i$ as a superposition of detector states in the form of Volkov
waves
\begin{equation}
  \varphi_i({\bf r},t) = \int {\rm d}{\bf p}\, b_i({\bf p},t) \chi_{\bf p}({\bf r},t)
  \label{eq:phitochi}
\end{equation}
and obtain the photoelectron probability in terms of the expansion coefficients
by summing up the contribution of all the orbitals: $\mathcal{P}({\bf p}) =
\lim_{t\rightarrow\infty}2/N\sum_{i=1}^{N/2} \vert b_i({\bf p},t)\vert^2$.

Using the continuity equation, t-SURFF allows us to express the coefficients $b_i$,
and thus $\mathcal{P}({\bf p})$, as a time integral of the photo-current flux
through the surface $S$ separating the domains of the two Hamiltonians. More
specifically
\begin{equation}
  b_i({\bf p},t) = -\int_0^t {\rm d }\tau \oint_S {\rm d } {\bf s}\cdot \langle \chi_{\bf p}(\tau) | \hat{\bf j} |\varphi_i(\tau) \rangle\,,
  \label{eq:bpt}
\end{equation}
with the single-particle current density operator matrix element given by
\begin{equation}
  \langle \chi_{\bf p}(\tau) | \hat{\bf j} |\varphi_i(\tau) \rangle\ = \frac{1}{2}\left\{ i \varphi_i({\bf r},\tau)\nabla\chi_{\bf p}^*({\bf r},\tau)- i \chi^*_{\bf p}({\bf r},\tau) \nabla\varphi_i({\bf r},\tau)  -2 \frac{{\bf A}(t)}{c}\chi^*_{\bf p}({\bf r},\tau)\varphi_i({\bf r},\tau) \right\}\,.
  \label{eq:Jop}
\end{equation}

\begin{figure}
  \resizebox{\textwidth}{!}{\includegraphics{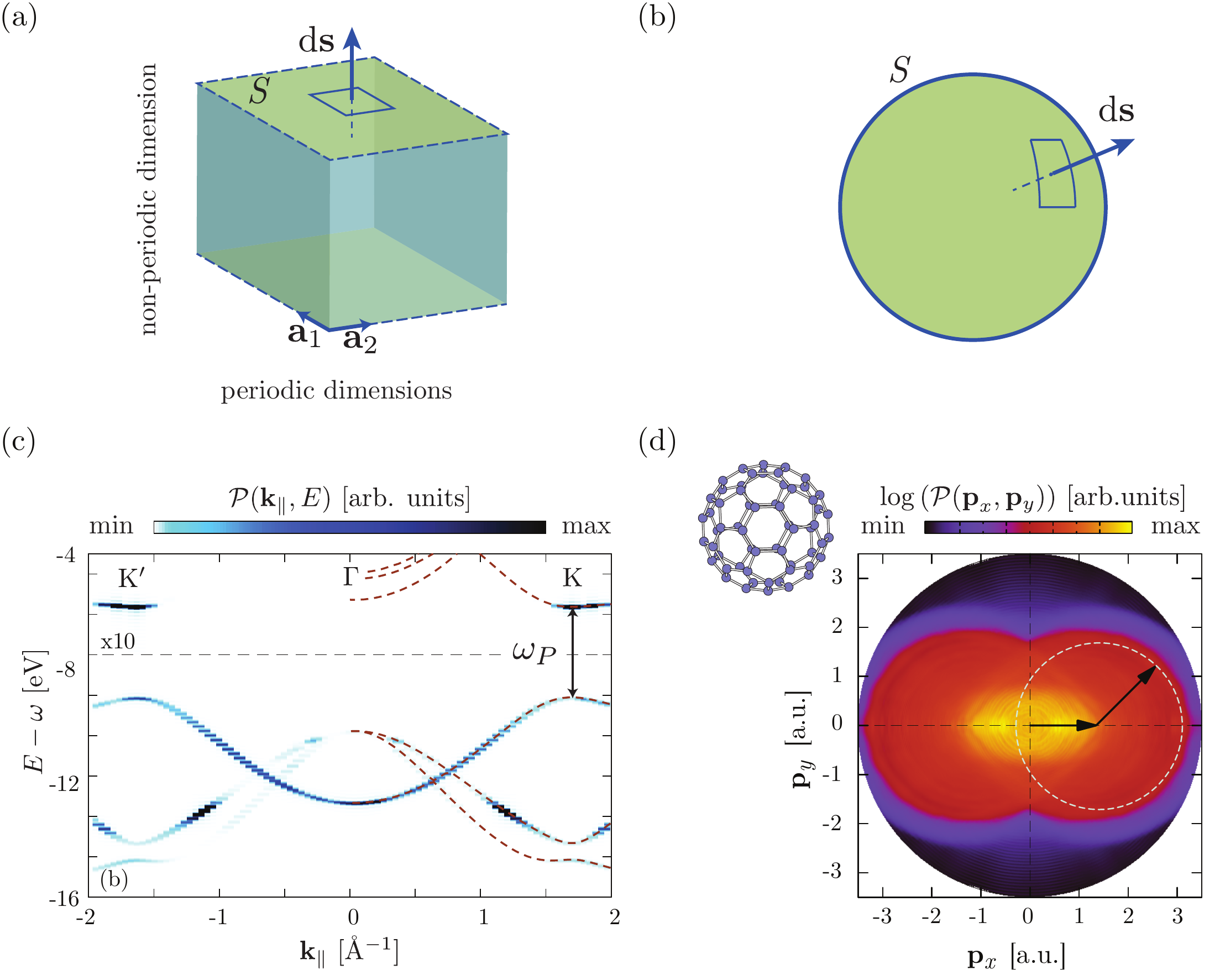}}
  \caption{\label{fig:tsurff} Calculated photoelectron spectrum with t-SURFF, as
    implemented in Octopus.~\cite{Wopperer:2017bm} The geometries employed to
    calculate the flux of the photoelectron current are depicted in (a) for
    periodic and (b) non-periodic systems. (c) Pump-probe ARPES spectrum of
    monolayer h-BN driven by a laser field resonant with the gap at K, adapted
    from Ref.~\onlinecite{DeGiovannini:2016bb}. (d) Photoelectron angular
    distribution obtained by strong-field ionization of C$_{60}$ with an IR
    field polarized along $x$, adapted from Ref.~\onlinecite{Wopperer:2017bm}.
    All the electrons rescattering at the same time in one period of the field 
    end up with final momenta forming a ring (dashed) centered at the value of 
    $\textbf{A}(t_r)$ at the rescattering time, $t_r$.
    The two arrows represent the graphical decomposition of the final momentum 
    of the photoelectron in the vector potential at the moment of rescattering 
    (horizontal arrow) plus the rescattering momentum. }
\end{figure}

The flexibility offered by the definition of the boundary surface $S$ allows us
to easily adapt t-SURFF to periodic and non-periodic systems. For periodic
systems we choose $S$ as a plane parallel to the material surface, while for the
non-periodic case the most natural choice is a sphere as shown in
Fig.~\ref{fig:tsurff}~(a) and (b).

For periodic systems one can apply the Bloch theorem to the Volkov Hamiltonian
and further reduce Eq.~\eqref{eq:bpt} to an expression where only the periodic
part of the Bloch waves are employed.~\cite{DeGiovannini:2016bb}

The present computation of the photoelectron spectrum relies on the condition
that the Volkov Hamiltonian can be used to describe electrons in a region of
space, which entails the following approximations: the ionized electrons i)
become non-interacting at some distance from the system and ii) they are free
with respect to the tail of the Coulomb potential of the system, i.e., it has to
be possible to neglect it at some distance. Both conditions are controlled by
the placement of the analyzing surface $S$ for the evaluation of the flux and in
practice this has to be converged by placing it at successively larger distances
form the system. This method needs an explicit description of the vacuum around
the probed system, which for solids requires an explicit construction of the
electronic structure of the surface layer and, if bulk properties are probed,
one needs to converge a slab of material. Compared to unit cell calculations of
solids, this puts the present method at a disadvantage. However, when aiming at
simulating specific experiments, it natively includes signals coming from the
surface layers and thus can capture more of the experimental reality of the
spectroscopic process compared with standard perturbative many-body approaches which 
are directly performed in the bulk unit cell. 

In practical calculations, in order to avoid spurious reflection of electrons
from the boundaries of the simulation box, one has to employ absorbing
boundaries and, depending on the condition on their transparency, this can add
up to the size of the simulation box.~\cite{DeGiovannini:2015jt} Other than the
increased size of the computational box, this method is not computationally
costly, as it requires only the evaluation of the gradient operator on surface
points $S$ and straightforwardly supports existing k-points, states, and mesh
parallelizations.

Pump-probe configurations can be naturally simulated since there is no
restriction on the functional form of the vector potential ${\bf A}(t)$ and,
therefore, we can accommodate any linear combination of pulses. As an example,
in Fig.~\ref{fig:tsurff}~(c) we show the result of a simulated pump-probe
time-resolved ARPES measurement of monolayer hexagonal boron nitride (hBN)
driven by a field resonant with the gap at the $K$-point in the Brillouin
zone. As it is apparent from the simulation, the excited state population
transfer to the conduction band is well observed in the resulting ARPES
spectrum.

The method is not limited to simple photoelectron spectroscopy, but can be used
to simulate complex experimental techniques, such as the reconstruction of
attosecond beating by interference of two-photon transitions
(RABBITT).~\cite{Sato:2018jw} Furthermore, due to the versatility of the
real-time description in Octopus, the underlying excitation does not need to be
coming from an optical pulse (i.e., an external vector potential) and, since the
ions can move according to the TDDFT-Ehrenfest dynamics,~\cite{Andrade:2009gaa}
one can also simulate features coming from the ionic or lattice motion,
specifically electron-phonon coupling signature in ARPES.~\cite{Hubener:2018id}
Pump-probe simulations are not limited to dynamical process of excitations, but
can also be employed to study steady-state modifications of driven electronic
states.~\cite{DeGiovannini:2016cb} This opens the possibility of studying
Floquet physics from an ab-initio perspective and to directly simulate the
effect that a periodic force would have on the dressed electronic structure --
an important aspect to underline given the growing interest in the field of
Floquet engineering~\cite{Oka:2018df} and Floquet
analysis.~\cite{DeGiovannini:2019hm}

Finally, t-SURFF is particularly suited to simulating strong field ionization
processes such as in laser-induced photoelectron diffraction
(LIED) experiments, where the ionization takes place by direct tunnelling into
the continuum and the field is so strong as to drive electrons in trajectories
recolliding with the parent system. In fact, since in Eq.~\eqref{eq:bpt} we
accumulate the flux through $S$ over the time of the propagation, with t-SURFF
electrons can seamlessly flow back and forth through the surface without
producing any artifact in the final spectrum. As an example, in
Fig~\ref{fig:tsurff}~(d) we present the photoelectron angular distribution of
C$_{60}$ ionized by a strong IR pulse capable of inducing rescattering
dynamics, which then gets imprinted in the photoelectron spectrum as
characteristic rings centered in the value of the vector potential at the
instant of rescattering.~\cite{Blaga:2013jo} In this regime, the result of
simulations obtained with Octopus are in excellent agreement with the
experiments.~\cite{Krecinic:2018fh,Trabattoni:2018vi}

\section{Electric and thermal conductivities}

Electrical conductivity in real materials can be described to a first
approximation by Ohm's Law, which is a linear relationship between the applied
electric field and the current density generated in response. The standard method to study electrical and thermal conductivity is to carry out
a DFT calculation at equilibrium and apply Kubo-Greenwood linear response theory
to evaluate the conductivities. Although this method has been used successfully in the past to calculate transport properties,~\cite{recoules2009ab,lambert2011transport} the application is limited to systems in the linear response regime. It is known that in many cases, especially in the presence of strong external fields, Ohm's law is no longer valid and the system exhibits non-linear behavior. In order to capture these complex behaviors in materials, one must go beyond simple, linear approximations and describe electron interactions in a non-trivial manner.

One promising route to go beyond linear response is to use density functional methods to \emph{directly} study thermoelectric transport. This topic has received less attention, but has recently been implemented in Octopus and applied to liquid aluminum \cite{andrade2018negative}. For detailed descriptions and derivations of non-equilibrium thermoelectric phenomena using density based methods, we refer the reader to Ref.~\onlinecite{eich2016functional}.



In Octopus, we are able to calculate the current density and heat current
density at each step during a time-dependent simulation. A current is induced in
this study by applying an electric field of the form $\mathbf{E}(t)=\mathbf{E_0} \delta (t)$,
where $\mathbf{E_0}$ describes the magnitude and direction of the field. The electric
field is induced through a time-dependent vector potential ($\mathbf{E}(t)=-c^{-1}
\partial_t \mathbf{A}(t)$, c is the speed of light) in the Hamiltonian. This vector
potential satisfies the periodic boundary conditions of an extended system. It
should be noted that there is no limitation on the form of the applied electric
field in general. We are able to evaluate the macroscopic current density as
\begin{equation} \label{eqn:current_density}
  \mathbf{J}(t) = -\frac{i}{\Omega} \int d\mvec{r} \sum_{j}^{N} \varphi_i(\mvec{r},t)[\hat{H}(t),\mvec{\hat{r}}]\varphi_i(\mvec{r},t)\,,
\end{equation}
where $\Omega$ is the volume of the unit cell. The energy current density, $\rm
\hat{\textbf{J}}_h(\textbf{r},t)$ is expressed as
\begin{equation} \label{eqn:energy_density}
  \hat{\textbf{J}}_h(\textbf{r},t) = \hat{\textbf{J}}_t(\textbf{r},t) + \hat{\textbf{J}}_v(\textbf{r},t) + \hat{\textbf{J}}_u(\textbf{r},t) + \hat{\textbf{J}}_f(\textbf{r},t)\,,
\end{equation}
where $\rm \hat{\textbf{J}}_t(\textbf{r})$ is the kinetic energy contribution
given as
\begin{equation} \label{eqn:KEcurrent}
  [\hat{\textbf{J}}_t(\textbf{r},t)]_i  = \frac{i}{8} \left( [\partial_i \hat{\varphi}^{\dagger}] [\nabla^2 \hat{\varphi}] - [\nabla^2 \hat{\varphi^{\dagger}}][\partial_i \hat{\varphi}] -  [\partial_i \nabla \varphi^{\dagger}] \cdot [\nabla \varphi] - [\nabla \hat{\varphi}^{\dagger}] \cdot [\partial_i \nabla \hat{\varphi}] \right)
\end{equation}
and $\varphi = \varphi(\mvec{r},t)$ are understood to be the time-dependent
states. The output of the time-dependent simulation can be further analyzed to
yield the frequency-dependent conductivity. The electrical (or thermal)
conductivity $\sigma$ can be found by Fourier transforming the corresponding
current as
\begin{equation} \label{eqn:FT}
  \sigma_{ij}(\omega) = \frac{1}{\mathbf{[E_0]}_i}\int^{\infty}_{0} dt e^{-i\omega t} \mathbf{J}_j(t)\,.
\end{equation}

In general, it is possible to study the conductivity of an extended system in
the presence of an applied electric field. Here, we illustrate the use of this
method on hydrogen at 1400 K and 400 GPa. At this temperature and pressure,
hydrogen is in its liquid metallic phase. A molecular dynamics simulation was
carried out in VASP to produce several ionic snapshots of the system. For each
ionic snapshot, we performed a TDDFT simulation with a time step of
0.05~a.u.\ (0.00121~fs) for a total time of 3~fs. The initial electric field
$E_0$ was 0.1~a.u. The calculation was carried out on a 3x3x3 k-point grid to
ensure that the current density decays to zero.

\begin{figure}
  \centering
  \includegraphics[width=0.49\linewidth]{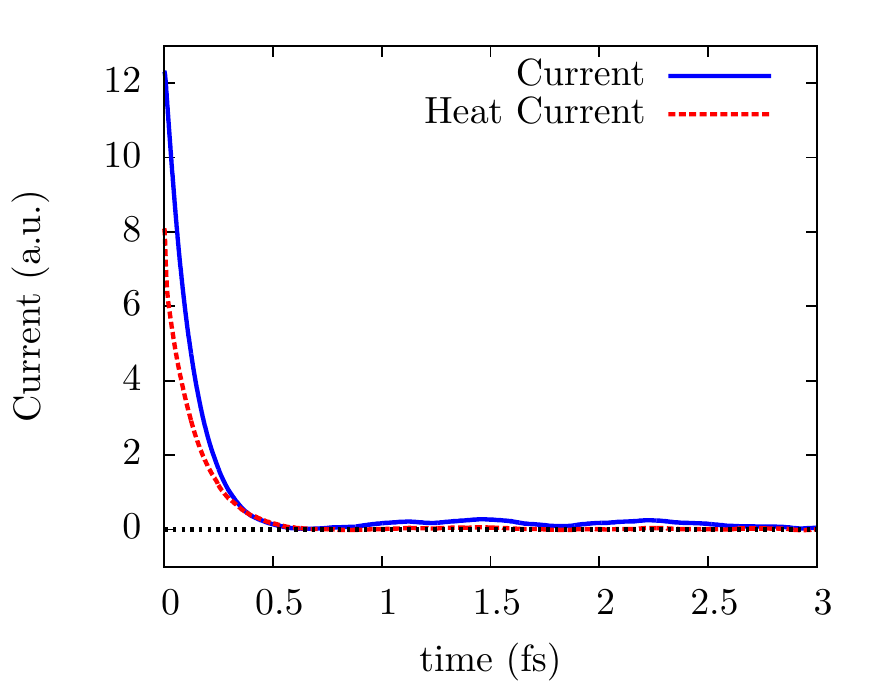}
  \includegraphics[width=0.49\linewidth]{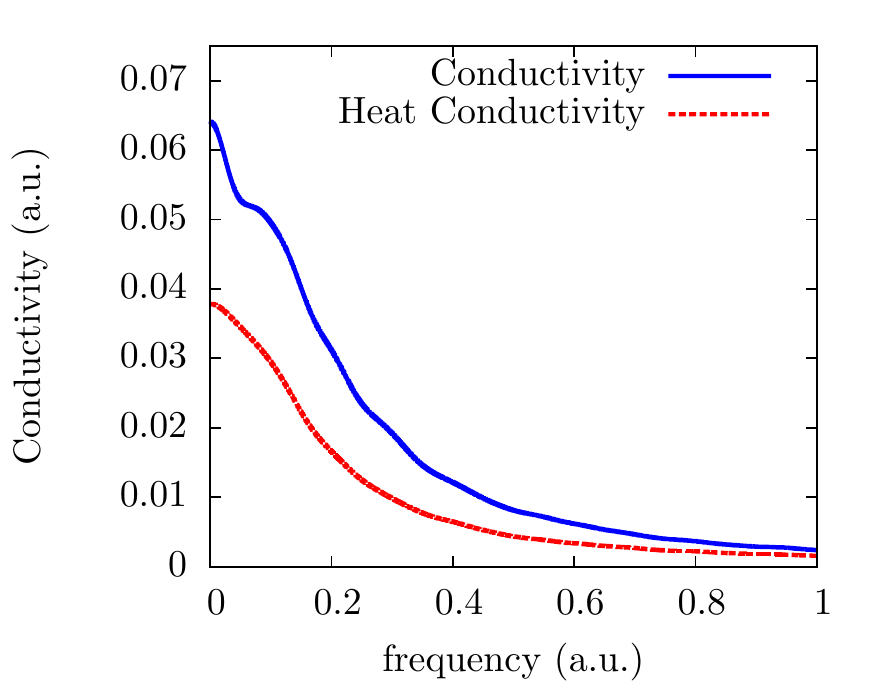}
  \caption{ (Left) Time-dependent current density for one ionic snapshot of a
    supercell of 128 atoms of hydrogen in the liquid phase at 1400~K and 400~GPa
    with an initial electric field $E_0$ of 0.1~a.u. The blue line represents
    the current density as given in Eq.~\eqref{eqn:current_density} and the red
    line is the heat current evaluated from Eq.~\eqref{eqn:KEcurrent}. Both
    currents are shown decaying to zero. (Right) The Fourier transform of the
    respective currents gives the frequency-dependent conductivities.}
    \label{fig:Current_conductivity}
\end{figure}

In Fig.~\ref{fig:Current_conductivity}, we have plotted the time-dependent
current density and heat current density as a function of time after an initial
electric field is applied at time zero. Both currents are shown to decay to zero
by the end of the simulation. It should be noted that the heat current
corresponds to the kinetic energy contribution given in
Eq.~\eqref{eqn:KEcurrent}. In general, this heat current is related to the
Peltier coefficient, since it describes the heat generated from the response to
an electric current. The remaining contributions in
Eq.~\eqref{eqn:energy_density} are not yet implemented in Octopus. Both $J_u$
and $J_f$ arise from electron-electron interactions and $J_v$ is the potential
energy. In future work it would be interesting to evaluate these remaining
contributions to determine if they are large or can reasonably be ignored for
some systems. The electrical conductivity and the heat conductivity can be
evaluated using Eq.~\eqref{eqn:FT}. These frequency-dependent conductivities are
shown in the right hand panel of Fig.~\ref{fig:Current_conductivity}. One may
also calculate the DC conductivity by taking $\omega=0$.



This suite of tools implemented in Octopus will allow for the study of
time-dependent thermoelectric phenomena using TDDFT. This approach to study
current and conductivity is more general and widely applicable than the standard
Kubo-Greenwood approach. While Kubo-Greenwood is a linear response theory, this
method can be applied to study materials where non-linear conduction effects can
appear. In a recent study, this TDDFT method was been used to illustrate
non-linear conductivity effects in liquid aluminum,~\cite{andrade2018negative}
which would not be possible by applying standard linear response theory. In the
future, we plan to extend these tools by implementing a thermal vector potential
that can induce a heat current in an extended system during a time-dependent
simulation.


%
%

\section{Local domain contribution to physical observables}
\label{sec:local_domains}

Usually, we are interested in studying systems consisting of different atoms,
molecules, regions, or domains. In such cases, we might want to understand the
different contributions from different parts of the system towards a specific
observable. Based on the Hohengberg-Kohn theorems for the
DFT,~\cite{Hohenberg1964} and its extension, the Runge-Gross
theorem,~\cite{Runge1984} for TDDFT, we can state that any physical observable
of the system is a functional of the electronic density, either static (ground
state) or time dependent. In other words, the expectation value of an operator
$\hat{\mathcal{O}}$ can be expressed as functional of the electronic density
\begin{equation}
 \mathcal{O}[n] = \langle \Psi[n] | \hat{\mathcal{O}} | \Psi [n] \rangle \,.
\end{equation}
Let's now consider the case where we write the total electronic density as a sum
of $N$ densities
\begin{equation}
  n(\mathbf{r}) = \sum_{i}^{N} n_{i}(\mathbf{r})\,,
\end{equation}
and no further constraints are imposed on $n$ nor on $n_i$. Such partitioning of
the density is at the basis of subsystem DFT, which allows to divide a system
into several Kohn-Sham subsystems that interact with each other in a
theoretically well justified manner (see Ref.~\onlinecite{NEUGEBAUER20101} and
references therein). However, in our case, we are only interested in this kind
of partition for post-processing purposes where we assign a density $n_i$ to a
specific part of the system in order to identify how it contributes to a given
observable. We are thus interested in operators for which the following
condition is true
\begin{equation}
  \mathcal{O}[n] = \sum_i^N \mathcal{O} [n_i]\,.
  \label{eq:op_additivity}
\end{equation}
Such operators are said to be additive. One example of an additive operator that
is particularly relevant in the context of TDDFT is the time-dependent dipole
\begin{equation}
  \mvec{d}(t) = \int \mvec{r} n(\mvec{r}, t) d\mvec{r} = \sum_i^N \int \mvec{r} n_i(\mvec{r}, t) d\mvec{r}\,,
\end{equation}
as this is the relevant observable for obtaining the optical absorption cross
section of finite systems.

The range of operators that fulfill Eq.~\eqref{eq:op_additivity} can be further
expanded if we now consider non-overlapping densities, that is, densities $n_i$
that are non-zero only in a given domain $V_i$ and that the domains do not
overlap. We refer to this type of partitioning as a local domain partitioning.
In such case, any (semi-)local operator that depends (semi-)locally on the
density should fulfill Eq.~\eqref{eq:op_additivity}. In this context, an example
of a relevant observable is the exchange-correlation energy within the LDA
approximation
\begin{equation}
  E_\mathrm{xc}^\mathrm{LDA}[n] = \int n(\mvec{r}) e_\mathrm{xc}^\mathrm{LDA} (n(\mvec{r})) d\mvec{r} = \sum_i^N \int_{V_i} n_i(\mvec{r}) e_\mathrm{xc}^\mathrm{LDA} (n_i(\mvec{r})) d\mvec{r}\,,
\end{equation}
where $e_\mathrm{xc}^\mathrm{LDA}$ is the exchange-correlation energy per unit
particle.

Currently, several different options can be used in Octopus to define specific
regions of the simulation box. These options include simple geometric shapes,
such as spheres centered on particular points of space, atom-dependent
domains, such as an union of spheres centred on the atoms, or the definition of
Bader volumes. The latter option follows the Quantum Theory of Atoms in
Molecules (QTAIM)~\cite{Bader1994} and associates a density region to a given
atom through a density gradient path, such that the boundaries of each volume
are defined as the surfaces through which the charge density gradient has a zero
flux. Note that some of these options allow for the user to specify overlapping
regions. In that case the overlap must be taken into account when analyzing the
results and extra care is required when comparing results from different
domains.

Let us exemplify the applicability of the local domain partitioning by
decomposing the optical response for a coupled chromophore system. Similar to
Frozen Density Embedding real-time TDDFT (FDE-rt-TDDFT),~\cite{Krishtal2015} we
decompose the total optical spectrum as a sum of the local response within each
domain
\begin{equation}
  \bm{\alpha}(\omega) = \sum_i \bm{\alpha}_{i}(\omega) \,,
\end{equation}
where $\bm{\alpha}$ is the dynamic polarizability. In addition, from each local
dynamic polarizabilty tensor we can compute for each fragment the corresponding
cross-section. This is justified by the fact that the relevant operator to
calculate $\bm{\alpha}$ is the dipole operator and, as shown above, this is an
additive operator that fulfills Eq.~\eqref{eq:op_additivity}.

As an example, we performed a series of optical spectrum simulations using
real-time TDDFT for a benzene-fulvene dimer with different $\pi$-stacking
separation, ranging from 4~\AA\ to 8~\AA.  We use the standard PBE
exchange-correlation functional~\cite{Perdew1996,Perdew1997} and the Optimized
Norm-Conserving Vanderbilt PBE pseudopotential (sg15) set.~\cite{Schlipf2015}
The real-space grid is defined as a parallelepiped box with length 16~\AA,
17~\AA\ and 20~\AA\ in the three Cartesian axes. The spacing
between points is set to be 0.13~\AA.

For each intermolecular separation we carry out a single ground-state
calculation and three time-propagations. For each time-propagation a dipolar
electric perturbation is applied along one of the Cartesian
axes.~\cite{strubbe2012} We let the perturbed Kohn-Sham states evolve for a
propagation time of $T = 24$~$\hbar$/eV (15.8 fs), with 
resolution of 0.26 eV ($2\pi/ T$).  The ground-state electron density is
fragmented following the Bader atomic decomposition and each molecular domain is
defined as the sum of these atomic volumes. Then, the corresponding dipole
operator is applied over each defined domain.  Finally, by Fourier transform of
the local time-dependent dipole moment, the local polarizability tensor is
recovered.

\begin{figure}[ht]
    \centering
    \includegraphics[width=0.3\textwidth]{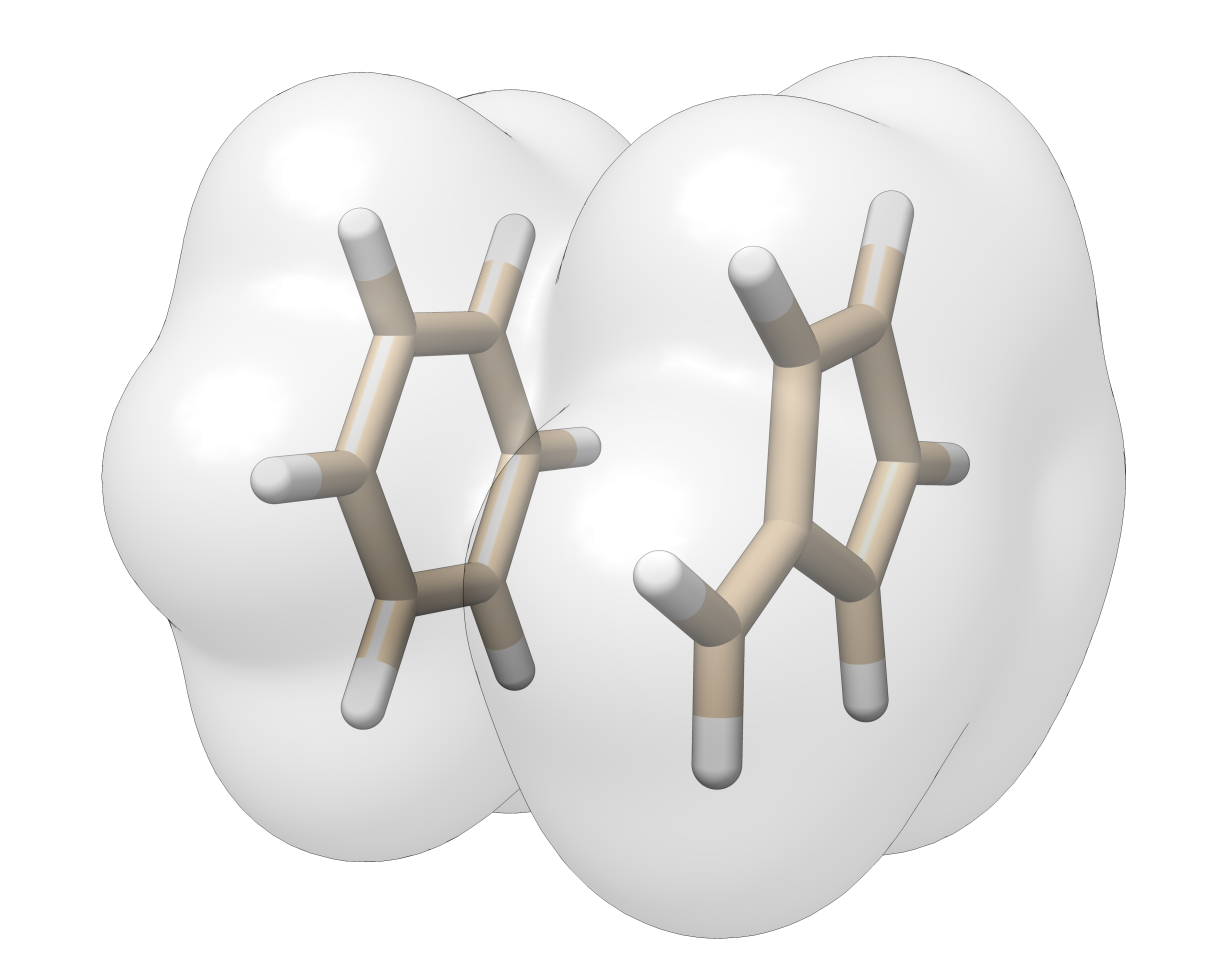}
    \label{fig:rho_global}\hfill
    \includegraphics[width=0.3\textwidth]{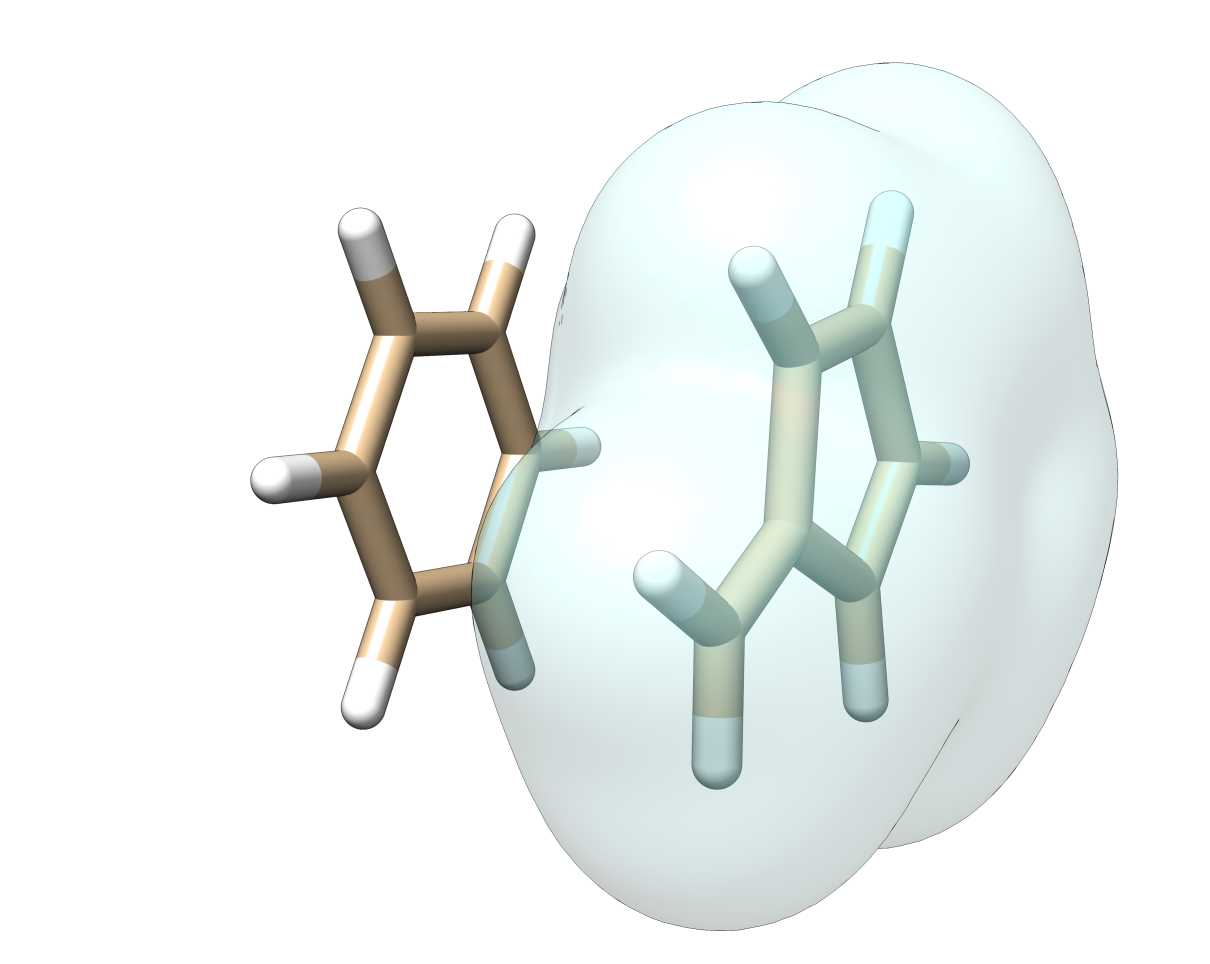}
    \label{fig:fulvene_global}\hfill
    \includegraphics[width=0.3\textwidth]{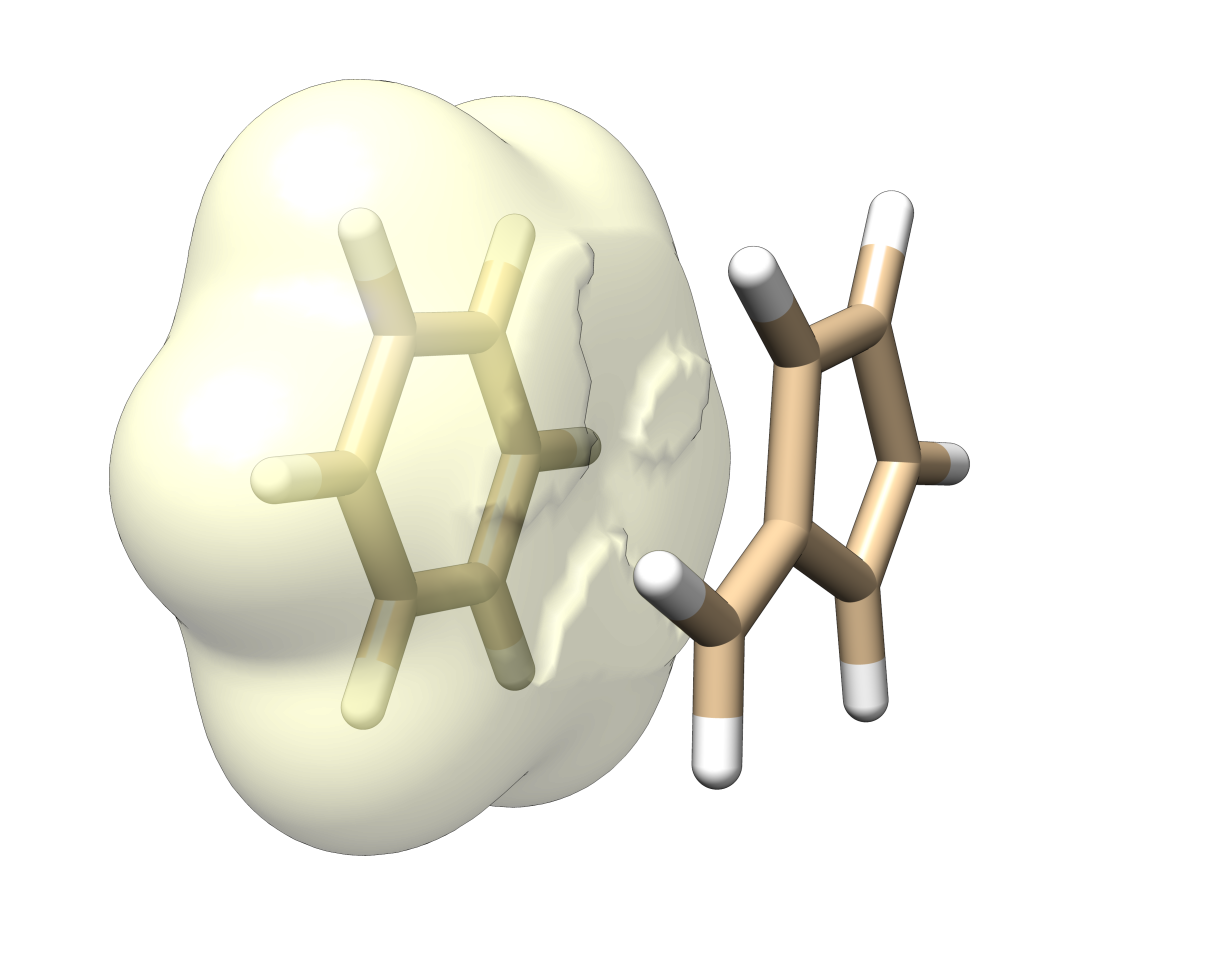}
    \label{fig:benzene_global}
    \includegraphics[width=0.22\linewidth, angle=-90]{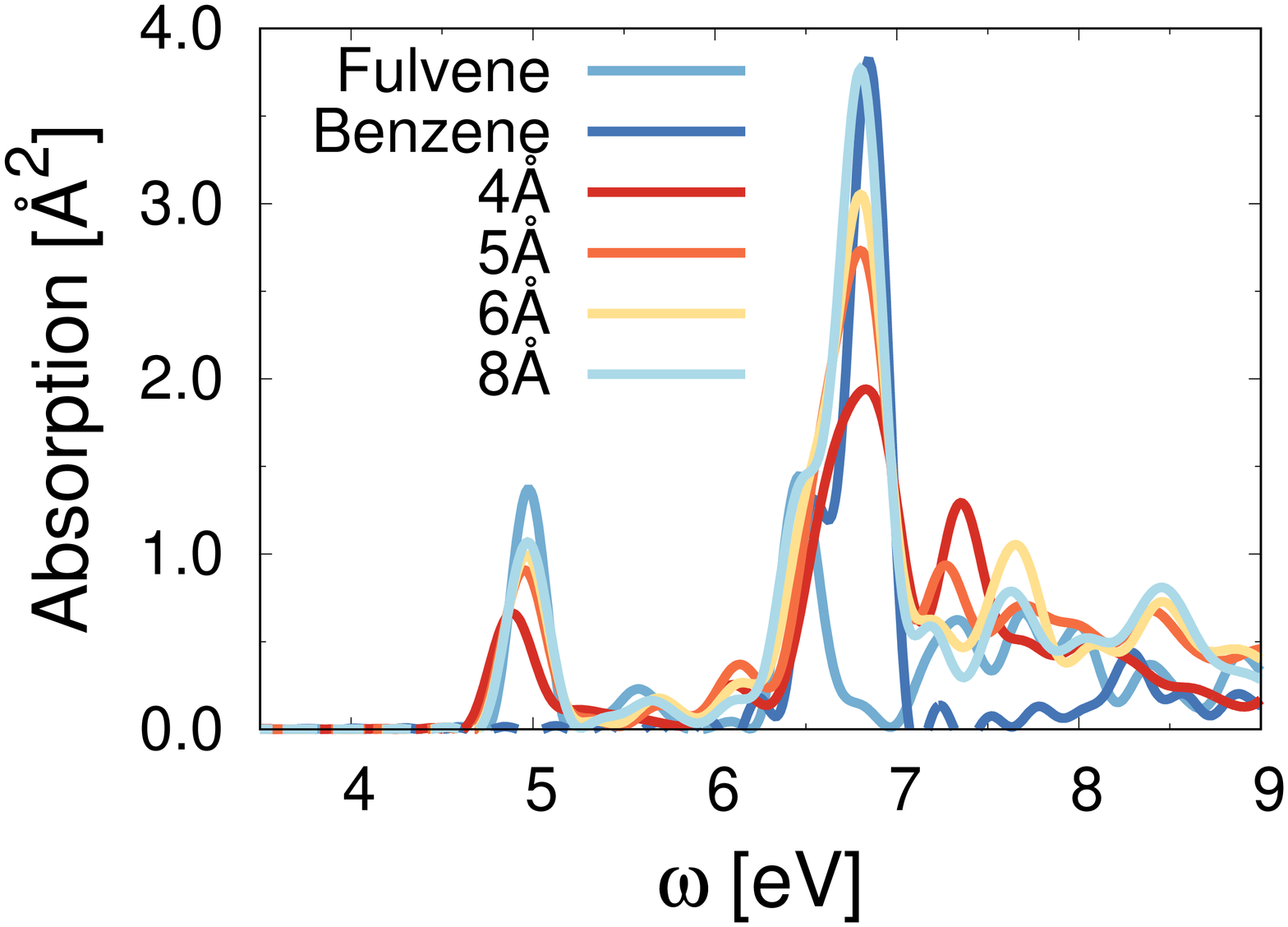}
    \label{fig:spc_global}\hfill
    \includegraphics[width=0.22\linewidth,angle=-90]{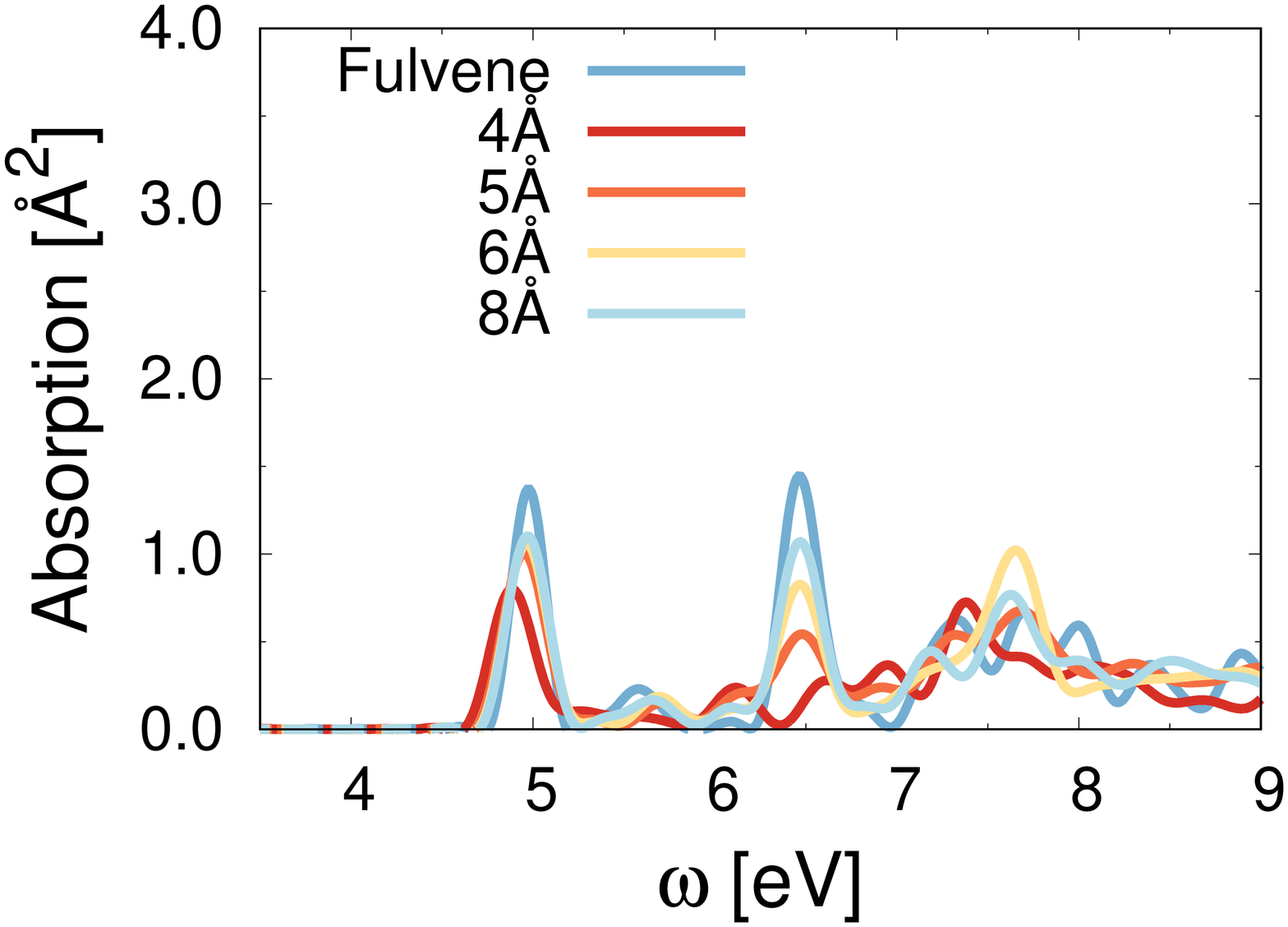}
    \label{fig:spc_fulvene}\hfill
    \includegraphics[width=0.22\linewidth,angle=-90]{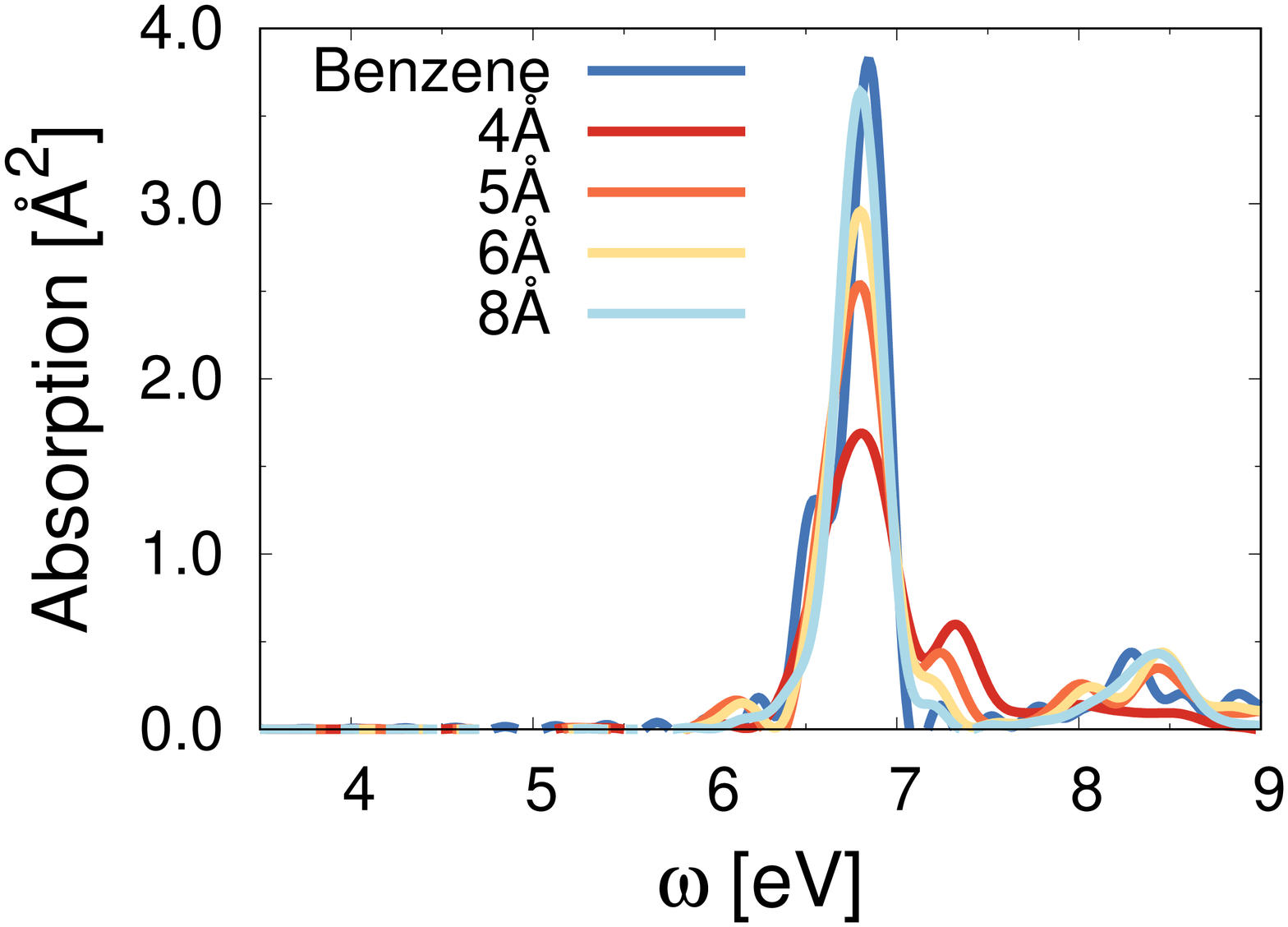}
    \label{fig:spc_benzene}

  \caption{Top: Schematic representation of the benzene-fulvene dimer with
    $\pi$-stacking along the $z$ axis and a intermolecular distance of
    4{\,\AA}. From left to right, the electronic density for the global system for
    an isosurface of 0.006 {\AA}\textsuperscript{-3} and the \textit{QTAIM}
    local density for the fulvene and benzene molecules respectively. Visualized
    with USCF Chimera software.~\cite{Chimera2004} Bottom: Photo-absorption
    cross section obtained from, from left to right, the full system density,
    the fulvene local density, and the benzene local density.}
  \label{fig:spc_local_multipoles}
\end{figure}

Figure~\ref{fig:spc_local_multipoles} shows the spectral decomposition of the
benzene-fulvene system as a function of the intermolecular separation. The full
system at large intermolecular separation (8~\AA) shows two major peaks located
at the same frequencies as for the isolated molecules. As the intermolecular
distance becomes smaller, the global spectrum changes due to electrostatic
effects induced by the electronic cloud of the neighbouring molecule. We can see
that the peak located at around 5~eV suffers a red-shift, while the stronger
peak between 6.5 and 7~eV reduces its intensity, giving rise to an oscillator
strength transfer to excited states at higher energy. The local domain analysis
reveals that the major change is caused by the reduction of the oscillator
strength of the fulvene peak located at around 6.5~eV.

These results are in agreement with the FDE-rt-TDDFT calculations of
Ref.~\onlinecite{Krishtal2015}, further validating our strategy. In addition,
the local domains methodology does not require any previous fragmentation,
selection, or localization of the basis set, allowing also to treat very large
systems, such as biological molecules like the major light harvesting complex
LHC-II.~\cite{Jornet-Somoza2015} More recently, we have shown that this
technique also allows for exciton coupling calculations when combined with a new
formulation of the transition densities from real-time
TDDFT.~\cite{Jornet-Somoza2019}

\section{New propagators for real-time TDDFT}
\label{sec:propagators}

At the core of most Octopus calculations lies the need to integrate the
time-dependent Kohn-Sham equations (TDKS). These are a set of non-linear
equations, given the dependence of the Kohn-Sham Hamiltonian with the electronic
density.  Upon the discretization of the electronic Hilbert space, they take the
generic form of a first-order ordinary differential equation (ODE) system
\begin{equation}
  \dot{\varphi}=f(\varphi(t),t)\,,
\end{equation}
\begin{equation}
  \varphi(t_{0})=\varphi^{0}\,,
\end{equation}
where $t_{0}$ is the initial time, $\varphi$ is an array containing all the
Kohn-Sham orbitals, $\{\varphi_{m}^N\}$ ($\varphi_0$ is its initial value), and
$f$ is a vector function ($f=(f_1,\dots,f_N)$) given by the action of the
Kohn-Sham Hamiltonian $\hat{H}[n(t),t]$
\begin{equation}
  f_i(\varphi(t),t) = -i\hat{H}[n(t), t]\varphi_i(t)\,,\;\;\;\;(i=1,\dots,N)
\end{equation}
Note that: (1) If the nuclei are also to be propagated, the state should be
supplemented with their position and momenta, and the equations with the
corresponding nuclear equations of motion. (2) Likewise, the formalism for
solids includes a polarization vector field~\cite{biry00:7998} that must also be included in the
system definition and propagation. (3) Strictly speaking, the TDKS equations are
not ordinary but ``delay differential equations'' (i.e. equations for which the 
derivative of the unknown function at a certain time depends on the function values at previous times), 
due to the dependence of the
exact exchange and correlation potential functional on past densities. This is
ignored in the adiabatic approximation, which is almost always assumed.

A myriad of numerical propagators for ODEs are available, all of them
theoretically applicable to the TDKS equations. Ideally, however, one should
choose a propagator that respects all the mathematical properties of the
equations that describe the problem at hand, for example, the preservation of
the norm of the orbitals. But also, in the case of the TDKS equations in the
adiabatic approximation, another such property is symplecticity (a fact that is
demonstrated in Ref.~\citenum{adrian2018}). This property is usually stated in
terms of the conservation of the volume of the flow of the system of ODEs in the
phase space, although the precise definition is as follows:

Any system of ODEs can be viewed as a ``flow'', a differentiable map $g:
\mathbb{R}^{p} \to \mathbb{R}^{p}$ that transforms the state $y(t_0)$ at some
point in time $t_0$ into the state at time $t$, $y(t) = g(y(t_0))$.  For systems
described with complex variables such as ours, since it can be split into a real
and an imaginary part, $p$ is even: $g: \mathbb{R}^{2N} \to \mathbb{R}^{2N}$.  A
map with an even number of variables such as this one is defined to be symplectic if and only if
\begin{equation}
    \frac{\partial g}{\partial x}^T J \frac{\partial g}{\partial x} = J\,,
    \qquad
    J = \left[\begin{array}{cc}
        0 & I
        \\
        -I & 0
      \end{array}\right]\,,
\end{equation}
where $I$ is the unit matrix of dimension $N$, and $x \in \mathbb{R}^{2N}$ (conventionally,
the first $N$ variables of $x$ are called the ``coordinates'', and the second half are the ``momenta'',
but no physical meaning should be assumed for them in this purely mathematical definition). 
A numerical propagator is also a differentiable
map that relates the solution $y(t)$ to $y(t+\Delta t)$. As such, it may respect
or not the symplecticity -- and other properties -- of the original flow.


The symplecticity also means that the system has to be ``Hamiltonian'': one may
find a set of coordinates $q\in\mathbb{R}^N$, $p\in\mathbb{R}^N$ (in this case,
the real and imaginary part of the orbitals coefficients), and some scalar
function $H(q,p)$ (the Hamiltonian expectation value), that permits us to rewrite
the system into the well known form of Hamilton's equations:
\begin{subequations}
\begin{align}
  \displaystyle \dot{q_i} & = \phantom{-}\frac{\partial H(p,q)}{\partial p_i}\,,
  \\
  \displaystyle \dot{p_i} & = -\frac{\partial H(p,q)}{\partial q_i}\,.
\end{align}
\end{subequations}
The relevance of the symplectic numerical propagators stems from the fact that
they present some features, such as a better conservation of the energy at long
times, where the value of the energy ends up oscillating around the true value
instead of diverging. For a more detailed discussion about symplecticity on
numerical propagators, check Ref.~\citenum{Hairer2006}.

The Octopus code has several propagator options, many of them already described
in Ref.~\citenum{castro:3425}, such as the Crank-Nicolson,
standard Runge-Kutta, exponential midpoint rule and variations (e.g., the
``enforced time-reversal symmetry'' scheme), split operator techniques, etc. In
a recent paper,~\cite{adrian2018} some of the present authors have studied
propagators of different families that had been scarcely (or not at all) tested
for the TDKS equations: multistep, exponential Runge-Kutta, and commutator-free
Magnus (CFM) expansions. After considering both the accuracy, stability, and the
performance of the propagators, the CFM techniques were identified as suitable
schemes for TDDFT problems and implemented in Octopus. In this section, we make
a brief description.

Developed in Ref.~\onlinecite{blanes2006} for linear non-autonomous systems, the
CFM expansion offer an alternative to the ``standard'' Magnus expansion, which
requires expensive application of nested commutators of the Hamiltonian with
itself at different times. In essence, a CFM expansions $\Gamma(t+\Delta t, t)$ consists of
substituting the propagator $\hat{U}(t+\Delta t,t)$ by a product of exponentials
\begin{equation}
\Gamma (t+\Delta t,t)=\prod\limits^{m}_{i=1} \exp(\hat{D}_{i})\,.
\end{equation}
The $m$ linear operators $\hat{D}_{i}$ are either the Hamiltonian at
different times within the interval $[t,t+\Delta t]$, or parts of it.

We have implemented in Octopus an order four ($q$=4) version, given for example in
Eq.~43 of Ref.~\onlinecite{blanes2006}, which we call hereafter CFM4. This
propagator requires two exponentials ($m=2$),
\begin{equation}
  \varphi(t+\Delta t) = \exp\left(-i\Delta t(\alpha_{1}\hat{H}[t_1]+\alpha_{2}\hat{H}[t_2])\right) \exp\left(-i\Delta t(\alpha_{2}\hat{H}[t_1]+\alpha_{1}\hat{H}[t_2])\right) \varphi(t)\,,
\end{equation}
where
\begin{align}
  \alpha_{1} & = \frac{3-2\sqrt{3}}{12},\quad\quad 
  & t_{1} & = t + \left(\frac{1}{2}-\frac{\sqrt{3}}{6}\right)\Delta t\,,
  \\
  \alpha_{2} & = \frac{3+2\sqrt{3}}{12},\quad\quad
  & t_{2} & = t + \left(\frac{1}{2}+\frac{\sqrt{3}}{6}\right)\Delta t\,.
\end{align}
$\hat{H}[t_1]$ and $\hat{H}[t_2]$ are the Hamiltonians at times $t_1$ and $t_2$,
which are in fact unknown because they depend on the Kohn-Sham states through
the density: we are dealing with a non-linear problem and the CFM expansions
were in fact developed for linear systems. We have various options to extend
them for our non-linear problem: for example, one could define $\hat{H}[t_1]$
and $\hat{H}[t_2]$ as interpolated Hamiltonians from $\hat{H}[t]$ and
$\hat{H}[t+\Delta t]$, in which case we end up with an implicit equation for
$\varphi(t+\Delta t)$ that we would have to solve at a substantial cost.

The alternative that we have implemented, however, is to approximate
$\hat{H}[t_i]$ via an extrapolation from the Hamiltonian at various previous
time steps (in practice, it is the Hartree, exchange, and correlation parts that
must be extrapolated). The resulting method is then explicit, i.e., no linear or
non-linear algebraic equations need to be solved. The fourth order accuracy is
preserved as long as the extrapolation is also done at order four.

\begin{figure}[t]
  \begin{center}
    \includegraphics[width=0.7\columnwidth]{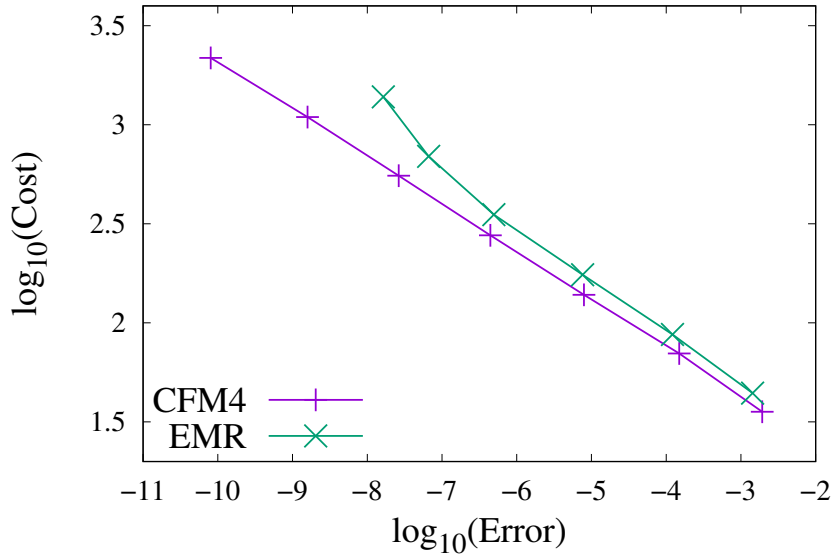}
    \caption{\label{fig:common} Wall-time computational cost of the method
      (seconds) against the error in the wave function (defined as ${\rm
        Error}(T,\Delta
      t)=\sqrt{\sum\limits_{m}\vert\vert\varphi_{m}(T)-\varphi^{exact}_{m}(T)\vert\vert^{2}}$,
      where the $\varphi^{exact}_{m}$ are the values of the Kohn-Sham orbitals
      computed using the explicit fourth-order Runge-Kutta integrator with a
      very small time-step) for the EMR and CFM4 propagators.}
  \end{center}
\end{figure}

As an example, Fig.~\ref{fig:common} shows the performance of this CFM4 method
against the well known exponential midpoint rule (EMR). The benchmark system
consists of a benzene molecule.  It is subject to an instantaneous perturbation
at the beginning of the propagation, and then it evolves freely for some fixed
interval of time.  The propagations are performed at varying values of $\Delta
t$.  The plot displays the cost of the propagation vs. the accuracy achieved in
each case. As we can see, the CFM4 outperforms the EMR for all the values of the
error examined, although the differences become more obvious when higher
accuracy is demanded. The key difference between the two methods is the
fourth-order accuracy of CFM4 scheme -- at the cost of requiring two
exponentials, instead of just one, while the EMR is only second-order
accurate. This is reflected in the different slopes of the curves as the error
becomes smaller, i.e., as $\Delta t \to 0$.


\section{Conjugate gradient implementation in RDMFT}

The RDMFT implementation in Octopus has been described in detail in a previous
paper.~\cite{andrade_real-space_2015} In this Section we briefly review the
existing implementation, providing some new insights, and introduce a recently
implemented method to solve the RDMFT equations that is better suited to the
real-space grids used in the code.

The optimization of the natural orbitals in RDMFT is subject to an
orthonormalization constraint for the orbitals. One minimizes the
functional
\begin{equation}
  E_\mathrm{total}-\sum_{j,k=1}^M \lambda_{jk}\left(\int d\mvec{r} \psi_j^*(\mvec{r})\psi_k(\mvec{r})-\delta_{jk}\right)\,,
\end{equation}
where $E_\mathrm{total}$ and $M$ denote the total energy and the number of
natural orbitals $\psi_j$ included in the calculation,
respectively. $\lambda_{jk}$ are the Lagrange multipliers which ensure the
orthonormality of the natural orbitals at the solution point. We note that, for
a RDMFT calculation, the number of natural orbitals has to be larger than the
number of electrons in the system (core electrons which are included in the
pseudopotential do not count). The exact number $M$ is system-dependent and
should be treated as an additional parameter with respect to which a convergence
study should be carried out, just like it is done for the basis set or the grid
parameters. Typically, the optimization is performed using the so-called Piris
method,~\cite{Piris2009} which was the method previously implemented in
Octopus. Within this approach one uses the orthonormality constraint of the
natural orbitals, which implies that a certain matrix constructed from the
Lagrange multipliers $\lambda_{jk}$ is diagonal at the solution point. As an
immediate consequence of the Piris method, the natural orbitals at the solution
point are linear combinations of the orbitals used as starting point for the
minimization. In other words, the initial orbitals serve as a
basis. Consequently, the necessary matrix elements for different energy
contributions can be calculated for the basis functions (initial orbitals)
before the iterative optimization of the natural orbitals and their occupation
numbers is started. In addition, the optimization of the occupation numbers can
be turned off completely, resulting in a Hartree-Fock calculation in the basis
of the initial orbitals.

\begin{figure}
  \includegraphics[width=0.45\textwidth]{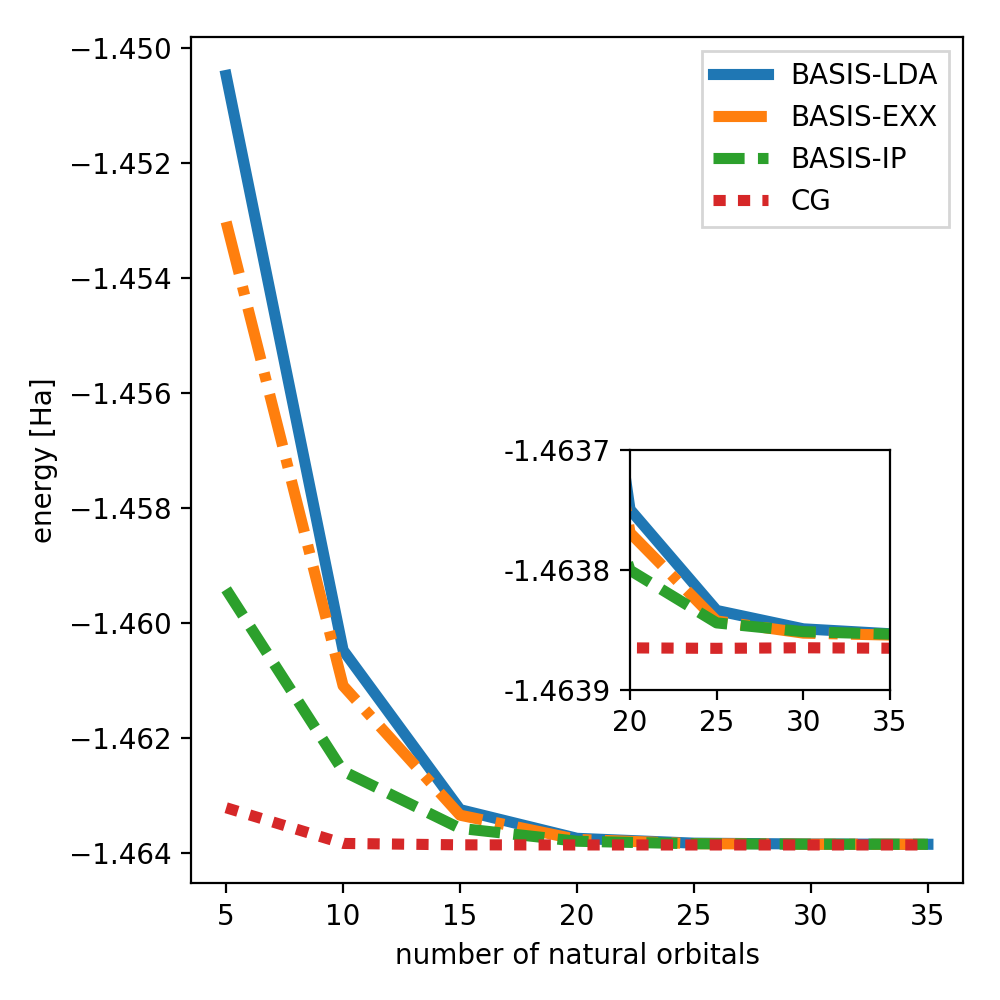}
  \caption{Total energy for the RDMFT calculation for one-dimensional
    H$_2$~\cite{Helbig2011} using the Piris method with different basis sets and
    using the conjugate gradient implementation. The inset shows a zoom into the
    area where convergence is reached. We employ the M\"uller
    functional~\cite{Mueller1984} for all calculations.}
  \label{fig:1dH2}
\end{figure}

For the existing implementation in Octopus of the Piris method, the initial
orbitals are taken to be the solutions obtained with a different level of
theory, like independent particles or DFT. In order to better understand the
effect of the choice of basis, we have tested the following choices: (i)
independent particles, density functional theory within (ii) the local density
approximation (LDA) or (iii) the exact exchange (EXX) approximation, as well as
(iv) the Hartree-Fock approximation. In all cases we have to ensure that the
number of unoccupied states in the calculation is sufficient to cover all the
natural orbitals which will obtain significant occupation in the the following
RDMFT calculation. The results for the convergence of the total energy of a
one-dimensional (1D) hydrogen molecule using the M\"uller
functional~\cite{Mueller1984} are given in Fig.~\ref{fig:1dH2}. The calculations
were performed on a 1D grid extending from $-12.0$ to $12.0$ bohr, with a grid
spacing of $0.03$ bohr. The nuclear potential for the 1D molecule reads
\begin{equation}
  v(x) = -\frac{1}{\sqrt{(x-d)^2+1}} -\frac{1}{\sqrt{(x+d)^2+1}}\,,
\end{equation}
with $d=1.628$ bohr, which corresponds to the equilibrium geometry. The
electron-electron interaction in one dimension is described by the soft-Coulomb
interaction
\begin{equation}
  w(x,x') = \frac{1}{\sqrt{(x-x')^2+1}}\,.
\end{equation}

Since Octopus performs all calculations on a finite grid, we typically obtain a
finite number of bound states in the calculations for the basis set and any
additional orbitals extend over the whole grid, i.e., they are unbound and
therefore delocalized. However, all natural orbitals with non-zero occupation,
because they decay with the ionization potential of the
system,~\cite{Asymptotics_exact_NOs} are localized on the system. Hence, the
extended basis states will only contribute with very small coefficients, if at
all, and their inclusion in the basis set does not lead to a significant
improvement of the results. In the past, this problem was addressed by
performing an additional step to localize the initial states before starting the
RDMFT calculation. However, further investigations showed that this only
improves the results for a small number of natural orbitals in the
calculation. Testing the convergence with respect to the number of natural
orbitals then showed that the additional localization step slows down the
convergence with respect to the number of basis functions. The fastest
convergence and lowest total energies are obtained by using the results from an
independent particles calculation, as those yield the largest number of
localized orbitals from all the different basis sets that were tested, as shown
in Fig.~\ref{fig:1dH2}. As the additional localization step proved to be
unnecessary and even hinders convergence, it has been removed from the new
version of the code.

Since the natural way of representing quantities in Octopus is directly on the
real-space grid, and to circumvent the limitations with the quality of the basis
sets available for the Piris method, we have decided to implement a conjugate
gradient optimization of the natural orbitals. This implementation follows the
procedure for DFT explained in Ref.~\onlinecite{Payne1992}, which we adapted for
RDMFT. This procedure allows us to take advantage of the full flexibility of a
real-space grid and provides a systematic way of improving the results by
enlarging the grid and reducing the spacing between grid points. The conjugate
gradient algorithm requires a set of initial orbitals to start the
self-consistent calculation, however, at convergence the results are independent
of that starting point. Therefore, while the calculation using the Piris method
requires a set of initial states which serve as the basis, the conjugate
gradient algorithm can be used starting from a initial set of random states. In
our tests of the conjugate gradient implementation, the quality of the initial
states only had an influence on the number of iterations necessary for the
convergence, but not on the final result. We suggest to use the orbitals
obtained from an independent particles calculation as initial states since they
can be obtained for small numerical costs and simultaneously can serve as a
basis set in the Piris implementation.

Since we are not solving an eigenvalue equation to obtain the natural orbitals,
they are not automatically orthogonal. As mentioned above, the orthogonality is
taken into account via a constraint. Compared to Ref.~\onlinecite{Payne1992}, the
non-diagonal Lagrange multipliers $\lambda_{jk}$ lead to two modifications in
the conjugate gradient procedure. First, the steepest-descent direction
(cf.\ Eq.\ (5.10) of Ref.~\onlinecite{Payne1992}) reads here
\begin{align}
  \zeta_i^m = -\hat{H}\psi_i^m + \sum_{k=1}^{M}\lambda_{ik}\psi_k^m\,,
\end{align}
where
\begin{align}
  \lambda_{ik}=\langle \psi_i^m|\hat{H}\psi_k^m\rangle\,.
\end{align}
Second, one parameter of the line-minimization (cf.\ Eq.\ (5.26) of
Ref.~\onlinecite{Payne1992}) is changed to
\begin{align}
  \frac{\partial E_\mathrm{total}}{\partial \Theta}
  = \braket{\phi_i'{}^m|\hat{H}\psi_i^m}+\braket{\psi_i^m|\hat{H}\phi_i'{}^m} - \sum_{k}\left(\lambda_{ki} \braket{\phi_i'{}^m|\psi_i^m}+\lambda_{ik}\braket{\psi_i^m|\phi_i'{}^m}\right)\,,
\end{align}
where $\Theta$ parameterizes the descend in the direction of the gradient and the single-particle Hamiltonian acting on a state $\ket{\phi}$ is defined
as
\begin{align}
  \hat{H}\ket{\phi}=\frac{\partial E_\mathrm{total}}{\partial \phi^*}.
\end{align}
For details of the notation, we refer the reader to the paper by Payne {\it et
  al}.~\cite{Payne1992}

The convergence study with respect to the number of natural orbitals for the
conjugate gradient algorithm is also included in Fig.~\ref{fig:1dH2}. As one can
see, a smaller number of natural orbitals needs to be included in the
calculation than for the basis set implementations with the Piris method. As the
number of natural orbitals equals the number of basis functions in these
calculations, this is mostly due to the fact that the available basis sets are
of rather poor quality. In addition, the converged total energy for the
conjugate gradient algorithm is slightly lower than for all the basis sets using
the Piris method (see inset of Fig.~\ref{fig:1dH2}), which shows that the
conjugate gradient algorithm exploits the full flexibility of the grid
implementation, allowing for contributions to the natural orbitals which are not
covered by any of the basis sets. We have also verified that the converged
result of the conjugate gradient method is indeed independent of the choice of
initial state for the M\"uller functional~\cite{Mueller1984} employed in our
calculations. The M\"uller functional is known to be convex when all infinitely
many natural orbitals are included.~\cite{FLSS2007} This property is most likely
not shared by all available RDMFT functionals. In addition, the number of
natural orbitals in any practical calculation is always finite. Consequently, in
practice one needs to test the convergence for the different starting points, as
the appearance of local minima cannot be excluded.

\section{Periodic systems and symmetries}

Electronic structure in periodic systems is usually described using plane waves,
but real-space grids have been shown to be a viable alternative when performing
DFT and TDDFT calculations.~\cite{PhysRevLett.72.1240,biry00:7998}
Unfortunately, the discretization introduced by the real-space grid often breaks
the direct connection between the physical system and the basis set, as
symmetries and translations are, in most of the cases, not compatible with the
discretized grid. However, real-space grids offer many advantages, such as
natural mixed periodic-boundary conditions for semi-periodic systems, and the
calculation of the exchange and correlation term of DFT is straightforwardly
obtained on the grid.

One of the main challenges for treating periodic systems in real space is the
generation of the real-space grid, and the corresponding weights for the finite
differences. This becomes relevant when the grid is generated along the
primitive axes of the Wigner-Seitz cell (primitive cell) of a solid, where the
generating axes are usually non-orthogonal.

In Octopus, the grid points are generated along the primitive axes, and the
calculation of the finite-difference weights follows the implementation
described in Ref.~\onlinecite{PhysRevB.78.075109}. Once the grid is generated,
and the weights for the finite differences (gradient and Laplacian) are
obtained, it is necessary to deal with the discretization of the reciprocal
space. Indeed, for periodic systems, the full crystal is replaced by the
primitive cell of the crystal in real-space, thanks to the Bloch theorem, but
is then complemented by the Brillouin zone, which also needs to be sampled. This
grid, usually called $\mathbf{k}$-point grid, is common to any type of basis sets,
as long as one decides to reduce the crystal to its primitive cell.

In order to reduce the numerical effort associated with the description of
periodic systems, we make use of the symmetries to reduce the Brillouin zone to
its irreducible Brillouin zone, which can drastically reduce the number of
$\mathbf{k}$-points. The space group of the crystal, as well as its symmetries,
are obtained thanks to the spglib library.~\cite{1808.01590} In the
present implementation, we restrict the symmetries to the symmorphic symmetries
(inversion, rotations and mirror planes), leaving for later the so-called non-symmorphic
symmetries, i.e., symmetries involving a fractional translation, as they are not
compatible with arbitrary real-space grids.

In order to assess the validity of our implementation, we used the so-called
Delta-factor test.~\cite{lejaeghere2016reproducibility} Using the Schlipf-Gygi
ONCVPSP 2015 pseudopotential set, we obtained the value of $1.50$ meV/atom,
which is very close to the value obtained by other codes using the same
pseudopotential set.

When investigating the electron dynamics driven by a laser field, or any type of
symmetry-breaking perturbation (vector potentials, kicks with a finite momentum,
strain, etc), some of the original symmetries are lost. To deal with that,
Octopus finds the small group of symmetries corresponding to the original
space-group of the solid, retaining only the symmetries that leave invariant the
perturbation direction. This defines the symmetries that are used for a
time-dependent calculation.

One important aspect when using symmetries, is the symmetrization (in
real-space) of the charge and current densities, as well as other observables,
such as kinetic energy density for instance. In complement to the reduction of
the Brillouin zone to its irreducible Brillouin zone, we implemented a
real-space symmetrization. We found that performing the real-space
symmetrization is very important to achieve good and stable numerical results
when taking into account symmetries.

For a comprehensive description of periodic systems, the ion dynamics has to be
considered in addition to the electron dynamics. For isolated systems, it is
often described with the TDDFT-Ehrenfest dynamics
method,~\cite{MARQUES200360,Alonso:2008fz} where the ions obey the Newton
equation with force fields computed by the TDDFT. In contrast, for periodic
systems, the ion dynamics has to be described with two sets of equations: One is
the Newton equation for ions on the reduced coordinates in the primitive
cell, and the other is the equation of motion for the primitive cell itself,
as the ionic coordinates of periodic systems are described by the combination of
the reduced coordinates with the lattice vectors of the primitive cell. The
lattice dynamics is often treated with the Parrinello-Rahman
method,~\cite{PhysRevLett.45.1196,doi:10.1063/1.328693} and the key ingredient
of the equation of motion is the stress tensor. Therefore, to realize the
\textit{ab initio} simulations for electron-ion-lattice dynamics based on the
TDDFT, the calculation of the stress-tensor has been implemented based on
Ref.~\onlinecite{martin2004electronic}, and the implementation of the lattice
dynamics is under way.

\section{Additional technical code improvements}
\label{sec:technical}

In order to make all the new developments and applications described in the
above sections possible, the code needs to be accurate, efficient, and
reliable. Also, when a code reaches the size and complexity of Octopus
(currently more than 200,000 lines of source code), the amount of time required
for maintenance and for adding new features becomes considerable. Therefore,
continuous efforts at optimizing, validating, and improving the source code
quality are needed. These efforts are essential, but often do not get the
attention they deserve. In this section we present some noteworthy developments
to improve the code reliability and efficiency, to make the project easier to
maintain, and to tackle new computer architecture challenges.

\subsection{Web application for analyzing regression tests}
In a complex code like Octopus, every new development may have unintended side
effects, possibly introducing errors to already existing features. To avoid such
a regression, every change to the code is required to pass a suite of tests
before being accepted. This suite is executed automatically using
Buildbot~\footnote{\url{http://buildbot.net}} and is integrated with the
continuous integration framework of the gitlab platform where Octopus is hosted
for several years now. The testsuite is executed with 30 different toolchains
spanning a variety of architectures, compilers, and MPI libraries. It contains
currently about 180 tests that execute Octopus roughly 740 times and that
contain about 12000 comparisons to reference values. The overall coverage of the
tests is determined using the \texttt{codecov.io} service and lies at about 71\%
at the moment.

Although the testsuite is efficient in avoiding regressions, analyzing why a
test failed has been difficult so far due to the large amount of data
generated. Thus, we have implemented an interactive analysis and visualization
of the testsuite results as a web application available at
\url{https://octopus-code.org/testsuite/}. This application makes it easier to
understand why a test failed: is it a problem with just one toolchain, \textit{e.g.}, a
particular architecture, or is it simply a larger numerical variation of the
results? Moreover, it facilitates updating the tests and improving the testsuite
itself.

The application has been implemented using the web framework Django coupled to a
Postgres database. The testsuite results are automatically uploaded by the
Buildbot service as soon as they are available. The application allows to
analyze single toolchain runs and single comparison matches and to compare all
toolchain runs for a single commit in git. It provides histograms to judge if
there are outliers or if it is a broad distribution; moreover this allows to
determine, \textit{e.g.}, if there is a difference between MPI and non-MPI toolchains.
This application has already helped to identify bugs causing regressions and
will continue to be useful to understand failed tests, update them, and to
improve the testsuite.

\subsection{Improving ground-state calculations}

To improve the reliability of ground-state calculations with Octopus, the
default eigensolver used inside the self-consistent field (SCF) cycle has been
improved and different real-space preconditioners for the eigensolvers have been
evaluated.

The default eigensolver, a conjugate-gradients algorithm, has been improved over
the previous implementation by now following closely
Ref.~\onlinecite{Payne1992}, which greatly improves the reliability of the
solver. The updated implementation differs from Ref.~\onlinecite{Payne1992} in
only one point: by default, the current band is not orthogonalized against all
bands, instead it is only orthogonalized against previously computed bands with
lower energies. According to our tests, in Octopus, this leads to a faster
convergence for most cases.

Preconditioners for eigensolvers are an integral part of SCF calculations
because they greatly accelerate convergence. They achieve this by applying an
approximate inverse of the Hamiltonian in each iteration that leaves the
solution invariant and brings the system closer to the real solution.  As
Octopus uses a real-space grid (as opposed to plane waves like in
Ref.~\onlinecite{Payne1992}), different preconditioners are needed. A range of
preconditioners has been compared for a wide variety of systems, comprising
molecules, semiconductors, metallic systems, and surfaces to test convergence
for very different cases.

The default preconditioner in Octopus is a low-pass filter obtained by adding
the neighbouring values in each dimension to the current value at a grid point,
weighted by a certain factor $\alpha$
\begin{equation}
  \psi'_{i,j,k} = \alpha \psi_{i,j,k} + (1-\alpha) \left( \psi_{i-1,j,k} +
  \psi_{i+1,j,k} + \psi_{i,j-1,k} + \psi_{i,j+1,k} + \psi_{i,j,k-1} +
  \psi_{i,j,k+1} \right)\,.
  \label{eq:filterpreconditioner}
\end{equation}
This preconditioner was first described by Saad \text{et al.}~\cite{saad1996}
with $\alpha=0.5$. It has proven to be the most effective preconditioner in our
comparison because it is quite cheap to apply and it nevertheless decreases the
number of SCF iterations noticeably. It can also be understood as two weighted
Jacobi iterations (up to a prefactor) to solve for the inverse of the kinetic
term of the Hamiltonian (i.e.,\ $0.5\Delta\psi'=\psi$). From the convergence
radius of the Jacobi iterations, we can conclude that the allowed values for
$\alpha$ are between 0.5 and 1. From theoretical considerations of damping of
different spatial wavelengths during Jacobi iterations (see
Ref.~\onlinecite{meister2011numerik}, Secs.~4.1.3, 4.2), the ideal value should
be 0.75 to most effectively damp high spatial frequencies.  In practice, we find
that this depends on the system. Moreover, increasing the number of Jacobi
iterations does not make the preconditioner more effective.  Also, using a
single Jacobi iteration (i.e., dividing by the diagonal of the Laplacian) does
not speed up the convergence significantly.

A multigrid method similar to the one implemented in GPAW~\cite{gpaw1_2005} also
uses Jacobi iterations to solve for the kinetic Hamiltonian, but employs
different grids to reach a faster decrease of the error. Although this method
reduces the number of SCF iterations, it is computationally more expensive
because the Laplacian is applied several times for each iteration and thus less
effective than the filter preconditioner.

A preconditioner specifically targeting real-space methods was proposed by
Seitsonen \textit{et al.}~\cite{seitsonen1995} It uses the ratio between the
difference of the energy and the potential to the kinetic energy (their Eq.~(3))
with a preconditioning function (their Eq.~(4)) originally from
Ref.~\onlinecite{Payne1992}. However, this preconditioner did not speed up
convergence significantly when used in Octopus.

Another way to get an approximate inverse of the Hamiltonian is to solve the
Poisson equation associated with the kinetic part of the Hamiltonian. This
reduces the number of iterations needed to converge the SCF calculation, but it
increases the total calculation time, as each iteration is much more costly.

In summary, we find that for our real-space code, the filter preconditioner is
most effective in reducing the total time needed to compute ground states
because it reduces the total number of iterations without being computationally
too expensive.

Nonetheless, sometimes certain states cannot be fully converged using a
preconditioner, especially for calculations with many unoccupied states. In this
case, restarting without preconditioner is needed to obtain full SCF
convergence. The reasons are still under investigation.

\subsection{Novel multi-system framework}
\label{subsec:multisystem}

In the way Octopus was originally designed, the entire code was structured
around the idea that there was only one system (albeit of arbitrary size and
complexity) with an associated Hamiltonian. This was typically a combined system
of electrons and nuclei, where the former were described using DFT and the latter
were treated classically. All possible sorts of interactions of this system with
some external source (e.g., external electromagnetic fields, solvents described
with the PCM, etc) were then added in an \textit{ad hoc} fashion. Although this approach
worked well for many features and applications, its limitations became evident
with several recent developments, in particular the coupled Maxwell-Kohn-Sham
equations described in Sec.~\ref{sec:maxwell}, where the Kohn-Sham orbitals need
to be time-propagated alongside the Maxwell fields.

This has prompted us to start a major refactoring of the code, where the full
physical system is treated as several subsystems interacting with each
other. Such subsystems can be electrons, nuclei, Maxwell fields, etc. While this
framework was mainly motivated by the Maxwell-TDDFT coupling, it has been
implemented very generically so that all the existing features can be converted
to it and that, in the future, many other developments which require the
coupling of several subsystems can be based on this.

\subsection{Memory layout}

A new memory layout for storing the orbitals was introduced in
Ref.~\onlinecite{andrade_gpu_2013}, where all states are stored in a number of
smaller batches with the innermost index being the state index instead of the
real-space grid index. Thus, for all states in a batch, exactly the same
operations can be executed when looping over the grid while accessing memory
contiguously. This allows efficient parallelization on GPUs, where all threads in a
warp need to execute the same instructions, as well as vectorization on CPUs,
where one instruction can operate on several data points. To fully utilize
these instructions, the kernel for computing finite differences (e.g., the
Laplacian) has been specialized to explicitly use SSE, AVX, or AVX512
instructions. Now, more parts of the code have been ported to use this new
layout to increase their performance and, whenever possible, the code will use
this layout by default.

\subsection{GPUs}

The first GPU implementation has been described in
Ref.~\onlinecite{andrade_gpu_2013}. This was based on OpenCL and was limited to
a small selection of numerical algorithms and calculation modes. These included
some of the most commonly used features of the code or algorithms that were
particularly suited for GPU porting, like the RMM-DIIS eigensolver for
ground-state calculations or the enforced time-reversal-symmetry propagator for
time-dependent calculations. Since then, the GPU implementation has been
expanded in several different ways. Most notably, it now supports CUDA through
an additional compatibility layer. The number of supported features and
algorithms has also been increasing steadily, such that most time-dependent
calculations can now be run efficiently on GPUs. The implementation has also
been expanded to support multiple devices per host. Using the packed storage
format described above, Octopus is able to store the states fully in the GPU
memory, provided the memory is large enough, thus reducing memory transfers to a
minimum. Recently, this was improved further by removing frequent allocations
and deallocations of temporary variables on the GPU, now using a custom memory
management for those. This proved very effective in improving the scaling to
several GPUs, also to several nodes with GPUs.

\begin{figure}[tbp]
  \centering
  \includegraphics[width=\textwidth]{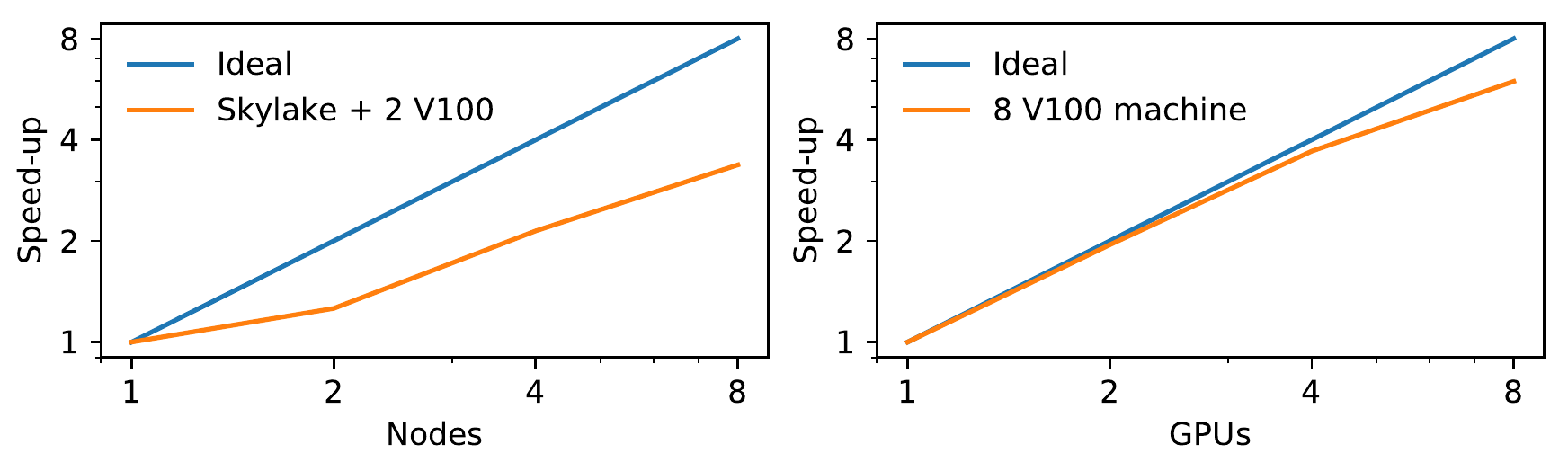}
  \caption{Scaling plots for the GPU implementation of Octopus. The left panel
    shows speed-up for a machine with two 20-core Intel Xeon 6148 Gold sockets
    (Skylake architecture) and two V100 GPUs (PCIe) per node. The right panel
    shows speed-up for a Supermicro server with 8 V100 GPUs (NVLink).}
  \label{fig:gpuscaling}
\end{figure}

We show two examples of time-dependent runs in Fig.~\ref{fig:gpuscaling}. For
the first example (left panel), $\beta$-cyclodextrine
was used as an input system and the simulation was run on the GPU island of the
COBRA supercomputer at the Max Planck Computing and Data Facility. Each node in
this island consists of two 20-core Intel Xeon 6148 Gold sockets (Skylake
architecture) with two Nvidia Volta 100 GPUs (PCIe); the nodes are connected with
an Omnipath fabric (100 Gbit/s). When executed on a full node with GPUs, the
average time step for the example is a factor of 4.8 faster than on one full
CPU node, i.e., the time to solution is reduced by a factor of 4.8. This also
means that the GPU version consumes less energy: although one GPU node draws
more power than a CPU node (ca. 950\,W vs 450\,W), the faster execution time
reduces the overall energy to solution by a factor of 2.3 when running on a GPU
node. When scaling to 2, 4, and 8 nodes, the speed-up is 1.26, 2.14 and 3.37,
respectively, and the corresponding parallel efficiency is 63\%, 54\%, and
42\%. Although the scaling is not yet perfect, it has improved considerably and
also hints at the need for improving inter-node communication. For the second example
(Fig.~\ref{fig:gpuscaling}, right panel), a time-dependent simulation of a part
of a chlorophyll complex was executed on a Supermicro server with two 8-core
Intel Xeon 6134 Gold CPUs (Skylake architecture) and eight Nvidia Volta 100 GPUs
interconnected with NVLink. Here, the speed-up of the average time step of the
simulation when using 2, 4, and 8 GPUs is 1.95, 3.70, and 5.97 which
corresponds to parallel efficiencies of 97\%, 93\%, and 75\%. These
efficiencies are much better than for the multi-node example, probably because
data is only communicated within one node. For both examples, all states were
stored in the GPU memory, and the parallelization was achieved by distributing
the states only.  Distributing the grid (i.e., parallelization over domain) is
not as effective, because it leads to more frequent communication, and thus has
a larger overhead.

Our current and future efforts regarding the GPU implementation are focused on
two aspects. First, we are planning to enhance the existing implementation by
improving the communication, possibly by overlapping computation and
communication, and by optimizing the existing kernels. Second, we want to port
more algorithms that are currently only implemented for the CPU version,
especially for spin-dependent calculations and ground-state runs. This is a
long-term effort and the ultimate goal is to have as many features and
algorithms as possible running efficiently on GPUs.

\section{Conclusions}
\label{sec:conclusion}

It has been almost 20 years since the development of Octopus started. During
this period the code has matured and expanded to cover an ever-growing range of
methods, theories, applications, and systems. It has also continuously adapted
to the available computer architectures and computing paradigms, allowing
researchers to tackle increasingly challenging problems in electronic structure
theory.

Although many of the code capabilities have been routinely used by many groups
around the world, mainly to study electronic excited states properties and
dynamics, which still remain to a large extent the core of Octopus, we believe
the main reason for the code's success lies elsewhere. From the beginning, the
code has been designed to take full advantage of the flexibility and versatility
offered by the use of real-space grids and provide developers with a framework
to easily implement and test new ideas and methods that can be later on adopted by
other codes (as it has already been the case with quite a few features
previously developed within Octopus). This is demonstrated by the number of new
theoretical methodologies and frameworks presented in this paper to deal with
non-equilibrium phenomena of complex systems and to address the combined
dynamics of electrons, phonons, and photons that go beyond what can be found in
other electronic structure codes. This includes several novel approaches to
treat the coupling of the electronic systems to the photons, like the coupled
Maxwell-Kohn-Sham equations, the OEP approach to the electron-photon coupling,
and the dressed RDMFT. Other examples include the description of magnons in
real-time and the orbital magneto-optical response in solids and molecules using
the Sternheimer approach. In the case of magnons, supercells need to be
employed, and the scalability of real-space grid methods make this approach very
promising for future applications in more complex correlated magnetic
systems. We expect that many of these methods and approaches will be integrated
into other electronic-structure codes and will become standard tools in the near
future.

The code flexibility is also demonstrated by the variety of systems that it can
efficiently treat, as the applications showcased in this paper include molecules,
nanoparticles, model systems, solvents, solids, monolayers, etc. Particularly
noteworthy is the efficient treatment of periodic systems, which traditionally
have been described using plane-wave basis-sets, and this has been achieved
without loss of accuracy, as shown by our results for the Delta-factor test. In
the end, it is our hope that combining this flexibility with a growing range of
methodologies will provide researchers with the necessary tools to study new
challenging phenomena, like novel correlated materials, out of equilibrium
physics, or coupled electron-boson systems.

We have also discussed recent improvements in performance and scalability, with
particular emphasis on the GPU support. These developments are crucial in the
view of the new challenges that electronic-structure applications are facing
with the upcoming exaflop supercomputers.

Finally, several other implementations are in the pipeline, such as Floquet and
cavity QED materials engineering, multitrajectory methods to deal with the
nonadiabatic electron-ion dynamics, treatement of open quantum dissipative
systems, spectroscopies with entangled photons, etc. All of these should be
available to the users in the next years, so we invite you to stay tuned to the
Octopus webpage at \url{https://octopus-code.org/}.

\begin{acknowledgments}
The authors would like to thank all the people that have contributed to the
development of Octopus over the last two decades. They would also like to thank
Lin Lin for useful and interesting discussions and acknowledge the open
discussions about real space methods with the group of Prof. Chelikowsky. This
work was supported by the European Research Council (ERC-2015-AdG694097), the
Cluster of Excellence ``Advanced Imaging of Matter''(AIM), Grupos Consolidados
(IT1249-19) and SFB925. The Flatiron Institute is a division of the Simons
Foundation. XA, AW and AC acknowledge that part of this work was performed under
the auspices of the U.S. Department of Energy at Lawrence Livermore National
Laboratory under Contract DE-AC52-07A27344. JJS gratefully acknowledges the
funding from the European Union Horizon 2020 research and innovation program
under the Marie Sklodowska-Curie Grant Agreement No. 795246-StrongLights. JF
acknowledges financial support from the Deutsche Forschungsgemeinschaft (DFG
Forschungsstipendium FL 997/1-1). DS acknowledges University of California,
Merced start-up funding.

\end{acknowledgments}

\bibliography{biblio}

\end{document}